%% file: main.tex
\documentclass[11pt,a4paper]{article}
\usepackage[T1]{fontenc}
\usepackage[utf8]{inputenc}
\usepackage[english]{babel}
\usepackage{lmodern,microtype,amsmath,amssymb,booktabs,tabularx,longtable,array}
\usepackage[margin=23mm]{geometry}
\usepackage{graphicx,placeins,capt-of}
\usepackage[round,authoryear]{natbib}
\usepackage{xurl}
\usepackage{listings}
\usepackage[colorlinks=true,linkcolor=blue,citecolor=blue,urlcolor=blue]{hyperref}
\newcommand{\ks}{\ensuremath{\kappa^1\,\mathrm{Ceti}}}
\newcommand{\lya}{\ensuremath{\mathrm{Ly}\alpha}}

\newcommand{\fluxunit}{\ensuremath{\mathrm{erg\,s^{-1}\,cm^{-2}}}}
\newcommand{\cgs}{\ensuremath{\mathrm{erg\,s^{-1}\,cm^{-2}}}}
\newcommand{\arcsec}{^{\prime\prime}}
\setcitestyle{authoryear,round,aysep={},yysep={;}}
\hypersetup{pdftitle={The PHAROS Spectral Library: Data Processing, Spectral Reconstruction, and Uncertainty Propagation},pdfauthor={Viktor Sumida}}
\title{The PHAROS Spectral Library: Data Processing, Spectral Reconstruction, and Uncertainty Propagation\\[5pt]\large A technical companion for fifteen solar-like stars}
\author{Viktor Sumida$^{1}$}
\date{}

\providecommand{\doi}[1]{\href{https://doi.org/#1}{doi:\nolinkurl{#1}}}
\begin{document}
\maketitle
\begin{center}\small
$^1$Center for Radio Astronomy and Astrophysics Mackenzie (CRAAM),\\
Universidade Presbiteriana Mackenzie, S\~ao Paulo, Brazil\\
\href{mailto:sumyda@proton.me}{\texttt{sumyda@proton.me}}
\end{center}
\begin{abstract}
We document the data reduction, reconstruction choices, and available
uncertainty propagation underlying fifteen PHAROS stellar spectral energy
distributions from 0.5 to 5000 nm. A catalogue timeline and a practical FITS guide introduce the available
products, their column and extension conventions, and examples of reading,
integration, and uncertainty display. Shared methods are presented once,
including flux geometry, support-aware integration, X-ray likelihoods,
the 2025 X/EUV luminosity relation, intrinsic Ly$\alpha$ inference,
airglow treatment, spectral junctions, and covariance-aware resampling.
Target sections record the adopted observations and alternatives,
including spatially separated binary components, unresolved system-level
EUV measurements, direct EUVE constraints, optical stellar-twin transfer,
and literature-anchored cases where new spectral inference is unsupported.
The distinction between native measurement errors, conditional model
intervals, empirical predictive scatter, and physical-method sensitivity
is maintained. Signed measurements and nonnegative irradiation products
are distinguished. Each target section concludes with the final nominal
spectrum and its available uncertainty bounds on common axes. This library
provides the detailed provenance and interpretation needed to accompany
the catalogue paper and its machine-readable products.
\end{abstract}
\tableofcontents
\clearpage
\input{sections/common}
\input{sections/18_Sco}
\input{sections/kappa1_Ceti}
\input{sections/EP_Eri}
\input{sections/pi1_UMa}
\input{sections/Xi_Boo_A}
\input{sections/10_Tau}
\input{sections/alpha_Cen_A}
\input{sections/beta_Com}
\input{sections/upsilon_And}
\input{sections/47_UMa}
\input{sections/iot_Hor}
\input{sections/tau_Boo}

\input{sections/70_Oph_A}
\input{sections/36_Oph_B}

\input{sections/lambda_Ser}
\clearpage
\phantomsection
\addcontentsline{toc}{section}{References}
\bibliographystyle{pharos}
\bibliography{references}
\end{document}

%% file: sections/common.tex
\section{Purpose, scope, and reading guide}\label{sec:scope}
PHAROS supplies panchromatic irradiation spectra for nearby solar-like stars.
This manual documents the construction and revision of the fifteen
released target products over 0.5--5000 nm. It complements the catalogue paper
by retaining the source-selection decisions, extraction tests, line-profile
assumptions, calibration comparisons, and uncertainty representations needed
to use the spectra. It describes the adopted products and consequential
alternatives rather than every intermediate exploratory calculation.

The spectra combine observations obtained with different instruments and at
different epochs. They represent specified composites, not simultaneous
snapshots or predictions of a future stellar state. A continuous delivered
SED can contain direct measurements, reconstructed intrinsic emission,
empirical band estimates, transferred templates, and atmosphere-model
extensions. These categories are retained in the FITS provenance. In
particular, filling an airglow mask with a proxy does not recover a new
observation of that line.

The practical FITS guide in Section~\ref{sec:data_guide} explains product
selection, columns, uncertainties, and example calculations before the shared
methods in Section~\ref{sec:common}. Each subsequent section gives the
target-specific decisions and ends with its released spectrum and available
uncertainty bounds. These figures use the same wavelength and surface-flux
limits for comparison. The notes preserve meaningful discrepancies with
published results and distinguish comparisons using the same photons from
independent observational tests. ``Released'' refers to a reproducible data
product and does not imply that every wavelength is directly observed or
that every physical source of model discrepancy has a measured distribution.

\begin{figure}[htbp]
\centering
\includegraphics[width=\textwidth,height=.64\textheight,keepaspectratio]{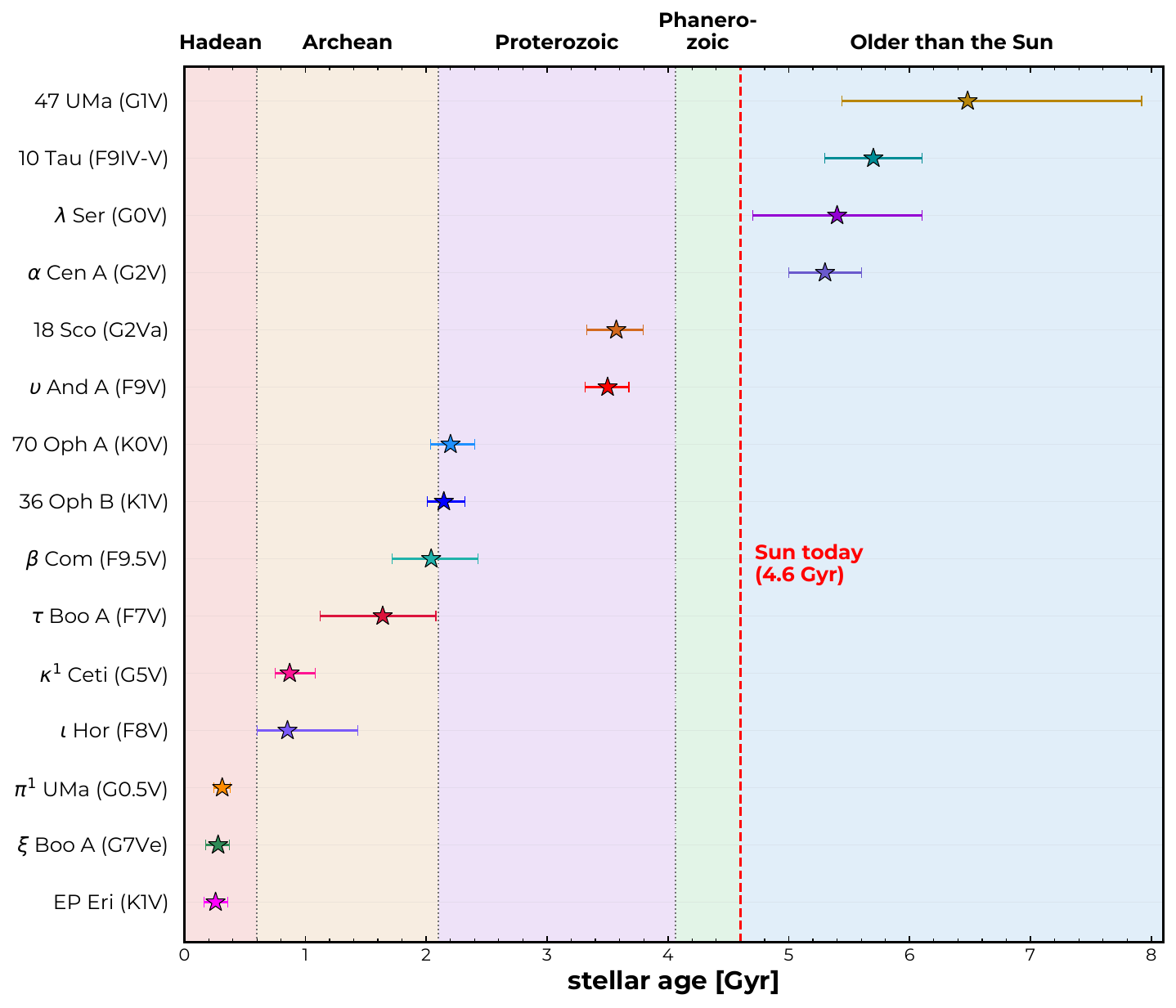}
\caption{Age context for the fifteen PHAROS targets, increasing from the lower
left to the upper right. Horizontal bars retain the catalogue age
uncertainties. The shaded bands show terrestrial eons on an approximate
solar-age coordinate, and the dashed line marks the 4.6 Gyr solar reference
used in this display. These are contextual labels, not universal stellar
activity stages or inferred planetary geological transitions. The sample
contains different spectral types, whose evolution need not follow the same
age dependence.}\label{fig:manual_timeline}
\end{figure}

Figure~\ref{fig:manual_timeline} places the fifteen targets on the catalogue's
age axis before the data-access and reconstruction details. The terrestrial
eon labels retain the original PHAROS organizational context: they do not
assign a geological history or atmospheric state to another planet. The
stellar ages shown are the catalogue display values; target-specific parameter
sets and any alternative age determinations are identified in the individual
sections. These labels do not enter the flux calibration or reconstruction.

\FloatBarrier
\clearpage
\input{sections/data_guide}
\clearpage
\section{Shared methodology and uncertainty conventions}\label{sec:common}
This section defines the conventions shared by the target reconstructions:
source identity, wavelength support and flux geometry, high-energy and
ultraviolet inference, spectral junctions, and uncertainty propagation.
The individual stellar sections identify which choices and data apply in
each case; the FITS guide above shows how those choices appear in the files.

\subsection{Identity, components, epochs, and quality}
Native target identifiers, coordinates, detector/aperture information, and
observing metadata establish source identity before spectra are combined.
For close binaries this includes the assignment of spatial traces and the
distinction between a resolved component and an unresolved system. An
aperture count fraction is not automatically a wavelength-independent
component flux fraction. When the data support spatial separation, source
and background profiles are fitted and their extraction dependence is tested.
When they do not, the system measurement is retained as A+B evidence and is
not silently relabelled as component A. Examples are shown in
Sections~\ref{star:Xi_Boo_A}, \ref{star:tau_Boo}, and \ref{star:36_Oph_B}.

Pipeline errors and native data-quality arrays are propagated through the
adopted selections. Flags for a model or repaired interval do not mean the
same thing as detector-quality flags. Poor acquisitions, missing support,
contaminated apertures, background excursions, and calibration differences
are investigated explicitly. No universal percentage is assigned to
variability between epochs. A comparison across epochs is identified as
such and is not treated as an independent noise term without a model.

\subsection{Units, bin support, and stellar geometry}\label{sec:geometry}
Wavelengths are in nm and flux densities in
$\mathrm{erg\,s^{-1}\,cm^{-2}\,nm^{-1}}$. Conversion from a per-Angstrom
density multiplies the numerical flux density by ten. Every adopted bin has
explicit lower and upper boundaries; output sampling is not a statement of
instrumental resolving power. Surface, Earth, and orbital-distance fluxes obey
\begin{equation}
 F_{\lambda,\star}=G F_{\lambda,\oplus},\qquad
 G=(d/R_\star)^2,\qquad F_\lambda(a)=F_{\lambda,\oplus}(d/a)^2,
 \label{eq:geometry}
\end{equation}
with consistent distance units. If an angular diameter is the primary
constraint, $G=4/\theta_{\rm rad}^2$. Radius and distance must then not be
perturbed independently. A geometrical realization is shared by all affected
wavelengths, including any atmosphere or proxy that depends on the same
stellar parameter. The target sections identify fixed-geometry bounds,
separate geometry modes, and band products that include geometry.

For a band $B=[\lambda_1,\lambda_2]$, supported integration uses
\begin{equation}
 F_B=\sum_{i\in\mathrm{valid}}w_i F_i,\qquad
 w_i=\max\{0,\min(\lambda_{i,+},\lambda_2)-\max(\lambda_{i,-},\lambda_1)\}.
 \label{eq:integral}
\end{equation}
The supported fraction is $\sum_iw_i/(\lambda_2-\lambda_1)$.
Missing support is not zero physical flux. Summing flux density times actual
bin width is preferable to trapezoids through the centres of unequal bins.

\subsection{X-ray inference and exact-band comparisons}\label{sec:xray}
Where target-specific count spectra support inference, absorbed thermal
plasma models are folded through the matching responses and fitted jointly
with their backgrounds. Poisson likelihoods are used in the low-count
regime. The W statistic profiles background intensities; sampling a profiled
likelihood with priors is not the same as fully marginalizing the background.
Grating orders 1--8 are included where LETG/HRC-S order overlap requires
them. Source/background regions and component leakage are checked before
interpreting plasma parameters.

The number of temperatures, abundance pattern and reference mixture,
absorbing column, APEC version, fitted energy band, and exported wavelength
band are target-specific. A lower statistic by itself is not sufficient to
adopt an unidentified component. Comparisons of models on identical data
are distinguished from tests that change extraction, masking, or grouping.
The adopted nominal spectrum is the selected fit or stated literature model;
an ensemble mean or median remains a distributional diagnostic unless
explicitly adopted in that release.

The exact 0.5--10 nm X band corresponds to approximately 0.124--2.480 keV.
It is different from the frequently reported 0.1--2.4 keV ROSAT band and
can extend outside the energies directly constraining a fit. Such portions
are thermal-model extrapolations. Imaging-only or very weak observations
may instead require a published band-flux anchor and a conditional shape;
these cases are identified rather than described as new resolved thermal fits.

\subsection{EUV normalization, shape, and observational constraints}\label{sec:euv}
Where an X-ray-based normalization is selected, we use equation (2) of
\citet{Sanz-Forcada.et.al.2025A&A...693A.285S}, with luminosities in $\mathrm{erg\,s^{-1}}$:
\begin{equation}
 \log_{10}L_{\mathrm{EUV}}=
 0.821\,[\log_{10}L_X-27.44]+28.16.
 \label{eq:sf25}
\end{equation}
Here $L_X$ is integrated over 0.5--10 nm and $L_{\mathrm{EUV}}$ over
10--92 nm. The reported slope and intercept uncertainties are 0.041 and
0.05, and the residual dispersion is 0.406 dex. The latter is represented
as one logarithmic predictive residual per complete EUV realization, not
independent pixel noise. Several early releases hold the coefficients at
their central values; later releases also sample their marginals,
conditionally independently at the published pivot because their covariance
is not supplied. Each target section specifies which convention was used.
This atlas documents those released conventions rather than retroactively
claiming that all products use an identical uncertainty prescription.

The X draw entering Eq.~\ref{eq:sf25} is the draw used for the X spectrum.
For slope $b=0.821$, the distance/radius dependence is
\begin{equation}
 F_{\mathrm{EUV},\oplus}\propto F_{X,\oplus}^{b}d^{2(b-1)},\qquad
 F_{\mathrm{EUV},\star}\propto F_{X,\oplus}^{b}d^{2b}R_\star^{-2}.
 \label{eq:euvgeometry}
\end{equation}
Repeating an independent geometrical perturbation after this construction
would count part of the same uncertainty twice. Extrapolation above 92 nm
is not included in the 10--92 nm normalization integral.

An integral relation does not determine the detailed EUV shape. The adopted
shapes include solar-based magnetic-scaling spectra
\citep{Namekata.et.al.2023ApJ...945..147N}, active/quiet FISM2 templates \citep{Chamberlin.et.al.2020SpWea..1802588C}, target-specific
emission-measure reconstructions from \citet{Sanz-Forcada.et.al.2025A&A...693A.285S}, and broad bands
conditioned on intrinsic Ly$\alpha$ \citep{Linsky.et.al.2014ApJ...780...61L}. The catalogue is therefore
not uniformly a Namekata or a FISM2 catalogue. A measured total unsigned
magnetic flux, an X-conditioned model input, and a large-scale polarimetric
field are distinct quantities. No universal unmeasured $10^{24}$ Mx value
is used to label all targets as magnetically measured.

Direct EUVE spectra are preferred within their supported observed bands
when source identity and component content permit. Source/background
extraction, higher-order contamination, and foreground H/He attenuation
are checked. EP Eri and $\pi^1$ UMa use EUVE-constrained short-EUV spectra;
unresolved binary spectra for other systems remain system-level diagnostics.
A different-epoch regression prediction is not forced onto a usable EUVE
measurement. Likewise, adopting a published target EMD does not require
applying a second normalization regression unless that is the explicitly
chosen component-specific procedure. Solar/EMD alternatives at a common
integral test spectral shape; their extrema are not automatically a
confidence interval.

\subsection{Intrinsic Ly\texorpdfstring{$\alpha$}{alpha} and neutral-line airglow}\label{sec:lya}
A reconstruction models an intrinsic emission profile, interstellar H I/D I
transmission and, where warranted, additional broad absorption, followed by
the appropriate instrumental line-spread function and pixel integration.
Self-reversal is allowed or tested within explicit bounds. Mg II can
constrain foreground velocities or inform plausible morphology without
fixing a unique hydrogen reversal. The saturated hidden core is often
degenerate with absorption. Mask, LSF, absorber, background and profile-family
tests therefore accompany statistical or local conditional intervals.

Orbital-night selection reduces airglow but does not guarantee its removal;
ACCUM data cannot be divided into night event lists. Two-dimensional local
sky tests, time-selected spectra, and spatial profiles are used where the
inputs permit. N I and O I are retained, subtracted, excluded, or replaced
according to the actual diagnostic, with the relevant flags preserved.
Weak sky evidence is not reported as a secure stellar/airglow decomposition.

Line and X-ray proxies from \citet{Linsky.et.al.2013ApJ...766...69L} are used either as explicit estimates
or as checks of a reconstruction, with the proper spectral class, line
definition, calibration band and flux-at-1-AU convention. Shared calibration
or a literature reconstruction of the same photons does not supply an
independent likelihood. A proxy disagreement does not by itself identify
which hidden profile is correct. Literature-adopted integrals and transferred
line shapes are labelled separately from new fits.

Where selected for missing short-FUV hydrogen emission, the higher-Lyman
prescription of \citet{Linsky.et.al.2014ApJ...780...61L} is
\begin{equation}
 F_{\mathrm{series}}=F_\alpha\,10^{-1.798}x^{0.351},\qquad
 x=F_{\alpha,\oplus}(d/\mathrm{AU})^2(R_\odot/R_\star)^2,
 \label{eq:lyseries}
\end{equation}
with $x$ expressed numerically in the calibration's cgs units. A solar
template partitions the total among missing windows. The original template
sampling is retained; a broad solar cell is not a recovered high-resolution
stellar line. The same Ly$\alpha$ and geometry draws must feed all dependent
hydrogen windows. The empirical uncertainty of an entire band must not be
misassigned as an independently measured error for each line.

\subsection{UV junctions and optical/IR completion}\label{sec:joins}
Independent observations are compared on identical wavelength support and
at an appropriate resolution. Native errors, DQ masks, line exclusions, and
the aperture/calibration history inform exposure selection. A constant
scale, additive contamination correction, or colour adjustment requires
evidence over the relevant interval; continuity alone is not a calibration
measurement. The uncertainty of a scale derived from the same anchor that
appears in the SED introduces cross-covariance with that anchor.

Available LOWLIB spectra supply their own reprocessed fluxes and errors
\citep{Pal.et.al.2023ApJS..266...41P}\footnote{LOWLIB version 1.0, MAST high-level science product: \url{https://doi.org/10.17909/kmdt-pw63}; \url{https://archive.stsci.edu/hlsp/lowlib}.}. Availability and target identity are checked rather than
substituting library stars indiscriminately. Gaia XP coefficient covariance
is propagated through its calibration and BP/RP combination when used.
Photospheric extensions employ the selected PHOENIX/NewEra model
\citep{Hauschildt.et.al.2025A&A...698A..47H}\footnote{Model-data identifiers retained from the reductions: NewEra V3,
\url{https://doi.org/10.25592/uhhfdm.17670}; the earlier grid used for cooler models,
\url{https://doi.org/10.25592/uhhfdm.16738}. Access software and model lists:
\url{https://doi.org/10.25592/uhhfdm.18108}. These identify data/software releases;
the scientific grid reference is Hauschildt et al. (2025).}
and an observational normalization; stellar-parameter derivatives are propagated
when supported. Missing near-UV intervals can require an explicitly labelled
empirical bridge. The optical spectrum of $\alpha$ Cen A is a calibrated
stellar-twin transfer with its own provenance, not a direct optical
measurement of $\alpha$ Cen A. The target section records that distinction.

\subsection{Covariance, calibration, and conditional model errors}\label{sec:uncertainty}
For a linear overlap or grouping operator $A$ and input covariance $C$,
\begin{equation}
 \boldsymbol f'=A\boldsymbol f,\qquad C'=ACA^{\mathsf T},\qquad
 \sigma_B^2=\boldsymbol w^{\mathsf T}C\boldsymbol w.
 \label{eq:bandcov}
\end{equation}
Splitting one native datum across output bins does not create independent
measurements. Sparse native covariance, low-rank shared modes, and paired
spectral realizations are retained in the representations used by each
release. A schematic decomposition is
\begin{equation}
 C=B+UKU^{\mathsf T}+\sum_r\operatorname{Cov}(D_r).
 \label{eq:covblocks}
\end{equation}
Here $D_r$ denotes the relevant correlated realization block, such as X/EUV
or Ly$\alpha$/higher-Lyman emission. Some cross-term metrics $K$ need not be
positive semidefinite by themselves; the complete covariance is the
meaningful object. The actual blocks and row mappings are target-dependent.

An applicable common calibration factor with fractional standard deviation
$s_a$ contributes, at first order,
\begin{equation}
 C^{\mathrm{cal}}_{ij}=s_a^2F_iF_j.\label{eq:cal}
\end{equation}
It does not decrease as independent noise when many pixels are integrated.
For a supplied interval bracketing the nominal flux, some releases display
the independent calibration contribution using
\begin{align}
 L'_i&=F_i-\sqrt{(F_i-L_i)^2+(s_aF_i)^2},\\
 U'_i&=F_i+\sqrt{(U_i-F_i)^2+(s_aF_i)^2}.\label{eq:pointcal}
\end{align}
These pointwise approximations do not reconstruct the full instrumental
response covariance. Already-propagated terms are not added a second time,
and an aperture-specific calibration accuracy is not transferred to a
different aperture without justification.

Native measurement errors, local covariance intervals, residual/bootstrap
intervals, posterior summaries, empirical predictive scatter, and named
physical alternatives have different meanings. The final plots show the
recommended available endpoints from the corresponding release; they are
not one homogeneous total-error confidence band. Method extrema remain
sensitivity ranges unless a probability model is specified. No artificial
linear or logarithmic symmetry is imposed. Unknown errors remain unavailable
rather than becoming numerical zero. Models can have parameter/normalization
uncertainty without having a quantified model-discrepancy distribution.

\subsection{Signed measurements, irradiation products, and plot conventions}\label{sec:display}
Signed background-subtracted observations are retained. Where a nonnegative
irradiation spectrum is supplied, adjacent bins are grouped with exact
integrated-flux conservation and the corresponding covariance operator.
This differs from replacing negative samples by zero, which would bias the
integral upward. The loss of resolution is encoded in the output edges.
Propagation is conditional on the data-selected grouping; alternative
groupings are separate method tests where provided.

The final panels show the adopted nominal spectrum as a solid target-colour
curve and the available bounds in the same colour with opacity 0.5. Their
common limits are 0.5--5000 nm and $10^{-2}$--$10^9$
$\mathrm{erg\,s^{-1}\,cm^{-2}\,nm^{-1}}$. A nonpositive lower endpoint
is shaded to the visible logarithmic floor when its upper endpoint is
positive; orange marks identify all valid intervals with lower endpoint
$\leq0$, including intervals entirely below zero. A positive bound below
the plot floor is not given an orange mark. These are rendering choices:
stored endpoints and nominal fluxes are unchanged. The black solar curve is
the same reference comparison used in the catalogue paper and its plotting
notebook, attributed there to \citet{Gueymard.2004SoEn...76..423G}. It is converted
from 1-AU irradiance to solar surface flux and is not a new
absolute-calibration constraint. No uncertainty array for that solar
input was supplied, so none is drawn.

The vertical guides at 12.4, 121.6, and 400 nm preserve the paper figure's
display convention. They are not the integration limits of
Eq.~\ref{eq:sf25}, the H-ionizing threshold, or the boundaries of individual
FITS components. Numerical analyses must use the explicit wavelength
windows in the relevant equation or table, not the display labels.

%% file: sections/data_guide.tex
\section{Using the released FITS products}\label{sec:data_guide}
This section is the practical entry point to the library. It explains which
file to open, how to identify its spectrum and uncertainty arrays, and how
to read, integrate, and display the data without changing their statistical
meaning. The examples use the released bin definitions and nominal fluxes.
Product names and column families below were checked against the fifteen
public target directories; the files retain target-specific extensions
rather than imposing a single identical schema on every reconstruction.

\subsection{Choose the spectrum and keep its supporting files}
Each target directory contains its main FITS product, a viewing notebook,
and a \path{README.txt}. Additional schema, uncertainty, provenance, and
band-integral files accompany individual targets. Begin with the exact
main filename in Table~\ref{tab:manual_files}, then consult its local README
and any \path{FITS_SCHEMA.txt}, \path{UNCERTAINTY_USAGE.txt}, or
\path{FINAL_UNCERTAINTY_GUIDE.txt}. Keep a covariance companion in the same
directory when the primary header specifies \texttt{COVFILE}. A file ending
in \texttt{.fits.gz} can be opened directly with Astropy.

For EP Eri, $\pi^1$ UMa, and $\xi$ Boo A, the table selects the separate
\texttt{irradiation} product: its nominal flux is nonnegative after
energy-conserving adjacent-bin grouping. The signed measurement composite
is retained separately. Several other targets keep the nominal irradiation
SED in \texttt{SED} and the pre-grouping measurements in \texttt{SED\_SIGNED}
within the same file. EP Eri also retains its original measurements in
\texttt{MEASUREMENTS}. These representations are alternatives, not
independent datasets to concatenate or coadd.

The main 18 Sco and $\kappa^1$ Ceti tables retain signed low-signal estimates;
they are not universally nonnegative irradiation tables. Use their supplied
support and uncertainties for measurement comparisons. A calculation that
requires nonnegative input needs a flux-conserving physical representation
of those intervals; deleting negative rows or clipping them to zero biases
the energy integral. For any target, a negative lower uncertainty endpoint
does not imply a negative nominal irradiating flux.

\input{sections/product_files}

The adopted nominal spectrum is the \texttt{F\_*} column specified below.
For 10 Tau, \texttt{F\_EARTH\_NM} and \texttt{F\_SURFACE\_NM} contain the
selected nominal solution; the additional mean and median columns are
diagnostics. Do not replace the nominal curve by either simply to center
an uncertainty band. The public plotting notebooks reproduce each target's
chosen convention. The small notebook supplied with this manual,
\path{examples/PHAROS_FITS_quickstart.ipynb}, demonstrates basic file use;
it does not rerun the astronomical reductions.

\subsection{What the FITS extensions contain}\label{sec:fits_hdus}
A FITS file consists of header--data units (HDUs). The primary header records
the target, version and applicable reconstruction or geometry metadata;
the spectrum is in the binary-table extension named \texttt{SED}. Access it
by name rather than assuming a fixed extension number. The overview in
Table~\ref{tab:manual_hdus} groups the main extension types. Not every
extension exists for every star; \texttt{hdul.info()} lists the actual
contents, and each extension's header and column definitions give its units
and dimensions.

\begin{table}[htbp]\centering\small
\caption{Main FITS extension families and their use. Names joined by commas
are examples, not interchangeable arrays.}\label{tab:manual_hdus}
\begin{tabularx}{\textwidth}{@{}>{\raggedright\arraybackslash}p{.35\textwidth}>{\raggedright\arraybackslash}X@{}}\toprule
Extension or family & Contents and intended use \\\midrule
\texttt{PRIMARY}, \texttt{SED} & Release metadata; adopted bin edges,
nominal density, provenance, flags and available uncertainty columns.\\
\texttt{SED\_SIGNED}, \texttt{MEASUREMENTS}, \texttt{REBIN\_MAP} & Original
signed estimates and, where supplied, the linear map to grouped output
bins. Native and grouped row numbers are not interchangeable.\\
\texttt{SEGMENTS}, \texttt{COMPONENTS}, \texttt{PROVENANCE} & Instrument,
model and reconstruction origins, spectral support, and input provenance.
The accompanying \path{spectral_components.csv}, when present, is a
convenient summary; the FITS bin edges define exact support.\\
\texttt{BANDS}, \texttt{BAND\_BUDGET}, \texttt{FINAL\_BANDS} & Integrated
band quantities and their stated uncertainty budgets. Check band edges,
units, geometry convention and whether a value refers to signed or
irradiation data before comparing it with a new integral.\\
\texttt{BASE\_COV}, \texttt{STAT\_COV}, \texttt{CAL\_MODES},
\texttt{CAL\_VECTORS}, \texttt{X\_JACOBIAN} & Sparse or factorized
covariance components. These may instead be in the companion named by
\texttt{COVFILE}. Their indexing and units are product-specific.\\
\texttt{POST\_DELTA}, \texttt{FINAL\_BAND\_DRAWS},
\texttt{NORMALIZATION\_DRAWS} & Shared realizations or deviations that
preserve correlations. Associated index, parameter and scope extensions
define the mapping and interpretation.\\
\texttt{METHOD\_SCENARIOS}, \texttt{METHOD\_DELTA},
\texttt{X\_SENSITIVITY} & Named physical or reduction alternatives;
they are not automatically posterior draws or confidence intervals.\\
\texttt{PARAMETERS}, \texttt{REFERENCES}, \texttt{README},
\texttt{UNC\_SCOPE} & Supporting parameters, literature, usage notes and
the scope of quantified errors, where provided.\\\bottomrule
\end{tabularx}
\end{table}
\FloatBarrier

The 70 Oph A extensions \texttt{EUVE\_AB}, \texttt{EUVE\_AB\_COV}, and
\texttt{EUVE\_AB\_METHODS} describe the unresolved binary. They are supporting
evidence, not component-A bins to insert into \texttt{SED}. More generally,
a diagnostic extension does not become an adopted spectral segment merely
because it is distributed with the product.

\subsection{Wavelengths, flux units, provenance, and validity}\label{sec:fits_columns}
\texttt{BIN\_LO\_NM} and \texttt{BIN\_HI\_NM} give the actual wavelength
support of each row in nm. \texttt{WAVELENGTH\_NM} is its representative
wavelength and must not substitute for the bin width. Flux densities are
energy flux per unit wavelength, in
$\mathrm{erg\,s^{-1}\,cm^{-2}\,nm^{-1}}$; they are neither integrated band
fluxes nor photon fluxes. Inspect \texttt{TUNIT} through the column metadata
before mixing these arrays with native instrument files. For the same
spectrum, $F_{\lambda,\mathrm{nm}}=10 F_{\lambda,\mathrm{\mathring A}}$.
Changing the wavelength labels alone is insufficient.

\begin{table}[htbp]\centering\small
\caption{Core SED column families. The aliases shown describe different
release schemas; inspect the column names in the selected file.}\label{tab:manual_columns}
\begin{tabularx}{\textwidth}{@{}>{\raggedright\arraybackslash}p{.36\textwidth}>{\raggedright\arraybackslash}X@{}}\toprule
Column(s) & Meaning \\\midrule
\texttt{F\_EARTH\_NM} & Nominal density diluted to the adopted Earth
distance. Reconstructed intrinsic segments at this geometry are not
necessarily the attenuated spectrum observed by the telescope.\\
\texttt{F\_SURF\_NM}, \texttt{F\_SURFACE\_NM} & Nominal stellar-surface
density. All fifteen main products supply at least one of these names;
18 Sco does not supply \texttt{F\_EARTH\_NM} in its main \texttt{SED}.\\
\texttt{F\_1AU\_NM} & Density at 1 au, where supplied. This is not a
luminosity and does not already refer to an arbitrary habitable-zone orbit.\\
\texttt{SOURCE}, \texttt{DATA\_TYPE}, \texttt{KIND} & Instrument/template
and observation, reconstruction, proxy or model identity, according to
the target schema. Mixed or repaired intervals keep that provenance.\\
\texttt{ORIGIN\_ID}, \texttt{ORIGIN\_BIN}, \texttt{SEGMENT\_ID} & Links to
original samples or source segments, where present; these are not general
wavelength indices. Use the supplied rebin map for grouped bins.\\
\texttt{VALID}; \texttt{QUALITY}, \texttt{QFLAG}, \texttt{QUALITY\_FLAG} &
Explicit validity, where present, and target-specific provenance/quality
flags. A nonzero flag may denote an adopted model or reconstruction, not
a rejected datum.\\
\texttt{NATIVE\_DQ}; \texttt{COV\_KNOWN}, \texttt{TOTALERR\_KNOWN} & Native
instrument flags, where preserved; separate indicators of uncertainty
coverage. Neither is a universal mask for the entire catalogue.\\\bottomrule
\end{tabularx}
\end{table}
\FloatBarrier

Use finite fluxes and positive, finite bin widths, and honor \texttt{VALID}
where supplied. Then inspect source and quality fields for the scientific
selection being made. Do not reject all nonzero quality flags: doing so
would remove deliberate X/EUV models and intrinsic line reconstructions.
The bit definitions vary between releases and are given in the relevant
schema or README. Retain NaN as unavailable information, not a zero flux
or an error-free datum. A component's minimum and maximum wavelengths can
span gaps; only the bin support establishes what is actually present.

To evaluate irradiation at orbital distance $a$, use
$F_\lambda(a)=F_{\lambda,\star}(R_\star/a)^2$ or
$F_{\lambda,\oplus}(d/a)^2$ with consistent distance units. A stored 1-au
density can be multiplied by $(1\,\mathrm{au}/a)^2$. Luminosity requires
$4\pi R_\star^2 F_{\lambda,\star}$, equivalently
$4\pi d^2 F_{\lambda,\oplus}$. Use the adopted parameter set, rather than
mixing a surface flux with a radius from a different calibration.
Section~\ref{sec:geometry} gives the corresponding geometry uncertainty
conventions. These are composite irradiation spectra, not simultaneous
light curves or forecasts of an activity state.

\subsection{Which uncertainty array should be used?}\label{sec:fits_errors}
The preferred plotting endpoints depend on the release schema
(Table~\ref{tab:manual_bounds}). A lower/upper pair stores \emph{flux
endpoints}, whereas an \texttt{ERR\_*} column normally stores a deviation
or standard deviation with the meaning stated by its schema. Do not
subtract an endpoint from the nominal flux a second time, or add a
statistical term again when it is already included in the recommended
budget. Inspect \texttt{BOUND\_TYPE}, \texttt{UNC\_KIND},
\texttt{UNC\_METHOD}, and the target's guide where available.

\begin{table}[htbp]\centering\small
\caption{Recommended available endpoint columns in the checked main SEDs.
Their statistical interpretation remains component-specific.}\label{tab:manual_bounds}
\begin{tabularx}{\textwidth}{@{}>{\raggedright\arraybackslash}p{.40\textwidth}>{\raggedright\arraybackslash}X@{}}\toprule
Targets & Endpoint columns and geometry \\\midrule
18 Sco; $\kappa^1$ Ceti & \texttt{UNC\_LO\_FIXED},
\texttt{UNC\_HI\_FIXED}: surface flux at fixed geometry, including the
quantified components described in the final uncertainty guide.\\
EP Eri; $\iota$ Hor; $\tau$ Boo A; 70 Oph A; 36 Oph B; $\lambda$ Ser &
\texttt{BOUND\_LO\_SURF}, \texttt{BOUND\_HI\_SURF}; corresponding
\texttt{\_EARTH} endpoints are also available. These may include
conditional errors, partial budgets or method envelopes.\\
$\xi$ Boo A; 10 Tau; $\alpha$ Cen A; $\beta$ Com;
$\upsilon$ And A; 47 UMa & \texttt{COND\_LO\_SURF},
\texttt{COND\_HI\_SURF}, with \texttt{\_EARTH} counterparts. The names
alone do not guarantee posterior percentiles for every component.\\
$\pi^1$ UMa & \texttt{COND\_LO\_EARTH}, \texttt{COND\_HI\_EARTH};
multiply both by the primary-header \texttt{GEOMETRY} factor for fixed-geometry
surface endpoints.\\\bottomrule
\end{tabularx}
\end{table}
\FloatBarrier

\texttt{ERR\_STAT\_*} describes available measurement/statistical
components; \texttt{ERR\_COND\_*} or \texttt{ERR\_AVAIL\_*} summarizes the
specified conditional or available covariance. Model families, calibration
terms and empirical prediction scatter have the coverage documented for
that target. An unquantified component does not become zero because a
Gaussian standard deviation cannot be supplied. In particular,
\texttt{ERR\_TOTAL}=NaN in early products does not mean that the recommended
endpoint columns are missing. Likewise, \texttt{COMPLETE=True} denotes
completed assembly over declared support, not universal knowledge of all
physical uncertainties.

For a band integral with weights $w_i$ equal to supported bin widths,
the covariance contribution is $\sigma_B^2=\boldsymbol{w}^{\mathsf T}
\mathbf C\boldsymbol{w}$. Do not integrate upper and lower pointwise
endpoints and call the results a band confidence interval. Use the
distributed band products for their exact bands, or propagate the shared
covariance/realizations for a new band. Surface bounds at fixed geometry
must not be interpreted as already including radius or distance uncertainty.
For 18 Sco and $\kappa^1$ Ceti, \texttt{FINAL\_BANDS} explicitly separates
fixed-geometry and propagated-geometry results; \texttt{FINAL\_BAND\_DRAWS}
preserves the common X/EUV/geometry draws.

Covariance arrays cannot safely be guessed from their names. For example,
$\pi^1$ UMa uses sparse \texttt{BASE\_COV} plus calibration, X-ray Jacobian,
and shared posterior terms. Its \texttt{CAL\_METRIC} includes cross-term
corrections and is not itself a stand-alone covariance to sample. EP Eri's
irradiation product retains covariance blocks on the original
\texttt{MEASUREMENTS} grid; the \texttt{REBIN\_MAP} must be applied to relate
them to \texttt{SED}. Many other products keep these components in a
separate \texttt{COVFILE}. Follow the local indexing and units, including
whether deviations are centered and whether scenarios store absolute
fluxes or changes. The zero-based sparse indices documented for these
products are array indices, not FITS extension numbers.

Named method scenarios remain separate from probabilistic uncertainty.
Treating them as equally probable draws, or adding their spread in
quadrature to every statistical error, would impose a new assumption.
Section~\ref{sec:uncertainty} explains the propagation model, and each target
section specifies the retained physical alternatives.

\subsection{A reproducible reading and integration example}\label{sec:fits_example}
The following cells are also supplied in the portable quick-start notebook.
They require Python with NumPy, Astropy and Matplotlib. Place the notebook
beside the selected FITS file or set \texttt{FITS\_FILE} to its full path.
The example selects the nonnegative EP Eri product; the same reader was
checked on the fifteen main products listed in Table~\ref{tab:manual_files}.
It uses surface flux so that the 18 Sco schema is supported directly.

\input{sections/fits_example}

The integral is an energy flux in $\mathrm{erg\,s^{-1}\,cm^{-2}}$ at the
stellar surface. It uses the piecewise-constant bin representation; a
boundary that splits a bin cannot recover unresolved sub-bin line shape.
Coverage below unity reports missing support rather than silently assigning
zero physical emission. No uncertainty is attached by this minimal code:
the correct calculation requires the target-specific correlated terms
described above. Regridding should conserve overlap-weighted integrated
energy and propagate the same linear map through those terms.

\subsection{Plotting bounds and preparing model inputs}
Use one nominal line and translucent endpoint shading. The quick-start
notebook respects bin edges and leaves unavailable bounds unshaded.
On a logarithmic axis, finite nonpositive lower endpoints extend to the
positive display floor and are marked in orange, without changing stored
values. Inspect signed nonpositive nominal values in the linear panel;
the plot floor is never an integration cutoff.

For atmospheric calculations, verify geometry, units, component identity,
bin edges and sign convention. Photon-density conversion uses
$F_\lambda\lambda/(hc)$ with consistent length units. Coarse-bin rates
with varying cross sections require an explicit within-bin assumption.
Preserve the signed reference, propagate shared uncertainty rather than
independent pixel noise, and cite the product version and provenance references.

%% file: sections/product_files.tex
\begingroup\small
\setlength{\tabcolsep}{3pt}
\begin{longtable}{@{}p{.17\textwidth}p{.79\textwidth}@{}}
\caption{Main products relative to the public \texttt{PHAROS\_v2} directory.
These filenames select the nominal SEDs discussed in this manual.}
\label{tab:manual_files}\\
\toprule Target & Directory and main FITS filename \\\midrule\endfirsthead
\multicolumn{2}{l}{\textit{Table~\thetable\ continued}}\\
\toprule Target & Directory and main FITS filename \\\midrule\endhead
\bottomrule\endfoot
18 Sco & \path{18_Sco_v2/}\newline\path{PHAROS_18_Sco_HD146233_v2.fits} \\[3pt]
$\kappa^1$ Ceti & \path{kappa1_Ceti_v2/}\newline\path{PHAROS_kappa1_Ceti_HD20630_v2.fits} \\[3pt]
EP Eri & \path{EP_Eri_v2/}\newline\path{PHAROS_EP_Eri_HD17925_irradiation_v2.fits} \\[3pt]
$\pi^1$ UMa & \path{pi1_UMa_v2/}\newline\path{PHAROS_pi1_UMa_HD72905_irradiation_v2.fits} \\[3pt]
$\xi$ Boo A & \path{Xi_Boo_A_v2/}\newline\path{PHAROS_Xi_Boo_A_HD131156A_irradiation_v2.fits} \\[3pt]
10 Tau & \path{10_Tau_v2/}\newline\path{PHAROS_10_Tau_HD22484_v2p0.fits} \\[3pt]
$\alpha$ Cen A & \path{alpha_Cen_A_v2/}\newline\path{PHAROS_alpha_Cen_A_HD128620_v2p0.fits} \\[3pt]
$\beta$ Com & \path{beta_Com_v2/}\newline\path{PHAROS_beta_Com_HD114710_v2p0.fits} \\[3pt]
$\upsilon$ And A & \path{upsilon_And_v2/}\newline\path{PHAROS_upsilon_And_HD9826_v2p0.fits} \\[3pt]
47 UMa & \path{47_UMa_v2/}\newline\path{PHAROS_47_UMa_HD95128_v2p0.fits} \\[3pt]
$\iota$ Hor & \path{iot_Hor_v2/}\newline\path{PHAROS_iot_Hor_HD17051_v2p0.fits} \\[3pt]
$\tau$ Boo A & \path{tau_Boo_v2/}\newline\path{PHAROS_tau_Boo_HD120136_v2p0.fits} \\[3pt]
70 Oph A & \path{70_Oph_A_v2/}\newline\path{PHAROS_70_Oph_A_HD165341_v2p0.fits} \\[3pt]
36 Oph B & \path{36_Oph_B_v2/}\newline\path{PHAROS_36_Oph_B_HD155885_v2p0.fits} \\[3pt]
$\lambda$ Ser & \path{lambda_Ser_v2/}\newline\path{PHAROS_lambda_Ser_HD141004_v2p0.fits} \\[3pt]
\end{longtable}\endgroup

%% file: sections/fits_example.tex
\par\noindent\begin{minipage}{\linewidth}
\begin{lstlisting}[language=Python]
from pathlib import Path
import numpy as np
from astropy.io import fits

FITS_FILE = Path("PHAROS_EP_Eri_HD17925_irradiation_v2.fits")
with fits.open(FITS_FILE, memmap=False) as hdul:
    hdul.info()
    header = hdul[0].header.copy()
    sed = hdul["SED"].data.copy()
    print(hdul["SED"].columns)

names = set(sed.names)
flux_key = ("F_SURFACE_NM" if "F_SURFACE_NM" in names
            else "F_SURF_NM")
f = np.asarray(sed[flux_key], dtype=float)
lo = np.asarray(sed["BIN_LO_NM"], dtype=float)
hi = np.asarray(sed["BIN_HI_NM"], dtype=float)
valid = np.isfinite(f) & np.isfinite(lo) & np.isfinite(hi)
valid &= hi > lo
if "VALID" in names:
    valid &= sed["VALID"].astype(bool)
# SOURCE and quality flags remain available for further selection.
\end{lstlisting}
\end{minipage}\par\medskip
The next cell computes a nominal band integral on valid support.
\par\noindent\begin{minipage}{\linewidth}
\begin{lstlisting}[language=Python]
# Example: 10--92 nm, on the released bin representation.
a, b = 10.0, 92.0
width = np.maximum(0.0, np.minimum(hi, b) - np.maximum(lo, a))
use = valid & (width > 0)
band_flux = np.sum(f[use] * width[use])
coverage = np.sum(width[use]) / (b - a)
print("Surface band flux [erg/s/cm2]:", band_flux)
print("Supported fraction:", coverage)
# This is a nominal integral, not a band uncertainty estimate.
\end{lstlisting}
\end{minipage}\par\medskip

%% file: sections/18_Sco.tex
\clearpage
\section{\texorpdfstring{18 Scorpii}{18 Scorpii}}\label{star:18_Sco}
This section follows the construction of the 18 Sco composite from its
adopted geometry to the X/EUV models, ultraviolet calibration, and photospheric
completion. Particular attention is given to the intrinsic Ly$\alpha$ profile,
its airglow and interstellar-absorption sensitivity, and the uncertainty
components retained in the released spectrum.

\subsection{Adopted geometry}
For 18 Sco, the adopted central values are $d=13.9$ pc and $R_\star=1.01\,R_\odot$. The geometrical draws use the limb-darkened angular diameter $\theta=0.6759\pm0.0062$ mas and parallax $71.94\pm0.37$ mas associated with the interferometric determination of \citet{Bazot.et.al.2011AA...526L...4B}. Since $R_\star=d\theta/2$, the measured-flux conversion is $G=4/\theta^2$ when $\theta$ is in radians. Radius and distance are consequently not perturbed independently. The 16th and 84th percentiles of $G/G_0$ are 0.9821 and 1.0178.

\subsection{X-ray selection and physical-model tests}
The X-ray reference uses the EPIC-pn spectrum of XMM-Newton observation
0303660101, with its associated source, background, redistribution and
effective-area products. A one-temperature absorbed APEC model was fitted
over 0.2--2.0 keV using the W statistic. The adopted atomic database is
APEC 3.1.3, with the \texttt{wilm} abundance convention, fixed solar
abundance, $N_{\rm H}=10^{18}$ cm$^{-2}$ and zero redshift. The optimum is
$kT=0.20186$ keV and XSPEC normalization $1.42761\times10^{-5}$,
with $W=16.222$ for 13 degrees of freedom. Approximately 10.9\% of the
exported 0.5--10 nm flux is outside the fitted energy interval and is
therefore a thermal-model continuation.

The joint temperature/normalization likelihood was evaluated on a
$241\times301$ grid with a uniform measure in the two linear parameters
and weights proportional to $\exp(-\Delta W/2)$. The background remains
profiled. Grid refinement and boundary-mass checks were performed before
exporting 16,384 paired realizations. The resulting temperature--normalization
correlation is approximately $-0.62$; independent error bars on those
parameters would not reproduce the spectral covariance.

A two-temperature alternative lowers $W$ by 4.981. In 400 parametric
source-and-background Poisson simulations under the fitted one-temperature
reference, refitting both models with multiple starts gives 20 improvements
at least this large. The plus-one tail estimate is 0.0524, with a Monte Carlo
standard error of about 0.011. This conditional experiment does not supply
the probability that the simpler model is true. It provides limited evidence
for an additional temperature in these data, so the one-temperature nominal
spectrum is retained. Absorption, abundance, and extra-temperature scenarios
remain separate physical sensitivities; their range is not appended as a
statistical confidence interval.

\subsection{EUV shape and photospheric completion}
The adopted EUV shape uses the solar empirical family of
\citet{Namekata.et.al.2023ApJ...945..147N}, with its activity coordinate chosen to reproduce the
stellar surface X-ray flux. The resulting solar-equivalent coordinate is
$1.2896\times10^{23}$ Mx. This is a coordinate of the template family,
not a measurement or inference of the stellar unsigned magnetic flux.
It replaces the unmeasured historical $10^{24}$ Mx input. Equation~\ref{eq:sf25}
sets the independent 10--92 nm normalization. The template is continued
to 115 nm, with the 92--115 nm part explicitly outside that normalization
band. About 4.77\% of the X-ray realizations fall below the family's basal
solar X-ray flux; for these draws the basal shape is retained while keeping
the original X-conditioned EUV luminosity. The adopted predictive interval
includes the 0.406-dex residual and holds the regression coefficients fixed;
coefficient-covariance assumptions were explored separately.

The infrared extension uses PHOENIX NewEra V3 \citep{Hauschildt.et.al.2025A&A...698A..47H}, interpolated
in linear surface flux at $T_{\rm eff}=5823$ K, $\log g=4.45$ and
$[\mathrm{Fe/H}]=+0.054$. The adopted parameter errors are 6 K, 0.02 dex
and 0.005 dex, respectively \citep{Melendez.et.al.2014ApJ...791...14M}; [Fe/H] is used as
the model metallicity coordinate with $[\alpha/\mathrm{Fe}]=0$.
Six 10-nm Gaia XP bands over 950--1010 nm constrain the model normalization
to $1.02875\pm0.00867$, conditional on the selected atmosphere family and
the propagated Gaia covariance. The extension begins at 950 nm. This
normalization is shared across the model spectrum rather than redrawn
independently at every wavelength.

\subsection{UV reconstruction and line diagnostics}
The near-UV audit compares the separately merged E230H and E230M
observations on identical supported intervals. Over 258--283 nm, excluding
279--281 nm around Mg II, 22 accepted 1-nm bands give
$s_{\rm H/M}=1.49912$. Its fixed-weight statistical error is 0.00101;
allowing additional independent band scatter gives a conditional scale
error of 0.00203. Tested window choices give 1.49566--1.50062.
The inferred change across 25 nm is approximately $-0.59\%$, with a
slope consistent with zero at the available precision. These tests support
an empirical relative scale over the overlap, rather than a 50\%
astrophysical change or proof of absolute accuracy outside it. The adopted
assembly preserves E230H where selected and uses the scaled E230M outside
that coverage, with native-quality edge selection and separate IUE support
near the short- and long-wavelength joins. G230LB scattered-light subtraction
was investigated as a diagnostic and is not an adopted component of this
SED.

The far-UV product combines COS G130M/G140L data with specific neutral-line
treatments. N I near 120 nm and O I near 135.6 nm use the fully orbital-night
exposure \texttt{lbiz01idq}, with about 900.13 s of good time and solar
altitude below zero throughout. These are observed night spectra, not
interstellar-corrected intrinsic lines. The 130-nm O I triplet uses a
joint stellar-plus-airglow description that also includes the Si II
130.437-nm blend and a local continuum; it remains explicitly an
airglow-subtracted observation. Native errors, supported intervals, and
residual-airglow qualifications are retained.

The intrinsic Ly$\alpha$ review allowed physically bounded self-reversal
and used Mg II as an auxiliary constraint on foreground velocity and
profile plausibility. The Mg II and COS observations are from 2021 May 27
and 2011 February 4, respectively. The flexible diagnostic refit gives
a median intrinsic flux of $8.266\times10^{-13}$ and a conditional
16--84\% interval of $[7.579,8.994]\times10^{-13}$ \fluxunit.
The C II and C IV relations of \citet{Linsky.et.al.2013ApJ...766...69L} predict approximately
$1.249\times10^{-12}$ and $[1.250,1.323]\times10^{-12}$ \fluxunit.
The revised profile therefore does not eliminate the proxy discrepancy.
Since these proxies do not determine the hidden line core uniquely, the
review did not replace the released finite-window reference. The further
airglow/absorption stress tests are reported below.

\subsection{Optical and near-ultraviolet calibration}
The optical calibration term is taken from the CALSPEC product \texttt{18sco\_stis\_006.fits}. Its header identifies the G430L exposure \texttt{odta07020} and G750L exposures \texttt{odta07030/040}, matching the observations used in the adopted optical segments. The file specifies a nominal broadband systematic uncertainty of 1\% and identifies both its statistical and systematic errors as one-sigma quantities. We therefore add a 1\% common calibration term to the adopted STIS CCD samples, without applying it to Gaia XP or the infrared model \citep{Bohlin.et.al.2020AJ....160...21B}\footnote{CALSPEC product \texttt{18sco\_stis\_006.fits}; the 1\% value and error convention are read from its \texttt{SYSERROR} column and FITS header: \url{https://ssb.stsci.edu/cdbs/calspec/18sco_stis_006.fits}.}. This comparison uses another reduction of the same observations and is not treated as independent repeated data.

For the E230H observations obtained with the $0.2''\times0.2''$ aperture, the STIS handbook gives a nominal absolute accuracy of 8\% at two sigma. We represent this as a 4\% one-sigma common scale \citep{Rickman.and.Brown.2025stii.book...25R}. The same calibration group applies to the E230M spectrum after its empirical normalization to E230H: the reference scale is transferred rather than independently redrawn. The previously fitted E230H/E230M factor of 1.49912 and its statistical scale covariance are retained. We do not add a separate white-dwarf-standard calibration term on top of these aggregate estimates.

The handbook accuracy applies to the stated wide echelle aperture; it is not a generic accuracy for all narrow-slit observations. The empirical E230M normalization and its wavelength-support information are therefore retained as part of the adopted model of the join. The Gaia XP coefficient covariance and its coupling to the normalized photospheric extension also remain in the release. The infrared interval constrains the observational normalization of the adopted atmosphere model; it is not a calibrated distribution for all possible atmosphere-model discrepancies.

\subsection{X-rays, EUV, and \lya}
The existing joint temperature/normalization realizations of the one-temperature X-ray reference are retained. The adopted integrated Earth flux is $F_X=2.55448\times10^{-14}$ \fluxunit. For EUV, the selected shape remains the X-conditioned solar-template construction, with equation~(\ref{eq:sf25}) fixing its amplitude. The adopted 10--92 nm Earth flux is $1.76608\times10^{-13}$ \fluxunit. The same X-ray realization continues to inform the template and the EUV amplitude.

For \lya, each of the 150 stored conditional profile realizations is integrated over the inserted 120.8--122.3 nm window before computing quantiles. The adopted window flux is $8.42830\times10^{-13}$ \fluxunit. The final geometry calculation samples this existing ensemble and the angular-diameter/parallax realizations; resampling does not increase the number of independent profile constraints. The reconstructed line and its interval remain conditional on the adopted intrinsic-profile, interstellar-absorption, and airglow treatment. No new intrinsic-line correction is imposed during uncertainty finalization.

The adopted N I/O I components retain their observational and airglow-treatment labels and quality flags. No change to their measured support or to their central flux is made in this step. The complete adopted 18 Sco spectrum contains 149,775 valid bins, all with recommended pointwise uncertainty bounds.

\subsection{Integrated results and final products}
Table~\ref{18_Sco:tab:earth} gives the adopted Earth fluxes and fixed-geometry band
intervals. Table~\ref{18_Sco:tab:surface} gives the surface quantities with joint
geometry propagated. These are quantiles of the specified statistical,
calibration and conditional/predictive ensembles. The adopted central spectrum
is retained even when it is not exactly the median of an ensemble.

\begin{table}[tbp]\centering\small
\caption{Earth band fluxes with the adopted fixed-geometry uncertainty components. Each entry gives the adopted value followed by the 16th and 84th percentiles in brackets, in erg s$^{-1}$ cm$^{-2}$.}\label{18_Sco:tab:earth}
\begin{tabular}{llll}\toprule Star & Band & Window (nm) & Flux and interval \\\midrule
18 Sco & X & 0.5--10 & $2.554\;[2.304,\,2.853]\times10^{-14}$ \\
18 Sco & EUV & 10--92 & $1.766\;[0.698,\,4.448]\times10^{-13}$ \\
18 Sco & \lya & 120.8--122.3 & $8.428\;[7.996,\,8.984]\times10^{-13}$ \\
\bottomrule\end{tabular}\end{table}
\begin{table}[tbp]\centering\small
\caption{Surface band fluxes with shared geometry included. Each entry gives the adopted value followed by the 16th and 84th percentiles in brackets, in erg s$^{-1}$ cm$^{-2}$.}\label{18_Sco:tab:surface}
\begin{tabular}{llll}\toprule Star & Band & Window (nm) & Flux and interval \\\midrule
18 Sco & X & 0.5--10 & $9.518\;[8.584,\,10.642]\times10^{3}$ \\
18 Sco & EUV & 10--92 & $6.580\;[2.592,\,16.568]\times10^{4}$ \\
18 Sco & \lya & 120.8--122.3 & $3.140\;[2.970,\,3.366]\times10^{5}$ \\
\bottomrule\end{tabular}\end{table}
The band calculation uses 16,384 joint realizations. The Ly$\alpha$ profile information remains that of the original 150 conditional realizations; Monte Carlo resampling does not create additional observed information. All three reported bands have complete supported integration windows.

 The final FITS products preserve all pre-existing spectral columns and append the integrated budget and its joint draws. The recommended pointwise columns are \texttt{UNC\_LO\_FIXED} and \texttt{UNC\_HI\_FIXED}; the new integrated products are \texttt{FINAL\_BANDS} and \texttt{FINAL\_BAND\_DRAWS}. Their units and fixed-versus-variable-geometry conventions are explicit. Native covariance, shared-scale terms, reconstructed profiles, and model-sensitivity products remain available for applications requiring more than pointwise plotting.

\subsection{Additional Ly$\alpha$ airglow and ISM sensitivity tests}
The released Ly$\alpha$ integral is retained after a separate investigation
of whether residual airglow or interstellar absorption could explain the
higher line-proxy estimate. The diagnostic reference fit gives
$8.1130\times10^{-13}$ \fluxunit, while the released inserted-window
reference remains $8.4283\times10^{-13}$ \fluxunit. The external proxy
comparison is $1.2489\times10^{-12}$ \fluxunit; it is not imposed as an
exact measurement.

Common and exposure-dependent airglow widths/wings, two interstellar clouds,
additional warm H I, and thermal/turbulent alternatives were explored.
Forcing the proxy-level flux raises the penalized objective by approximately
19.3 in the baseline and by roughly 19--23 in several conventional
airglow/absorber tests. A more flexible exposure-dependent airglow model
reduces that cost to about 6.2 but requires an unvalidated deformation of
the night template. The night-only diagnostic rejects 99 pixels affected
by quality problems over approximately 1214.93--1215.91 Angstrom; its weak
constraint is therefore not evidence for a clean measurement of the hidden
core. Different masks and covariance assumptions are not combined into a
single likelihood-ratio significance.

Conditional injection/recovery tests give median downward biases of about
8--14 percent for the tested constructions. These are estimator tests under
specified synthetic inputs, not a measured correction to the stellar flux.
Differences of two same-visit day exposures constrain changing airglow but
cannot identify a common stellar/airglow offset. No detection of extra
physical absorption, validated calibration correction, or uniquely preferred
proxy-level reconstruction follows. The original SED is therefore preserved,
with the profile/airglow dependence stated explicitly.

\FloatBarrier
\begin{minipage}{\linewidth}
\subsection{Diagnostic figure}
\begin{center}
\includegraphics[width=\linewidth]{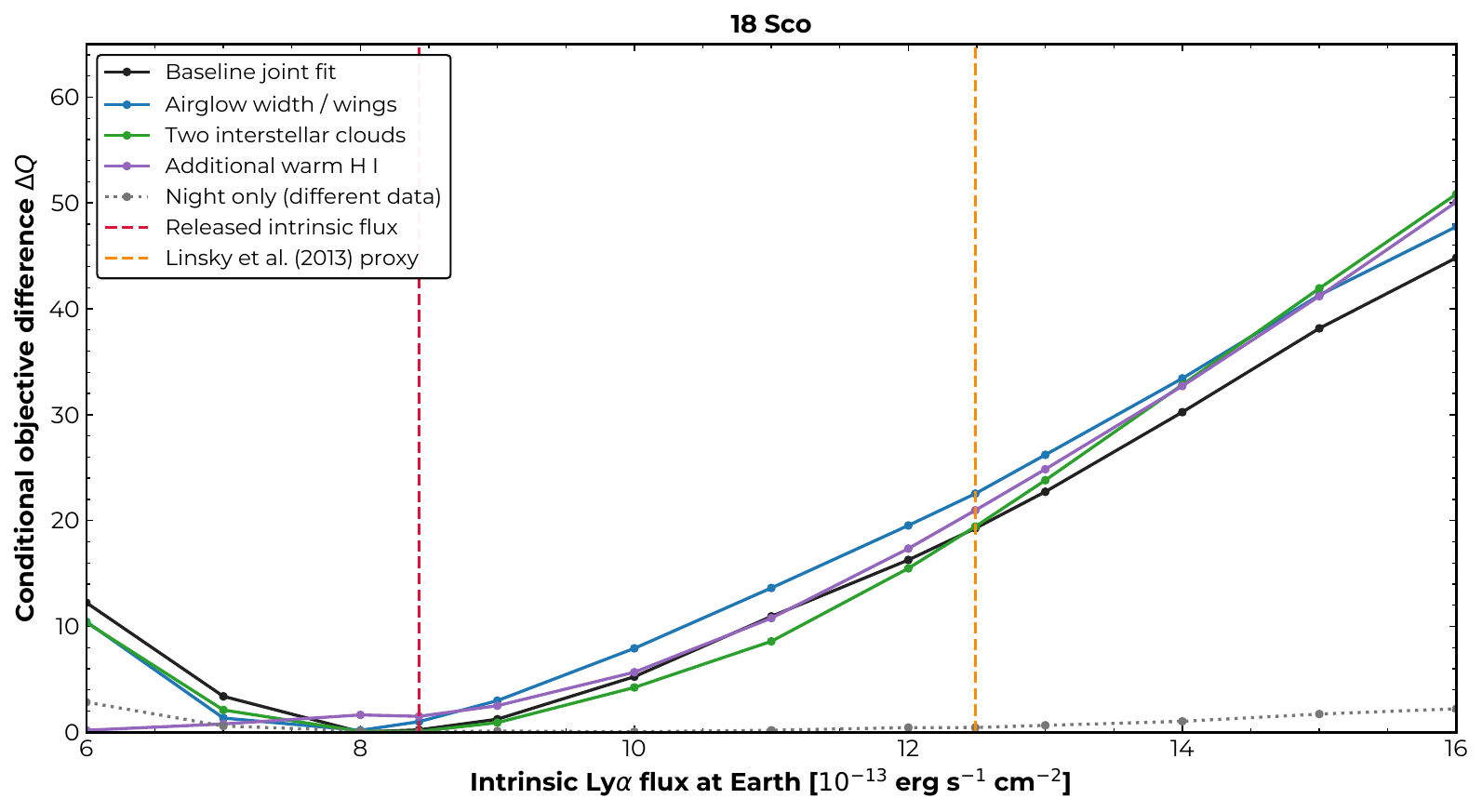}
\end{center}
\captionof{figure}{Conditional Ly$\alpha$ flux-profile tests for 18 Sco. Each curve shows the recorded penalized objective relative to its own unrestricted reference fit; the alternatives have different nuisance descriptions and are not probabilities for the physical models. Vertical lines mark the released finite-window reference and the external line-proxy comparison. The night-only curve is a diagnostic with fewer usable pixels. These tests did not change the adopted spectrum.}\label{fig:diagnostic:18_Sco}
\end{minipage}\par

\FloatBarrier
\begin{minipage}{\linewidth}
\subsection{Final spectrum with available uncertainties}
Figure~\ref{fig:sed:18_Sco} shows the adopted full SED and its available uncertainty bounds.

\begin{center}
\includegraphics[width=\linewidth]{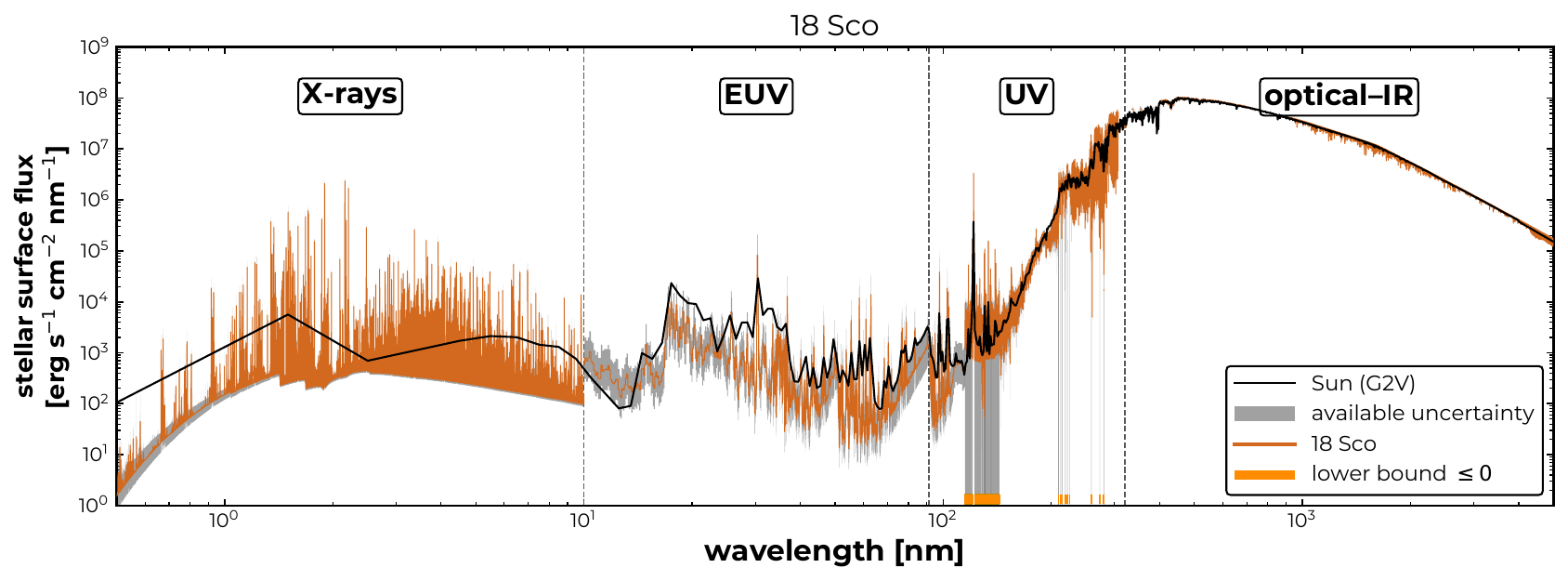}
\end{center}
\captionof{figure}{18 Scorpii: final nominal surface-flux spectrum and available uncertainties. Target colours, shading, solar comparison and nonpositive-bound markers follow Section~\ref{sec:display}. The component origins and uncertainty scope are documented above.}\label{fig:sed:18_Sco}
\end{minipage}
\FloatBarrier

%% file: sections/kappa1_Ceti.tex
\clearpage
\section{\texorpdfstring{$\kappa^1$ Ceti}{kappa1 Ceti}}\label{star:kappa1_Ceti}
This section documents the RGS re-extraction and coronal inference for
$\kappa^1$ Ceti, followed by Ly$\alpha$, the EUV estimate, and the ultraviolet
and photospheric junctions. It explains how shared normalization, line-profile,
and geometry terms enter the released uncertainties and integrated products.

\subsection{Adopted geometry}
For \ks, the adopted distance is 9.27624 pc from the selected Gaia DR3 parallax, and $R_\star=0.95\,R_\odot$. We retain the formal parallax draws and the radius uncertainty used in the adopted parameter set \citep{Gaia.Collaboration.2023A&A...674A...1G,Namekata.et.al.2023ApJ...945..147N}. The latter is implemented explicitly as $R_\star/R_\odot=0.95+s(z)z$, where $z$ is standard normal, $s=0.03$ for $z\geq0$, and $s=0.02$ otherwise. The parallax and radius perturbations are independent in this approximation. The resulting percentiles of $G/G_0$ are 0.9406 and 1.0434.

\subsection{RGS re-extraction and line-residual tests}
RGS1 and RGS2 were re-extracted using their own source-excluded CCD9
background light curves. The nominal threshold is 0.1 count s$^{-1}$,
selected from quiet rates of approximately 0.03--0.05 count s$^{-1}$
before spectral fitting. Light-curve bins require at least 90\% frame
coverage, and the SAS event selection requires complete frames inside
the accepted good-time intervals. Thresholds of 0.05 and 0.2 count
s$^{-1}$ and a shared-time selection were checked as extraction
sensitivities. The adopted MOS1/RGS fit uses the independently screened
RGS observations and the reference abundance pattern of
\citet{Hamaguchi.et.al.2023ApJ...944..163H}, with the relevant mixture conversion applied before
linking abundance ratios.

Local line-window and Poisson tests examined residuals rather than
removing discrepant bins to improve the statistic. A local excess near
15.26 Angstrom corresponds to approximately
$5.11\times10^{-14}$ \fluxunit, or 0.82\% of the integrated X-ray flux.
It is not added to the SED as an empirically measured extra plasma line.
A third thermal component reduces $W$ by only 3.67 for two additional
parameters and introduces a temperature near the 0.08-keV lower bound.
Its extrapolated flux is retained as a sensitivity, not adopted as a
secure cool component.

A deficit in RGS1 near the 17-Angstrom detector-gap region persists on
shared wavelength/time support and is not removed by wavelength shifts
of $\pm0.005$ Angstrom. RGS2 provides full coverage of the line complex.
An edge-guard extraction changes the integrated flux by about 5.30\%.
These results motivate keeping the extraction dependence distinct from
the local parameter covariance below; they do not establish an atomic
or stellar correction to the affected bins.

\subsection{Ly$\alpha$ reconstruction and sampling}
The Ly$\alpha$ reconstruction uses three STIS E140M exposures totaling
7805 s, a self-reversed intrinsic emission family, three interstellar
absorbers, instrumental convolution, and per-exposure wavelength and
background nuisance terms. The exploratory ensemble MCMC mixed slowly
and was not used for the released sampling. Two independent importance
runs used 30,000 proposals each from equal mixtures of Student-$t$
distributions with five degrees of freedom and covariance matrices
$C$ and $3C$. Bounded Gaussian background terms were marginalized
analytically, preserving the stated uniform log-column prior.

The effective sample sizes are approximately 2105 and 1734; the largest
normalized weights are 0.0091 and 0.0142, and the reported importance-tail
diagnostics are about 0.34 and 0.33. The independent mean-flux estimates
agree to 0.004 of the conditional posterior standard deviation. The full
analytic profile has a mean flux of $1.1178\times10^{-11}$ \fluxunit,
about 32.4\% above the $8.44\times10^{-12}$ \fluxunit\ literature
reconstruction of \citet{Wood.et.al.2005ApJS..159..118W}. This is a comparison of reconstruction
choices, not an independent observation of the hidden core. The
121.15--122.00 nm insertion grid contains 98.80\% of the full analytic
profile; the adopted finite-window flux and its calibration-propagated
interval are reported below. The profile is not renormalized to conceal
the omitted far wings.

\subsection{EUV, UV junctions, and recovered neutral lines}
The EUV shape uses the \citet{Namekata.et.al.2023ApJ...945..147N} construction with the adopted
Table-4 unsigned-flux input
$\Phi=1.39^{+0.24}_{-0.22}\times10^{25}$ Mx. The nominal stellar
parameters are $T_{\rm eff}=5742$ K, $\log g=4.49$ and
$R_\star=0.95\,R_\odot$; the LOWLIB photospheric metallicity is
$[\mathrm{Fe/H}]=+0.04$. The integrated amplitude uses the 2025 relation
in Eq.~\ref{eq:sf25}, with the shared X-ray and geometrical realizations.
The primary predictive normalization keeps the regression coefficients
fixed and includes its 0.406-dex residual. Magnetic-input alternatives
are shape sensitivities, not additional independent radius draws.
The continuation over 92--115 nm is outside the normalization band.

Six selected STIS files are combined with unique-exposure handling and
response/time weights. E230M is multiplied by approximately 1.611,
anchored to the unsaturated IUE LWR04858 spectrum and checked against
the simultaneous E140M coverage. The fitted scale and its IUE-anchor
covariance are propagated. LOWLIB replaces the older optical-library
reduction, and the NewEra V3 infrared spectrum shares its observational
normalization with that optical anchor.

The N I/O I audit recovered 211 valid STIS bins in previously masked
regions while preserving 31 natively invalid bins. Native covariance
is retained, and a residual-sky flag marks the recovered observations.
STIS spatial profiles were compared with re-extracted COS full,
orbital-night, and stricter-night spectra. The near coincidence of the
stellar and geocoronal velocities prevents a unique spectral separation,
and the COS comparison also involves a different epoch. Consequently,
no airglow-free or interstellar-corrected line reconstruction is claimed
for these recovered neutral-line samples. The subsequent uncertainty
finalization adds the applicable common calibration contribution without
discarding the native measurement errors.

\subsection{Joint X-ray shape uncertainty}
The X-ray reference is the adopted MOS1 plus background-screened RGS fit to XMM-Newton observation 0822791001. It uses a two-temperature \texttt{vapec} spectrum, APEC 3.1.3, and the abundance convention and fixed absorption of that fit. O, Ne, and Fe are free; the remaining linked abundance ratios follow the adopted pattern based on \citet{Hamaguchi.et.al.2023ApJ...944..163H}, expressed in the fit's abundance convention. The two temperatures are approximately 0.2807 and 0.6564 keV. The fit uses MOS1 over 0.3--2.4 keV and RGS over 0.35--1.7 keV; the exported 0.5--10 nm band therefore includes model continuation beyond the fitted energy ranges. The analysis uses the covariance of this extraction and fit rather than a chain from an earlier reduction.

There are nine varied quantities: two temperatures, two emission normalizations, three abundances, and two relative RGS scale constants. The MOS1 scale is fixed at unity. The RGS constants are nuisance parameters: their covariance with the plasma parameters is retained even though they do not directly multiply the exported stellar spectrum.

Let $\mathbf q$ contain the natural logarithms of these positive fit parameters, $\mathbf C_q$ their local covariance, and $\mathbf J$ the spectral Jacobian with elements $J_{ij}=\partial F_i/\partial q_j$. The propagated covariance is
\begin{equation}
 \mathbf C_F=\mathbf J\mathbf C_q\mathbf J^{\mathsf T}.
 \label{kappa1_Ceti:eq:xcov}
\end{equation}
We obtain the parameter covariance from the local likelihood curvature supplied by XSPEC \citep{Gordon.and.Arnaud.2021ascl.soft01014G}\footnote{PyXspec \texttt{FitManager} API, including the \texttt{covariance} attribute and \texttt{error} method: \url{https://heasarc.gsfc.nasa.gov/docs/software/xspec/python/html/fitmanager.html}.}, transform it into logarithmic parameters, and calculate central finite differences of the unabsorbed spectrum on the exported grid. The covariance is not rescaled by a reduced goodness-of-fit statistic. Reducing the finite-difference step from 0.001 to 0.0005 changes the propagated pointwise standard deviations by at most $1.3\times10^{-7}$ in relative terms.

For positive pointwise display we use
\begin{equation}
 [L_i,U_i]=F_i\exp\left[\mp\frac{\sqrt{(C_F)_{ii}}}{F_i}\right].
 \label{kappa1_Ceti:eq:xlog}
\end{equation}
These are local log-flux limits, not percentiles of a sampled posterior. Direct likelihood tests along covariance directions give departures from the quadratic prediction of approximately 0.77--1.56 times the predicted change; the local approximation is therefore stated explicitly, particularly for weak extrapolated spectral bins. Correlated changes of line strengths are retained through equation~(\ref{kappa1_Ceti:eq:xcov}).

The integrated 0.5--10 nm flux remains $6.22648\times10^{-12}$ \fluxunit. Equation~(\ref{eq:bandcov}) gives $\sigma_X=1.80032\times10^{-13}$ \fluxunit, or 2.89\%. The independently retained $\Delta W=1$ flux-profile limits are $[6.04946,6.41216]\times10^{-12}$ \fluxunit; their half-width is $1.81351\times10^{-13}$ \fluxunit. Agreement at the 0.73\% level in the widths supports the integrated local propagation. The pointwise relative error is not uniform: its median is about 9.0\%, with substantially larger values in some weak features. An integrated 2.89\% error must therefore not be assigned independently to every spectral bin.

\subsection{Combining X-ray shape and the existing X/EUV draws}
For the full X-ray band, define $\mathbf g=\mathbf J^{\mathsf T}\mathbf w$ and $v=\mathbf g^{\mathsf T}\mathbf C_q\mathbf g$. The parameter covariance conditional on the integrated normalization is
\begin{equation}
 \mathbf C_{q\mid X}=\mathbf C_q-
 \frac{\mathbf C_q\mathbf g\mathbf g^{\mathsf T}\mathbf C_q}{v}.
 \label{kappa1_Ceti:eq:conditional}
\end{equation}
The executable implementation uses
\begin{equation}
 \delta\mathbf q=\frac{\mathbf C_q\mathbf g}{v}\,\delta F_X+\boldsymbol\eta,
 \qquad\boldsymbol\eta\sim\mathcal N(0,\mathbf C_{q\mid X}),
 \label{kappa1_Ceti:eq:conditionaldraw}
\end{equation}
where $\delta F_X$ comes from the existing joint X/EUV normalization ensemble. The residual shape contribution integrates to zero to first order over the normalization band. Consequently, the old normalization variance is not added again to equation~(\ref{kappa1_Ceti:eq:xcov}). This construction reproduces the stored integrated X-ray draws while allowing covariance-aware sub-band calculations. The final full-band table uses the retained flux-profile ensemble for X and its associated Sanz-Forcada prediction for EUV. The central EUV Earth flux remains $1.86024\times10^{-11}$ \fluxunit.

\subsection{E140M, \lya, and the photospheric anchor}
The E140M observations use the $0.2''\times0.2''$ aperture. We adopt a nominal 4\% one-sigma echelle calibration representation, shared by the E140M observations and the normalization of their intrinsic \lya\ reconstruction \citep{Rickman.and.Brown.2025stii.book...25R}. This common factor is sampled once per realization. The final \lya\ band calculation uses the 2,000 stored conditional flux realizations for the inserted 121.15--122.00 nm window. Its adopted flux is $1.10430\times10^{-11}$ \fluxunit; it is explicitly a finite-window integral, rather than the infinite-profile line flux.

The E230M scale remains tied to its selected IUE anchor, including the previously propagated scale/anchor covariance. The unscaled E230H data use the narrower $0.2''\times0.09''$ aperture; the wide-aperture 4\% term is not transferred to that observation. The optical component retains the supplied errors of the slit- and scattered-light-corrected LOWLIB spectrum \citep{Pal.et.al.2023ApJS..266...41P}. The infrared model shares its fitted normalization with the LOWLIB anchor rather than acquiring independent noise in every modeled pixel.

The restored N I/O I regions remain observed fluxes with the existing residual-airglow quality flag, not reconstructed intrinsic profiles. The adopted X-ray model likewise remains the reference for the fitted continuum and line spectrum; its parameter interval does not constitute a new empirical correction of individual line residuals. All 114,948 valid adopted bins in the \ks\ composite have recommended uncertainty bounds.

\subsection{Integrated results and final products}
Table~\ref{kappa1_Ceti:tab:earth} gives the adopted Earth fluxes and fixed-geometry band
intervals. Table~\ref{kappa1_Ceti:tab:surface} gives the surface quantities with joint
geometry propagated. These are quantiles of the specified statistical,
calibration and conditional/predictive ensembles. The adopted central spectrum
is retained even when it is not exactly the median of an ensemble.

The Ly$\alpha$ band interval includes the common E140M calibration factor.

\begin{table}[tbp]\centering\small
\caption{Earth band fluxes with the adopted fixed-geometry uncertainty components. Each entry gives the adopted value followed by the 16th and 84th percentiles in brackets, in erg s$^{-1}$ cm$^{-2}$.}\label{kappa1_Ceti:tab:earth}
\begin{tabular}{llll}\toprule Star & Band & Window (nm) & Flux and interval \\\midrule
\ks & X & 0.5--10 & $6.226\;[6.052,\,6.411]\times10^{-12}$ \\
\ks & EUV & 10--92 & $1.860\;[0.726,\,4.679]\times10^{-11}$ \\
\ks & \lya & 121.15--122 & $1.104\;[1.050,\,1.158]\times10^{-11}$ \\
\bottomrule\end{tabular}\end{table}
\begin{table}[tbp]\centering\small
\caption{Surface band fluxes with shared geometry included. Each entry gives the adopted value followed by the 16th and 84th percentiles in brackets, in erg s$^{-1}$ cm$^{-2}$.}\label{kappa1_Ceti:tab:surface}
\begin{tabular}{llll}\toprule Star & Band & Window (nm) & Flux and interval \\\midrule
\ks & X & 0.5--10 & $1.168\;[1.092,\,1.229]\times10^{6}$ \\
\ks & EUV & 10--92 & $3.489\;[1.354,\,8.767]\times10^{6}$ \\
\ks & \lya & 121.15--122 & $2.071\;[1.911,\,2.204]\times10^{6}$ \\
\bottomrule\end{tabular}\end{table}
The band calculation uses 32,768 joint realizations. The Ly$\alpha$ profile information remains that of the original 2,000 conditional realizations; Monte Carlo resampling does not create additional observed information. All three reported bands have complete supported integration windows.

 The final FITS products preserve all pre-existing spectral columns and append the integrated budget and its joint draws. The recommended pointwise columns are \texttt{UNC\_LO\_FIXED} and \texttt{UNC\_HI\_FIXED}; the new integrated products are \texttt{FINAL\_BANDS} and \texttt{FINAL\_BAND\_DRAWS}. Their units and fixed-versus-variable-geometry conventions are explicit. Native covariance, shared-scale terms, reconstructed profiles, and model-sensitivity products remain available for applications requiring more than pointwise plotting.

\FloatBarrier
\begin{minipage}{\linewidth}
\subsection{Final spectrum with available uncertainties}
Figure~\ref{fig:sed:kappa1_Ceti} shows the adopted full SED and its available uncertainty bounds.

\begin{center}
\includegraphics[width=\linewidth]{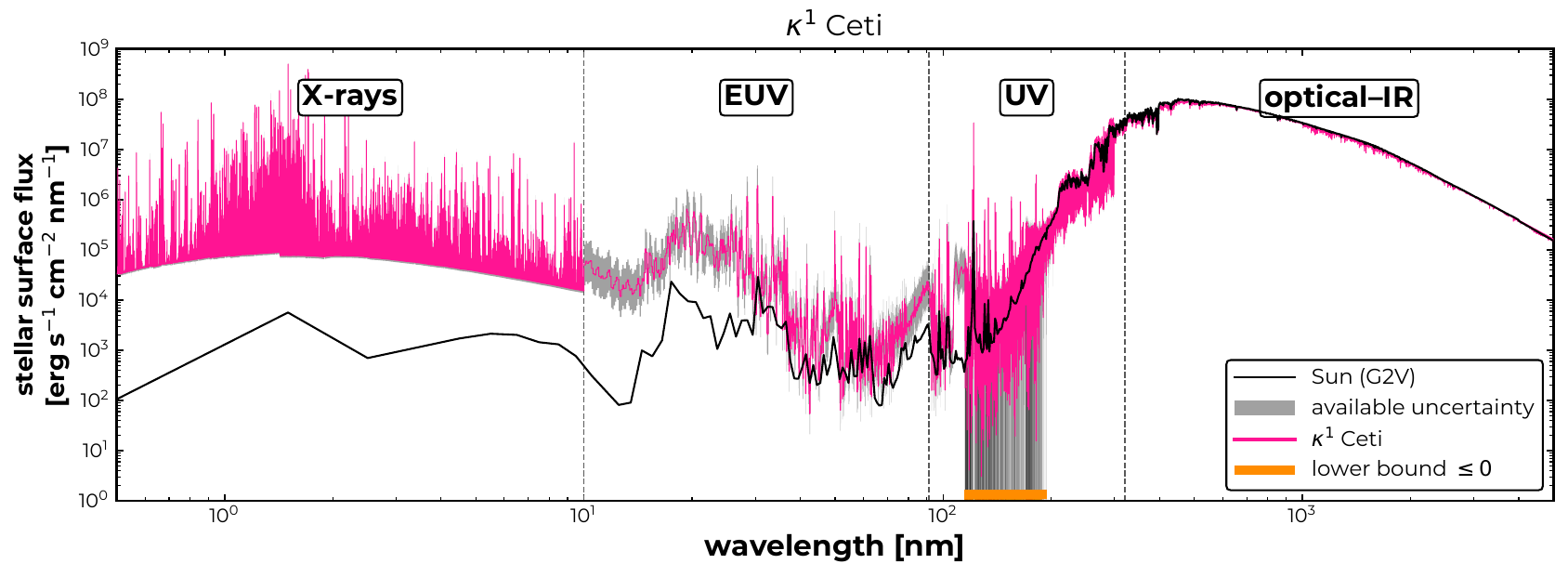}
\end{center}
\captionof{figure}{$\kappa^1$ Ceti: final nominal surface-flux spectrum and available uncertainties. Target colours, shading, solar comparison and nonpositive-bound markers follow Section~\ref{sec:display}. The component origins and uncertainty scope are documented above.}\label{fig:sed:kappa1_Ceti}
\end{minipage}
\FloatBarrier

%% file: sections/EP_Eri.tex
\clearpage
\section{\texorpdfstring{EP Eridani}{EP Eridani}}\label{star:EP_Eri}
\noindent This section describes the new revision of EP Eri (HD 17925). It records
the adopted 0.5--5000 nm composite, its quantified uncertainty components,
and the model estimates used where contaminated or unobserved data cannot
uniquely determine stellar flux. The composite is not a simultaneous observation.

\subsection{X-ray model and recoverable uncertainties}
We recovered the original XMM-Newton/MOS1 spectrum, background, redistribution
matrix and ancillary response for observation 0203060501 (2004 January 20,
48.2 ks). We reproduced the previous two-temperature, solar-abundance APEC
3.0.9 solution, obtaining $W=353.660$ for 55 degrees of freedom. Because the
background was retained as a Poisson measurement, XSPEC's \texttt{cstat}
selection evaluated the W statistic. Its ratio to the degrees of freedom was
not treated as a calibrated goodness-of-fit probability.

We compared one-, two- and three-temperature descriptions using APEC 3.1.3
and the Wilms reference abundance table. In addition to a common metallicity,
we tested models allowing O, Ne, Mg, Si and Fe to vary independently, while
sharing each elemental abundance between temperature components. We retained
$N_{\rm H}=10^{18}$ cm$^{-2}$ in the reference comparison. Although the nominal
selection was 0.2--2.4 keV, the native quality flags and grouping left 59 valid
groups spanning approximately 0.360--2.395 keV. The softer portion of the
reported X-ray band therefore remains model extrapolation.

The two-temperature model with fitted elemental abundances gave $W=73.683$
for 50 degrees of freedom. The corresponding three-temperature model gave
$W=47.852$ for 48 degrees of freedom, with temperatures near 0.192, 0.380 and
0.742 keV. We generated 100 Poisson realizations under each fitted model,
including the measured background as a plug-in expectation, and refitted
each realization. Zero of the 100 two-temperature simulations had W as large
as observed, whereas 54 of 100 three-temperature simulations did. We selected
the latter as the adopted X-ray reference. This test establishes conditional
compatibility with this spectrum, not uniqueness of the plasma description.

We evaluated the intrinsic flux over exactly 0.5--10 nm and obtained
\begin{equation}
 F_X=7.1258\times10^{-12}\ {\rm erg\,s^{-1}\,cm^{-2}}.
\end{equation}
Profiling the integrated flux with \texttt{cflux}, while reoptimizing the
remaining free parameters, gave $[7.0500,7.2788]\times10^{-12}$ in the same
units at $\Delta W=1$. We also propagated the joint local parameter covariance
through a numerical spectral Jacobian. The integrated local standard
deviation was $1.05\times10^{-13}$, or 1.47\%. These local spectral intervals
are approximations rather than posterior quantiles, particularly for the
weakly constrained cool component. The full Jacobian and covariance were
retained to avoid treating wavelength bins as independent.

We tested the hydrogen-column range suggested by \citet{Wood.et.al.2000ApJ...542..411W},
removal of photons below 0.5 keV, and a different abundance pattern tying C/N
to O and the remaining fixed metals to Fe. The latter changed the integrated
X-ray flux by about 6.7\%, exceeding the conditional statistical interval.
These changes were retained as physical sensitivity tests, rather than
combined into an unjustified Gaussian error.

\subsection{Sanz-Forcada 2025 broad-band comparison}
We used equation 2 of \citet{Sanz-Forcada.et.al.2025A&A...693A.285S} (Eq.~\ref{eq:sf25}),
where the X-ray and EUV bands are 0.5--10 and 10--92 nm, respectively.
For the Gaia DR3 distance of 10.36054 pc, the central Earth-flux
prediction is $1.998\times10^{-11}$ erg s$^{-1}$ cm$^{-2}$.
We propagated an asymmetric logarithmic approximation anchored to the X-ray
profile limits, together with the 0.406-dex predictive scatter. With 32768
draws, the 16th--84th percentile range was approximately
$[0.800,5.111]\times10^{-11}$ erg s$^{-1}$ cm$^{-2}$.
The regression coefficients were fixed at their published central values;
no unavailable coefficient covariance was invented. This uncertainty is a
common amplitude contribution and was not tapered toward spectral joins.
This prediction is retained as a different-epoch comparison, not imposed
on the observed EUVE flux. The adopted EUV spectrum and stellar geometry are
specified below; no universal unmeasured magnetic flux of $10^{24}$ Mx is
used to determine the released EUV spectrum.

\subsection{Ly$\alpha$: literature adoption and an archival calibration discrepancy}
We recovered the public GHRS observations \texttt{z2vq0409t} (G160M) and
\texttt{z2vq0407t} (G270M), taken on 1996 January 9, with calibrated fluxes,
statistical errors, wavelengths and data-quality arrays. Each contains four
FP-SPLIT groups. We constructed diagnostic coadds using the archived
heliocentric wavelength solutions and retained the covariance introduced by
flux-conserving rebinning. Additional relative alignment and fixed-pattern
processing have not yet been reproduced.

\citet{Wood.et.al.2000ApJ...542..411W} reported an intrinsic Ly$\alpha$ Earth flux of
$6.05\times10^{-12}$ erg s$^{-1}$ cm$^{-2}$ with an estimated 25\% error from
profile experiments. The original work did not identify that percentage as
a Gaussian 68\% confidence interval. The value compiled by \citet{Linsky.et.al.2013ApJ...766...69L} is not an independent reconstruction. Our diagnostic
archival coadd integrates to approximately $8.16\times10^{-12}$ over
120.8--122.3 nm before an ISM correction, already above the published
intrinsic value. The individual groups also differ. The processing trailer
confirms that these files underwent the final 2006 CALHRS reprocessing;
the original header date alone would give a misleading impression of their
processing history.

We checked the GHRS continuum outside the Mg~II emission cores against
STIS/E230H observation \texttt{ocd810010} and LOWLIB. The latter two agree
within approximately 3--6\% over the valid, common portions of 278--279 and
280.7--281.5 nm, whereas the GHRS continuum is about seven times larger.
The inconsistency therefore extends beyond the Ly$\alpha$ absorption model.
We did not infer a uniform correction applicable to different GHRS gratings.
The precise pipeline cause has not been established, and these archived
fluxes were not used as the adopted absolute calibration.

Instead, we adopted the published intrinsic line flux and its estimated
25\% reconstruction uncertainty from Wood et al. We recovered the 168
vertices of the intrinsic model in their original vector Fig.~4, available
in the arXiv source archive. The integral over the plotted interval,
approximately 121.4--121.7 nm, is $5.9512\times10^{-12}$ in Earth-flux units.
A factor of 1.0166 normalizes this finite-window template to the tabulated
$6.05\times10^{-12}$. This is an explicit representation of the literature
reconstruction, not a new fit to the inconsistent archival fluxes. The
estimated integrated-flux range is
$[4.5375,7.5625]\times10^{-12}$ erg s$^{-1}$ cm$^{-2}$.
Scaling the template by these limits illustrates the normalization
uncertainty; it does not supply an independently measured pointwise
confidence interval or a profile-shape covariance.

\subsection{IUE--LOWLIB transition and residual scattered light}

We selected the LOWLIB product corrected for scattered light and slit
position, retaining its supplied errors and not automatically applying the
additional dust correction \citep{Pal.et.al.2023ApJS..266...41P}. In 10-nm diagnostic bands,
LOWLIB and the legacy NGSL segment agree closely in the optical, but the
blue UV differs substantially. Large-aperture IUE spectra are typically only
about 0.2--0.34 of LOWLIB in the tested 180--195 nm intervals. We therefore
did not force the IUE scale to that blue excess.

We inventoried 18 local IUE files and their native error and quality arrays.
Checking the MAST target classifications revealed that LWR11877, LWR11878
and SWP15357 are wavelength-calibration or flood exposures, not stellar
spectra; they were excluded regardless of their numerical fluxes or flags.
Eight SWP observations from December 1994 were combined as a distinct epoch
block with exposure/support weights and rebinning covariance retained.
LWP07362 and LWP14454 were retained as low-dispersion checks. We also
retrieved eight public high-dispersion LWP/LWR files, including four LWP
observations contemporaneous with the 1994 SWP campaign. For MXHI spectra,
the wavelength origin was applied at \texttt{STARTPIX}, and the native
noise vector in flux numbers was converted with
$|\mathrm{ABS\_CAL}/\mathrm{NET}|$ \citep{Garhart.et.al.1997IUENN..57....1G}\footnote{NEWSIPS manual, Sections 10.2.5 and 12.9; MXHI layout: \url{https://archive.stsci.edu/iue/manual/newsips/node155.html}.}. Neither the low- nor
high-dispersion LWP data determine a useful blue-continuum normalization.

The documented G230LB scattered-light mechanism is an approximately
ramped count-rate pedestal \citep{Worthey.et.al.2022arXiv221105043W}. We used the difference
between the supplied LOWLIB processing stages to isolate the existing
subtraction template, transferred to the same slit-corrected flux frame:
\begin{equation}
 T_\lambda=(F_{\rm uncorr}-F_{\rm scatter})
 \frac{F_{\rm scatter+slit}}{F_{\rm scatter}},\qquad
 F_{\rm revised}=F_{\rm scatter+slit}-aT_\lambda.
\end{equation}
We fitted an additional amplitude in thirteen 2-nm intervals spanning
170--196 nm, keeping the native SWP scale fixed. The result is
$a=0.3275$ with a conditional amplitude uncertainty of 0.0122, including
an estimated excess band scatter. This is an additional 32.8\% of the
pedestal already subtracted by LOWLIB, not a uniform fractional change
to the stellar spectrum. Blue-half, red-half, and alternative-window fits
give amplitudes near 0.322--0.333. Substitution of the 1985 SWP exposure
gives 0.306 as a separate sensitivity case, not an epoch-error term.

The revised selection uses SWP below a complete bin boundary at 197.45 nm
and the corrected LOWLIB spectrum above it. The additive template vanishes
outside its original G230LB support, leaving the optical spectrum unchanged.
In the withheld 196--197.45 nm interval, revised LOWLIB/SWP is 1.210:
the residual is about 21\%, or 0.80 conditional standard deviations when
the empirical scattered-light model component is included. We interpret
this as consistency within the estimated correction uncertainty, not exact
continuity or independent proof of the extrapolated pedestal.
Independent LWP comparisons use identical valid wavelength intervals;
their blue uncertainties are substantial. Several-percent LWP/LOWLIB
differences in the red UV remain larger than native statistical errors,
and were not removed by forcing a cross-epoch multiplicative scale.

The fitted amplitude is correlated with the data used to estimate it. We
therefore retained this covariance explicitly. Writing $a=k^T y$ and
$v=\mathrm{Cov}(y_{\rm selected},a)$ gives
\begin{equation}
 C_{\rm revised}=C_{\rm native}-vT^T-Tv^T
                 +\mathrm{Var}(a)TT^T.
\end{equation}
The excess band scatter, approximately 0.0408 in amplitude, also defines
a separate empirical model component. For propagation we assume an
exponential correlation length of 2 nm, equal to the diagnostic band width;
this correlation length is not independently measured. The corresponding
envelope is a conditional approximation to uncertainty in the residual
pedestal and its extrapolation, not additional measured detector noise.
No generic instrumental or epoch percentage is added.

The subsequent UV assembly incorporates the intrinsic Ly$\alpha$ template
and the spatial airglow analysis described below. The following sections document the adopted EUV shape constrained by the
recovered EUVE data, the stellar-parameter checks and the infrared extension.

\subsection{Spatial airglow test and ultraviolet assembly}

We retrieved the two-dimensional NEWSIPS SILO images for the eight selected
December 1994 SWP exposures. These images distinguish the cross-dispersion
distribution of a point source from an extended signal within the aperture
\citep{Garhart.et.al.1997IUENN..57....1G}\footnote{NEWSIPS manual, Sections 9 and 12.6; SILO layout: \url{https://archive.stsci.edu/iue/manual/newsips/node152.html}.}. We fitted each wavelength column with
\begin{equation}
 D(y,\lambda)=A_\lambda P(y)+B_\lambda Q(y)+c_\lambda+d_\lambda(y-51),
\end{equation}
where the normalized narrow profile $P$ was estimated from the stellar
C~II and Si~IV regions. The broader profile $Q$ was an empirical blurred
oval-aperture family constrained by the extended Ly$\alpha$ signal.
The calibration from flux numbers to physical flux was recovered from
the corresponding NEWSIPS \texttt{FLUX/NET} vectors. The FITS pixel scale
was applied once, by the FITS reader. Original files were preserved.

We retained the larger of the spatial residual noise and a noise floor
inferred from the supplied NEWSIPS errors under an equal-spatial-variance
approximation. Local profile-parameter errors were propagated as shared
modes, and wavelength rebinning retained the resulting covariance.
Alternative profile widths, centres, diffuse shapes and spatial backgrounds
were evaluated separately. Their range is a method sensitivity, not a
Gaussian confidence interval. The unknown SILO resampling covariance and
its noise cross-covariance with retained NEWSIPS bins were not invented.

Over 119.5--120.4 nm, the original NEWSIPS integral is
$7.850\times10^{-14}$ erg s$^{-1}$ cm$^{-2}$. The fitted point-source
integral is $(3.769\pm1.023)\times10^{-14}$ in the same units, with the
quoted conditional error. The method alternatives span
$[3.156,4.567]\times10^{-14}$. An extended component is present in this
region, consistent with airglow contamination. We adopt the point-source
extraction in the affected N~I bins. These integrals include the local
continuum and any unresolved blends, and are not interstellar-deabsorbed
line fluxes.

For 130.1--130.8 nm, the point-source result is
$(7.544\pm0.435)\times10^{-14}$, compared with the original
$8.346\times10^{-14}$. For 135.3--136.0 nm the corresponding values are
$(2.171\pm0.311)\times10^{-14}$ and $2.105\times10^{-14}$.
Neither oxygen interval establishes a significant diffuse component in
this test. We therefore retain the original NEWSIPS fluxes and supplied
errors there. The first interval is an O~I/Si~II blend at this resolution.
The held-out C~IV region agrees with the original extraction within the
tested method range; C~II and Si~IV were training controls.

We also retrieved the calibrated GHRS/G140L observations
\texttt{z2vq040ht} and \texttt{z2vq040gt}, including their error and quality
arrays. Their fluxes were retained on the archive scale. The N~I-region
integral in the former is $(2.105\pm0.233)\times10^{-14}$, useful as a
separate-epoch check. These large-aperture observations are not assumed
to be perfectly free of airglow, and were not rescaled to force agreement.

The contaminated IUE Ly$\alpha$ bins were replaced by the Wood template
plus an interpolated local point-source continuum. Exact integration of
the piecewise-linear template preserves the adopted
$6.05\times10^{-12}$ line integral. Its common 25\% estimated normalization
envelope remains separate from conditional statistical errors. Seven
negative low-signal UV bins are retained in the signed product. The nonnegative irradiation companion and the final plotting convention are described below and in Section~\ref{sec:display}.

\subsection{Direct EUVE constraint on the proposed EUV shape}

The stellar atlas of \citet{Craig.et.al.1997ApJS..113..131C} provides separate
SW, MW and LW spectra of GJ~117/EP~Eri from 1994 December 2--8, concurrent
with the selected IUE block. Each FITS product includes spectral flux
and sigma planes. We converted the documented photon fluxes per angstrom
to energy fluxes per nm, retained negative measurements, and compared
identical wavelength support. Our diagnostic joins use SW at 10--17 nm,
MW at 17--33 nm and LW above 33 nm.

As a processing check, integration of the emission-line windows used by
\citet{Poppenhaeger.2022MNRAS.512.1751P} gives
$(9.97\pm0.83)\times10^{27}$ erg s$^{-1}$ at 10.355 pc, consistent with
the published $9.5\times10^{27}$. This uses the same atlas and is not
an independent observation, nor the total 20--50.4 nm luminosity.

We evaluated the proposed Namekata spectrum with the recorded magnetic
input and the Sanz-Forcada 2025 normalization, and attenuated the model
before comparing it with observed EUVE fluxes. The nominal absorption
scenario uses $\log N({\rm H\,I})=18.05$ from Wood et al., assumed
$N({\rm He\,I})/N({\rm H\,I})=0.1$ and zero He~II, with the cross-sections
of \citet{Verner.et.al.1996ApJ...465..487V}. Alternative hydrogen columns and
helium fractions were retained as sensitivity cases, not fitted intervals.

The observed-to-proposed ratios vary substantially with wavelength.
Even allowing a free common amplitude gives approximately 0.187 and
$\chi^2=103.4$ for eight degrees of freedom in the selected broad bands.
This is a conditional diagnostic based on the atlas errors and fixed
absorption scenario, not a test including all model and calibration
uncertainties. Nevertheless, a global rescaling does not reproduce the
relative band strengths. We therefore do not promote that N23+S25 shape
to the final SED. The 2025 relation remains a broad-band proxy for
comparison; the observed EUVE spectrum should constrain the shape.

The 30.4 nm band is excluded from this free-amplitude diagnostic, and
the long-wavelength excess above 50 nm is not adopted as stellar flux.
The archive documentation identifies geocoronal He~I 58.4 nm and
higher-order He~II contamination\footnote{MAST EUVE FAQ, question 4, instrumental and geocoronal spectral features: \url{https://archive.stsci.edu/euve/faq.html}.}. The following analysis
addresses the order correction and spatial contamination, and replaces
the initial FUSE preview inspection with exposure-level quality checks.

\subsection{EUVE order correction and spatial validation}

We retrieved the original night-image and event products and the official
\texttt{egodata1.17} effective areas for spectral orders one through eight.
Following the bootstrapping principle of the EUVE \texttt{suborder}
software, we predict each higher-order contribution from shorter
wavelengths before correcting interstellar absorption. The redistribution
conserves photon number, including the wavelength Jacobian and the
conversion between photon and energy flux density. Written as
$y=(I+K)f$, the correction gives
\[
 f=(I+K)^{-1}y,\qquad
 C_f=(I+K)^{-1}C_y(I+K)^{-T}.
\]
We retain the resulting covariance when integrating or rebinning the
spectrum. The shortest EUVE wavelengths initialize the first-order
spectrum. Adding the different-epoch 2004 X-ray model below 6.65~nm as a
sensitivity case changes the inferred 10--40~nm integral by only
$2.3\times10^{-15}$ erg s$^{-1}$ cm$^{-2}$; this is not used to assert
constant activity between epochs.

The automatically extracted SW spectrum requires particular care.
Its aperture is centered on spatial pixel 137, whereas two separate
coronal intervals locate the repeated stellar trace at
$145.31\pm0.27$ and $144.89\pm0.34$ pixels. The image and original events
show an isolated concentration near apparent wavelength 11.4~nm and
pixel 137. A two-component extraction gives a 10--17~nm pre-order flux
of approximately $(4.81\pm0.78)\times10^{-13}$ erg s$^{-1}$ cm$^{-2}$,
close to the atlas value; the automated aperture gives a substantially
larger value. We therefore retain the atlas scale and reject the
automated SW spectrum as an absolute-flux reference. This decision is
based on the spatial separation and repeated stellar trace. We do not
assign a physical origin or stellar line identification to the isolated
feature.

For He~II 30.4~nm, orbital night alone does not eliminate diffuse emission.
We infer the narrow source profile from MW coronal emission at 20--29~nm
and fit it together with a smooth spatial background. In the
30.15--30.60~nm window the narrow component contains approximately
$80\pm39$ counts, a weak detection. Its pre-order energy flux is
$(6.39\pm3.23)\times10^{-14}$ erg s$^{-1}$ cm$^{-2}$.
The adopted 30.0--30.8~nm interval uses this signed source estimate with its own
conditional covariance. Background degree, extraction-window size and
spatial-profile parameter shifts are retained as method bounds rather
than independent Gaussian errors. The release does not set weak or
negative measurements to zero. Unprovided covariance between the custom
extraction and adjacent atlas pixels is not fabricated.

After order correction and nominal H/He transmission correction, the
10--40~nm integral is
$(3.32\pm0.26)\times10^{-12}$ erg s$^{-1}$ cm$^{-2}$, where the quoted
uncertainty is conditional statistical uncertainty only. Hydrogen-column
and helium-fraction alternatives are supplied separately. Fixed 1~nm
measurement bins retain four negative noise estimates; these are not
interpreted as negative intrinsic emissivity.

\subsection{Unobserved EUV bands and native FUSE checks}

We evaluate the 40--91.2~nm solar-model band relations of
\citet{Linsky.et.al.2014ApJ...780...61L}. Their Section~IV uses the
Ly$\alpha$ flux at 1~AU scaled by $(R_\odot/R_\star)^2$; this factor,
omitted in an earlier diagnostic, is included here. The present values
use the Gaia distance 10.36054~pc and the final DUNES-consistent radius
0.76718~$R_\odot$, derived below. The estimated Wood Ly$\alpha$ envelope is propagated
through the nonlinear band relations. The empirical dispersions
published for the observed short-EUV bands are not assigned to the
unobserved solar-model bands as though they were measured prediction
errors.

In the three 10~nm EUVE comparison bands, the measured integrals are
$(1.08,1.17,1.08)\times10^{-12}$ erg s$^{-1}$ cm$^{-2}$, compared with
$(1.00,1.00,0.79)\times10^{-12}$ from the Ly$\alpha$ relations. These
differences are consistent with the reported empirical dispersions.
EP~Eri and this same EUVE observation are included in their Table~3 and
short-EUV calibration, so this is an in-sample consistency and processing
check, not independent validation. Their adopted radius was 0.93 solar
radii. We reevaluate the relations with our internally consistent stellar
parameters rather than copying their radius into the new surface conversion.
Combining the EUVE constraints and the long-band estimates gives a
final 10--91.2~nm integral of
$6.36\times10^{-12}$ erg s$^{-1}$ cm$^{-2}$. Long-band densities are
reported as band averages, without inventing a resolved line spectrum.
The Sanz-Forcada 2025 prediction remains a separate comparison in its
exact 10--92~nm band; it is not used to force a normalization onto the
1994 EUVE measurements.

For FUSE observation E1610201, we inspected all nine exposures in each
of eight channels, including the supplied errors and \texttt{QUALITY}
arrays. This quantity is an extraction-window percentage, not an HST
data-quality bit mask. We also retrieved the native night-only
\texttt{ANO} products \citep{Sonnentrucker.et.al.2009FUSE}. Night selection reduces the C~III
97.7~nm SiC1B/SiC2A window fluxes from
$(1.85,5.09)\times10^{-13}$ to
$(0.983\pm0.091,0.814\pm0.071)\times10^{-13}$ erg s$^{-1}$ cm$^{-2}$.
The night measurements differ by approximately 1.5 combined statistical
standard deviations, while the day data
show the strong channel-dependent contamination expected from scattered
sunlight. For O~VI 103.2~nm, the night LiF1A and LiF2B fluxes are
$(1.197\pm0.034)$ and $(1.176\pm0.047)\times10^{-13}$ in the same units.
LiF1B gives a lower C~III 117.5~nm flux than LiF2A; we select LiF2A for
this region and retain both as a processing check, without forcing a
relative scale correction.

\subsection{FUSE completion and the IUE junction}

The final selection uses one complete-support night channel per 0.01~nm
bin. A fixed preference order and the quality masks choose channels without
conditioning on measured flux or its sign. Overlapping channels are not
averaged as independent measurements. When a preferred channel lacks valid
support, another calibrated night channel supplies the bin if possible.
The supplied errors are propagated through the explicit overlap operators,
including covariance when a native sample contributes to adjacent bins.

Night selection does not make the absorbed H~I profiles uniquely invertible.
We therefore identify their masked windows as model completions. We use the
Lyman-series relation in Table~5 of \citet{Linsky.et.al.2014ApJ...780...61L} (Eq.~\ref{eq:lyseries}),
with fluxes in the units specified by that calibration. The relative
partition among H-dominated masked regions follows the quiet-Sun WHI/SIRS
reference spectrum \citep{Woods.et.al.2009GeoRL..36.1101W}; a local background estimated from the
adjacent solar bins is removed before normalizing the line excess.
The replaced intervals are 91.2--93.3, 93.7--93.9, 94.9--95.1,
97.1--97.4 and 102.4--102.9~nm. These coarse regions can include blended
metal emission, which is not separately resolved by this approximation.
Other geocoronal/neutral-line, second-order or remaining quality gaps use
a Ly$\alpha$-scaled solar template. The actual 0.1~nm template resolution is
retained rather than presented as a high-resolution stellar reconstruction.

The estimated Wood normalization range is propagated coherently, with
the nonlinear exponent 1.351 for the Lyman-series contribution. Direct
solar Ly$\alpha$ scaling and the empirical full-FUV normalization provide
additional method alternatives. Their span is a sensitivity envelope,
not a measured uncertainty on individual stellar line ratios. The three
WHI columns do not provide independent short-wavelength activity spectra:
the same rocket EVE spectrum supplies 6--106~nm. No spurious inter-epoch
variance is estimated from those identical values.

Unmasked FUSE bins remain observed night flux. We do not claim that every
metal line has been corrected for interstellar absorption. Strong usable
C~III and O~VI windows retain their measured fluxes and supplied errors.
The final 91.2--117~nm composite integrates to approximately
$1.16\times10^{-12}$ erg s$^{-1}$ cm$^{-2}$; its known statistical component
is $3.22\times10^{-14}$, while 5.2~nm of the interval contains labelled
proxy completion. That statistical value is therefore not the total
prediction error of the integrated composite.

FUSE extends to 118.1~nm before the adopted IUE segment. The common broad
117.0--118.1~nm measurement interval gives
$(9.21\pm0.66)\times10^{-14}$ and
$(9.66\pm1.63)\times10^{-14}$ erg s$^{-1}$ cm$^{-2}$ for FUSE and IUE,
respectively. Their difference is only 0.26 combined statistical standard
deviations. Broad common windows avoid comparing line peaks at different
resolutions. We do not apply a fitted scale to force agreement between
the 2004 FUSE and 1994 IUE epochs.

\subsection{Consistent parameters and the optical--infrared extension}

Gaia DR3 source 5160075762132996992 has parallax
$96.52004\pm0.02584$ mas and RUWE 1.015, giving
$d=10.36054\pm0.00277$ pc from direct inversion. The quoted uncertainty is
the formal catalog value; no unverified zero-point adjustment is applied.
\citet{Eiroa.et.al.2013A&A...555A..11E}, online Table~4, give
$T_{\rm eff}=5217$~K, $\log g=4.57$, $[\mathrm{Fe/H}]=+0.10$ and
$L=0.392L_\odot$ at 10.35~pc. Rescaling that luminosity to Gaia and applying
the Stefan--Boltzmann relation with the same temperature gives
\[
 L=0.39280L_\odot,\qquad R=0.76718R_\odot.
\]
This avoids combining the older $0.85R_\odot$ estimate with an incompatible
temperature--luminosity pair. The adopted $0.85\pm0.05R_\odot$ range from
\citet{Ambruster.et.al.2003csss...12..912A} is retained as a method alternative,
not reinterpreted as a Gaussian error on the DUNES-derived radius.
The DUNES activity ages, 0.14 and 0.40~Gyr, are method estimates, not a
confidence interval. The literature multiplicity status is uncertain;
the product represents the unresolved target and does not invent a
secondary contribution or component decomposition.

We replace the legacy infrared normalization with the official PHOENIX
NewEraV3 grid \citep{Hauschildt.et.al.2025A&A...698A..47H}. Eight corners bracket 5100--5300~K,
$\log g=4.5$--5.0 and $[\mathrm{Fe/H}]=0$--0.5. The downloaded files are
checksum-verified, and integration of each low-resolution model reproduces
its $\sigma T_{\rm eff}^4$ bolometric flux. Trilinear interpolation gives
the adopted atmospheric parameters. Before fitting a scale, we integrate
the high-resolution models over the original LOWLIB bins, rather than
comparing samples at unequal resolutions.

The 800--950~nm LOWLIB anchor gives a geometric normalization
$a=2.74057\times10^{-18}$, with conditional statistical uncertainty
$3.95\times10^{-22}$. The corresponding model-dependent radius is
$0.76073R_\odot$, close to the DUNES-consistent value. This is a useful
consistency check, not a second independent radius measurement. The
reserved 950--1000~nm interval has integrated LOWLIB/model ratio 0.97970.
Alternative anchor windows change the scale by approximately one percent.
The extension begins at the actual LOWLIB boundary 1000.15034~nm and
continues to 5000~nm; its bins have resolving power approximately 5000.
Grid-corner and anchor-window spectra are retained as model sensitivities,
not assigned an unsupported Gaussian atmosphere error.

\subsection{Uncertainty representation and final product}

The final FITS product is contiguous from 0.5 to 5000~nm and stores explicit
bin edges, central fluxes at Earth and at the adopted stellar surface,
source/estimate flags, covariance components and method scenarios. The
surface conversion is $(d/R)^2=3.58781\times10^{17}$. Since the adopted
radius is derived from a luminosity that scales as $d^2$, its formal
distance dependence cancels from $d/R$. Treating those two quantities as
independent draws would count the parallax contribution incorrectly.
Radius alternatives nevertheless act coherently on every surface bin and
on the radius-scaled EUV and Lyman-series estimates.

The X-ray Jacobian and parameter covariance, EUVE order-correction
covariance, FUSE resampling covariance, and the UV statistical and
conditional LOWLIB correction covariances are supplied. For the infrared
extension, write $a=w^T f_{\rm optical}$ and $f_{{\rm IR},i}=a m_i$.
For a band-integration vector $z$, the effective optical weights become
\[
 z_{\rm optical}\longrightarrow z_{\rm optical}
       +(z_{\rm IR}^Tm)w.
\]
Evaluating the optical covariance once with these effective weights retains
the shared infrared normalization and optical--infrared cross terms.
The complete matrices pass symmetry and positive-semidefinite checks to
numerical precision, and the public notebook reproduces the released band
integrals and their known covariance components directly from the FITS.

The reconstructed Ly$\alpha$ normalization, proxy alternatives, plasma
assumptions, atmospheric grid and geometry choices remain identifiable.
They are not converted into independent pixel errors or equally probable
posterior samples. No arbitrary instrumental or epoch percentage is added.
The release closes the budget of quantified components while retaining
the distinction between measurement uncertainty and model sensitivity.
Unknown universal total errors are not encoded as zero. In particular,
the masked FUSE estimates and long-EUV band averages do not become direct
measurements simply because the composite has full wavelength coverage.

The public package contains the final FITS, its schema and provenance,
integrity hashes, and a small executed viewing notebook. Full reduction
notebooks and inputs remain in the analysis workspace. The viewer shows
the full spectrum with and without uncertainty shading and selected
spectral regions. Central curves are solid, uncertainty regions are
translucent, and optional PDF export is disabled by default. Signed noisy
data remain in the FITS and in linear-axis diagnostic panels; logarithmic
displays do not turn negative measurements into positive flux.

\subsection{A non-negative companion spectrum for irradiation calculations}
Negative background-subtracted flux estimates were retained in the reviewed
observational composite. They contribute to unbiased signed band sums, but
are not physically admissible as incident stellar radiation. Setting each
negative EUVE bin to zero would increase the measured 10--40 nm integral by
4.84\%, and the full 10--91.2 nm integral by 2.53\%. We therefore constructed
a separate irradiation product through adaptive signed-energy-conserving
grouping, motivated by the negative-bin prescription in the MUSCLES data
documentation\footnote{\url{https://archive.stsci.edu/hlsps/muscles/hlsp_muscles_readme.md}}.
The original composite is unchanged and its signed measurements are also
embedded in the companion FITS.

The algorithm sums the signed integrated flux of a negative bin with that
of an adjacent bin, and divides by their combined wavelength width. At each
step the adopted neighbour minimizes the resulting wavelength span, with
ties resolved toward the blue. We require the same grouping to yield a
non-negative spectrum for the reference SED and all 33 existing named method
scenarios. Protected boundaries include the spectral joins, selected strong
FUSE-line windows, standard EUV band boundaries and the H/He ionization
thresholds used in the cross-section calculation. A native bin intersecting
a protected boundary is divided assuming its tabulated mean density, with
the shared parent covariance retained. No independent spectral information
is created by this division. An X/EUV interface discrepancy of
$2.68\times10^{-9}$ nm, arising from numerical rounding, is reconciled with
an energy-preserving weight. A negative total within a protected interval
would cause an explicit failure rather than clipping or inventing emission;
no such failure occurred for the adopted scenarios.

Writing the sparse grouping matrix as $A$, the flux-density vector and its
conditional covariance obey
\begin{equation}
 \boldsymbol f_{\rm irr}=A\boldsymbol f_{\rm native},\qquad
 C_{\rm irr}=A C_{\rm native} A^{\mathsf T}.
\end{equation}
The output has 42,418 bins instead of 43,583, with 296 output bins combining
multiple native measurements. The 724 negative central estimates are
eliminated. The largest adopted EUVE bin is 4.0 nm wide and the largest bin
in 91.2--118.1 nm is 1.94 nm wide. This loss of resolution is explicit in
the bin edges; the coarsened flux is not interpolated onto a nominally fine
grid. The total energy and the protected band integrals are conserved to
numerical accuracy, including $F_{10-91.2}=6.3630\times10^{-12}$
erg s$^{-1}$ cm$^{-2}$. The adopted Ly$\alpha$ line remains
$6.05\times10^{-12}$ erg s$^{-1}$ cm$^{-2}$.

The FITS stores $A$, the original covariance factors, the native measurements,
and the transformed non-negative method scenarios. For any output integration
weights $\boldsymbol z$, the native covariance is evaluated using
$A^{\mathsf T}\boldsymbol z$, preserving the shared optical/IR normalization
and other correlations. Known statistical and LOWLIB model components remain
separate from method sensitivities. This propagation is conditional on the
chosen grouping, which is selected from noisy data. To examine that selection,
we generated 1,000 correlated Gaussian EUVE realizations using the native
30-bin covariance and selected a new grouping for each realization. None
failed the protected-interval requirement. Their sampling summaries and
alternative blue-neighbour/red-neighbour groupings are supplied as additional
diagnostics, not interpreted as a universal posterior confidence interval.
Gaussian measurement bounds can still extend below zero; such bounds are
not physical incident spectra. The central irradiation spectrum and all
named physical alternatives are non-negative.

Energy conservation alone does not guarantee unchanged ionization rates.
Using the cross sections of \citet{Verner.et.al.1996ApJ...465..487V}, we integrated
the bin-averaged spectrum against the photoionization kernel
$\sigma(\lambda)\lambda/(hc)$ and the primary excess-energy kernel
$\sigma(\lambda)[1-E_{\rm th}\lambda/(hc)]$. The latter represents energy
available per target atom before a specified atmospheric heating efficiency;
it is not a hydrodynamic mass-loss calculation. All quadrature intervals are
split at relevant atomic thresholds. Increasing the quadrature order from
20 to 48 changed the checked rates by less than $10^{-8}$ fractionally.
Relative to the original signed estimator, the unattenuated rates change as
follows:
\begin{center}
\begin{tabular}{lrr}
\toprule
Species & Photoionization change & Excess-energy change\\
\midrule
H I   & $+0.086\%$ & $+0.499\%$\\
He I  & $+2.564\%$ & $-0.120\%$\\
He II & $+0.908\%$ & $+0.623\%$\\
\bottomrule
\end{tabular}
\end{center}
The changes are smaller than the corresponding propagated known statistical
standard deviations for these integrated unattenuated rates. We also provide
diagnostic neutral-screen calculations for $N_{\rm HI}=10^{17}$, $10^{18}$
and $10^{19}$ cm$^{-2}$, with $N_{\rm HeI}=0.1N_{\rm HI}$ and
$N_{\rm HeII}=0$. These screens test sensitivity to spectral redistribution;
they do not prescribe a planetary atmosphere. The treatment follows the
usual photon and excess-energy integrals, as used, for example, by
Caldiroli et al.\ (2021, A\&A, 655, A30;
\href{https://doi.org/10.1051/0004-6361/202141497}{doi:10.1051/0004-6361/202141497}).

For a planet at distance $a$ AU, the companion's 1-AU flux-density column
is divided by $a^2$. All wavelengths are in nm and energy-flux densities
are in erg s$^{-1}$ cm$^{-2}$ nm$^{-1}$. The formal distance error is retained
as a separate, fully correlated
normalization component $2\sigma_d/d\simeq0.0536\%$ at fixed orbital distance.
Band integrals use the actual bin widths,
not a trapezoidal approximation through bin centres. A portable notebook
demonstrates this scaling, covariance-aware band integration, the complete
spectrum with and without translucent uncertainty shading, and the altered
weak-flux regions. The original EUV proxy assumptions and masked-region
provenance remain applicable; grouping solves the sign constraint without
claiming to recover unresolved intrinsic line structure.

\FloatBarrier
\begin{minipage}{\linewidth}
\subsection{Final spectrum with available uncertainties}
Figure~\ref{fig:sed:EP_Eri} shows the adopted full SED and its available uncertainty bounds.

\begin{center}
\includegraphics[width=\linewidth]{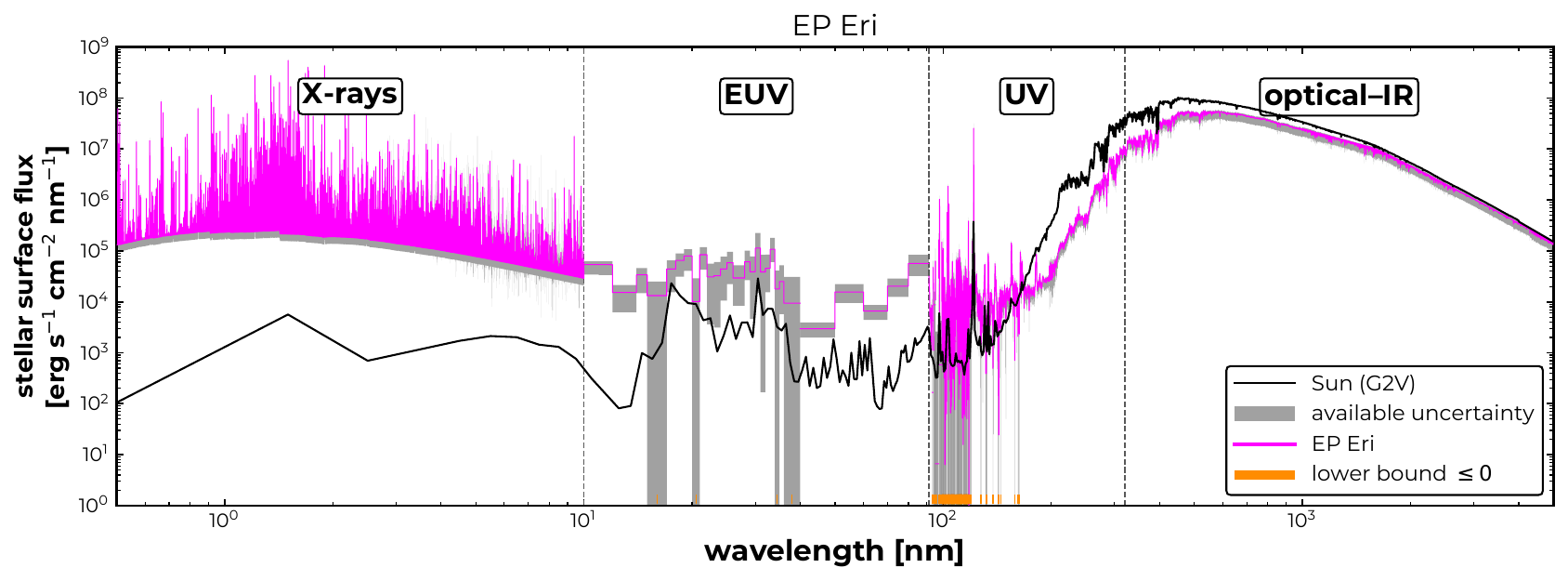}
\end{center}
\captionof{figure}{EP Eridani: final nominal surface-flux spectrum and available uncertainties. Target colours, shading, solar comparison and nonpositive-bound markers follow Section~\ref{sec:display}. The component origins and uncertainty scope are documented above.}\label{fig:sed:EP_Eri}
\end{minipage}
\FloatBarrier

%% file: sections/pi1_UMa.tex
\clearpage
\section{\texorpdfstring{$\pi^1$ Ursae Majoris}{pi1 Ursae Majoris}}\label{star:pi1_UMa}
This section describes the adopted observations and reconstruction of
$\pi^1$ UMa, with emphasis on the X-ray fit, direct EUV constraints, and the
ultraviolet lines that constrain unobserved intervals. It then connects these
components to the photospheric spectrum and explains the signed and
nonnegative products and their correlated uncertainties.

\subsection{Parameters and observing epochs}
The target is identified with GaiaDR 3 source 1092545710514654464. Its parallax
gives $d=14.43885\pm 0.01011$pc (formal uncertainty, without an unverified
zero-point correction). We use $R=0.95 R_\odot$, $T_{\rm eff}=5873$K and
$\log g=4.44$ from the stellar compilation used by \citet{Namekata.et.al.2023ApJ...945..147N}; the reported
radius uncertainty is 0.01$R_\odot$. The atmosphere uses [Fe/H]$=-0.11$ from
the DUNES compilation \citep{Eiroa.et.al.2013A&A...555A..11E} as a scaled-solar [M/H] approximation.
An alternative internally consistent DUNES temperature/luminosity solution,
rescaled to the Gaia distance, gives $R=0.9784 R_\odot$ and is documented as a
geometry alternative. Radius and distance are not assumed independent in a
claimed total uncertainty. The adopted 500$\pm$100 Myr group age follows the
discussion in \citep{Wood.et.al.2014ApJ...781L..33W}; it does not enter the empirical spectral joins.

The SED combines EUVE observations in 1998, XMM--Newton in 2000, FUSE in 2003,
STIS E230H in 2010, STIS E140M in 2012, IUE observations in 1980--1994, and GaiaDR 3.
It is consequently a multi-epoch representation. The source identifiers and
selection rules are distributed with the products. No stellar component
decomposition is imposed. The supplied LOWLIB source list contains no exact
match to this target; another star is not substituted for it.

\subsection{X-rays: re-extraction, model selection and uncertainties}
We re-extract MOS1 imaging observation 0111400101 from its calibrated event
list. A 10--12 keV, single-event background light curve is constructed in 100 s
bins. The nominal threshold is 0.35 counts\,s$^{-1}$; thresholds 0.2 and 0.5 are
tested separately. Source and background regions follow the target-identified
PPS geometry, and source/background spectra, area factors and responses are
rebuilt with SAS 22.1 and the calibration files available locally. This is
re-extraction of calibrated events, not a claim of new ODF calibration. The
legacy pn Timing soft-band fit is not used as the reference spectrum.

The nominal MOS1 fit spans 0.2--2.4 keV and uses TBabs with three vapec
components (APEC 3.1.3; Wilms abundance reference). O, Ne, Mg, Si and Fe vary,
with common abundances across temperatures. C and N track O; Al, S, Ar, Ca and
Ni track Fe, while He is fixed. These ties are physical modelling assumptions,
not separate abundance measurements. They improve the W statistic from 291.36
to 279.06 at 278 degrees of freedom relative to the solution fixing the remaining
metals at solar values. Multiple starts and the flux-profile search are used
to avoid adopting a local minimum. The final temperature components are
approximately 0.20,0.40 and 0.75 keV.

We retain a Poisson source/background likelihood rather than interpreting
$W/\nu$ as reduced chi-square. In 60 Poisson simulations refitted with the same
procedure, half have a larger W statistic than the observed spectrum. This
plug-in goodness calculation supports adequacy at this sample size; it is not
the probability that the model is physically true. The exact, unabsorbed
0.5--10 nm flux is $4.5704\times 10^{-12}$\,\fluxunit. Its conditional
$\Delta W=1$ profile interval is
$[4.5142,4.6541]\times 10^{-12}$\,\fluxunit, re-optimizing all free temperatures,
abundances and relative emission measures. The released spectral covariance
uses a checked local log-parameter Jacobian; its integrated standard deviation
is $7.55\times 10^{-14}$\,\fluxunit.

Under the adopted abundance grouping, changing the background threshold or
the lower fitted energy gives 0.5--10 nm fluxes near
$(4.49$--$4.57)\times 10^{-12}$\,\fluxunit. The alternative solar-other-element
pattern permits a much larger cool-component extrapolation, up to
$1.10\times 10^{-11}$\,\fluxunit\ in the tested 0.3 keV-cut case. These are
named model alternatives, not extra independent Gaussian errors. The exact
APEC version, fitted band, exported band and parameter constraints accompany
the products.

\subsection{Intrinsic \lya\ and neutral ultraviolet lines}
Two native STIS E140M exposures are fitted jointly with their pipeline ERROR
arrays and serious-quality masks. A self-reversed Voigt emission profile is
attenuated by interstellar H I and D I and an additional broad H I absorber,
then convolved with the tabulated 0.2-arcsec STIS line-spread function and
integrated over native pixels. The self-reversal strength varies. Constraints
on stellar velocity and Mg II interstellar velocity are used, but neither the
published H I column nor an empirical \lya\ flux is imposed as a flux prior.
The geocoronal interval is masked using the heliocentric-correction sign
appropriate to the native STIS products; its centre is near$+20$km\,s$^{-1}$.

Two independent 40-walker ensembles, using stretch and differential-evolution
moves, yield 25000 retained steps per walker after the final burn-in. The shortest
retained-length/autocorrelation-time ratios exceed 82. Their flux estimates
agree. We obtain an intrinsic 120.8--122.4 nm flux
$F_{\mathrm{Ly}\alpha}=(5.813\pm 0.200)\times 10^{-12}$\,\fluxunit\ (posterior mean and
conditional standard deviation). The 16 th--84 th percentiles are
$[5.621,6.008]\times 10^{-12}$\,\fluxunit. The interstellar column is
$\log N({\rm H\,I})=18.119^{+0.013}_{-0.013}$, consistent with 18.12 in
\citep{Wood.et.al.2014ApJ...781L..33W}, which analysed the same 2012 data. The broad absorber is
supported within this profile family; its parameters are not a new unique
astrospheric-wind solution. Line-spread function, airglow-mask width, D/H and
turbulence alternatives give integrated fluxes between 5.19 and 6.04 in the
same units. These method tests are separate from posterior percentiles.

As external consistency checks, continuum-subtracted C II 1335.7\AA\ and the
C IV 1548+1551\AA\ doublet predict \lya\ fluxes of 6.66 and
$7.82\times 10^{-12}$\,\fluxunit, respectively, using the F 5--G 9 relations of \citet{Linsky.et.al.2013ApJ...766...69L}. The reconstruction is about 13\% and 26\% lower, compared with
the published empirical RMS values of 22.1\% and 22.2\%. The C II relation is
applied to 1335.7 alone, not the sum with 1334.5. The two proxies share calibration
information and are not combined as independent measurements.

N I and O I source/sky diagnostics are evaluated relative to the
wavelength-dependent STIS trace. The O I spatial wings resemble stellar
control lines; no unmeasured diffuse-sky subtraction is imposed on these
narrow-slit spectra. Native errors are retained. Tiny rejected-pixel gaps use
flagged linear estimates with propagated covariance; larger echelle gaps use
the measured coarse IUE donor, with its shared-bin covariance. This does not
recover unresolved line structure or imply an interstellar correction to
every metal line.

\subsection{Observed EUV and the unobserved interval}
Both night EUVE visits show a source at the target trace. Counts in a
19-pixel source aperture and symmetric background strips are propagated as
Poisson source/background measurements. Trace width, aperture, background
strip and centring alternatives are checked. The coadded spectrum uses live
time weights, the detector wavelength solution, and the supplied effective
area/order calibration. Higher-order subtraction, H I/He transmission and
flux-conserving rebinning are linear operators with retained covariance.
The reference uses $\log N({\rm H\,I})=18.12$, He I/H I$=0.1$ and no He II;
column and helium alternatives are explicit. The different-epoch X model
below 7 nm is tested as a higher-order sensitivity rather than imposed on the
1998 observations.

The intrinsic 10--40 nm measured flux is
$(3.332\pm 0.377)\times 10^{-12}$\,\fluxunit\ at fixed transmission. Tested
extraction/ISM choices span$(2.897$--$3.871)\times 10^{-12}$\,\fluxunit.
Signed noise estimates are preserved. The two visits are consistent within
their measurement uncertainties. The 40--91.2 nm region is represented by five
band estimates of \citet{Linsky.et.al.2014ApJ...780...61L}, conditioned on the common \lya\
posterior, rather than by an invented resolved emission-line spectrum. The
short-EUV Linsky predictions provide a consistency test against EUVE; earlier
Sun-in-Time observations also informed this literature, so this is not an
independent validation of the calibration.

We test Namekata's relation using the published hemispheric magnetic flux
$\Phi=1.64\times 10^{25}$Mx \citep{Namekata.et.al.2023ApJ...945..147N}, without a universal
$10^{24}$Mx fallback. Its short-EUV prediction exceeds the measured flux by
roughly an order of magnitude in some bands. Similar discrepancies and
non-simultaneous magnetic/XUV observations are discussed by those authors.
The magnetic estimate is therefore a method comparison rather than a reason
to renormalize direct EUVE measurements.

The latest adopted X--EUV comparison is \citet{Sanz-Forcada.et.al.2025A&A...693A.285S} (Eq.~\ref{eq:sf25}),
with $L_X$ in 0.5--10 nm, $L_{\rm EUV}$ in 10--92 nm, luminosities in
erg\,s$^{-1}$, and 0.406 dex residual scatter. The X flux profile and physical
alternatives are propagated through this relation. The nominal predicted
Earth flux is$1.232\times 10^{-11}$\,\fluxunit, while the hybrid 10--92 nm
integral is$6.687\times 10^{-12}$\,\fluxunit, about 0.27 dex lower. The relation
is a comparison; its scatter is not independently added to every EUV bin or
double-counted in the adopted Linsky/EUVE uncertainty.

\subsection{FUSE, near-UV and photospheric completion}
Three archival FUSE observations are inspected. The long 2003 D9100201 night
product is selected after auditing 128 exposure/channel QUALITY arrays and
the native coadd shifts. QUALITY measures the usable extraction-aperture
percentage, not an HST-style bit mask. A fixed channel preference is used;
bins are not selected by positive flux or by the smallest noisy error.
The selected LiF channels agree in O VI, and the red LiF channels agree in
C III 1175 within their errors. Night selection strongly suppresses Ly$\beta$
and neutral airglow, while O VI remains stable. The observed C III 1175 flux
is$7.50\times 10^{-14}$\,\fluxunit\ in 2003 and$1.18\times 10^{-13}$\,\fluxunit\
in 2012 STIS. No whole-spectrum scale is inferred from this single line.

Strong H I and neutral-airglow windows are labelled estimates. WHI/SIRS
supplies their coarse relative shape \citep{Woods.et.al.2009GeoRL..36.1101W}; the H-dominated
windows use the Linsky 2014 Lyman-series relation, and other masked intervals
use \lya-scaled solar flux. The same \lya\ draws propagate through all
affected intervals. Alternative solar normalizations are retained as method
sensitivities. These proxies are not recovered stellar observations.

IUE large-aperture, positive-exposure, target-confirmed SWP/LWP observations
are coadded after sample-level quality and saturation rejection. LWR17268 is
excluded following its large red calibration discrepancy. The SWP continuum
scale is 0.6696$\pm$0.0111 relative to STIS, using 160--163 and 166--170 nm to
avoid He II 164 nm. The LWP scale is 1.0451$\pm$0.0031 from 257--277 nm against
E230H. Errors are conditional scale errors; alternate windows and unscaled
spectra are explicit sensitivities. No wavelength taper is used to conceal
a join or to assert a unique cause for epoch/calibration differences.

Gaia XP is calibrated with its full coefficient covariance. The adopted
339--1011 nm interval avoids strong blue-edge ringing and the red-edge
downturn. PHOENIX NewEra \citep{Hauschildt.et.al.2025A&A...698A..47H}, interpolated in temperature,
gravity and metallicity, supplies 335--339 nm and 1011--5000 nm. One scale is
derived from the 950--1010 nm Gaia integral. Its uncertainty and its cross
covariance with Gaia are retained. Atmosphere and anchor alternatives are
separate sensitivities; a synthetic photosphere is not assigned an invented
measurement error.

\subsection{Distributed uncertainty and nonnegative irradiation products}
The statistical/conditional covariance is stored as a sparse base term,
low-rank calibration corrections, an X-ray Jacobian/parameter-covariance
term, and shared \lya\ posterior deviations. In schematic form,
\begin{equation}
C=C_{\rm base}+UKU^{\mathsf T}+JC_XJ^{\mathsf T}+C_{\rm posterior}.
\end{equation}
The metric $K$ includes cross terms and need not itself be positive definite.
Only the complete sum is a covariance. Integrated variances are evaluated
as $w^{\mathsf T}Cw$, with the actual wavelength-bin widths $w$.
Observational and conditional error columns must not be added to one another.
Named method scenarios are not equiprobable posterior draws. The geometry is
held fixed for surface-flux errors, and no unmeasured instrumental or epoch
variance is fabricated.

The signed product is retained for inference. A second product pools adjacent
bins until central flux densities are nonnegative, preserving signed energy
at the protected broad-band boundaries. The same linear map transforms all
covariance factors and method scenarios and is supplied as REBIN\_MAP.
This grouping is data dependent: the quoted propagated errors condition on
the frozen grouping and do not constitute a positivity-constrained posterior.
Negative lower error bounds can still occur and are shown on linear plots.
Logarithmic plots do not force such bounds to an arbitrary positive floor.
The complete computational record remains in the working notebook; the public
directory contains only the selected spectra, covariance, documentation and
a portable plotting notebook.

\subsection{Final spectrum with available uncertainties}
Figure~\ref{fig:sed:pi1_UMa} shows the adopted full SED and its available uncertainty bounds.

\begin{center}
\includegraphics[width=\linewidth]{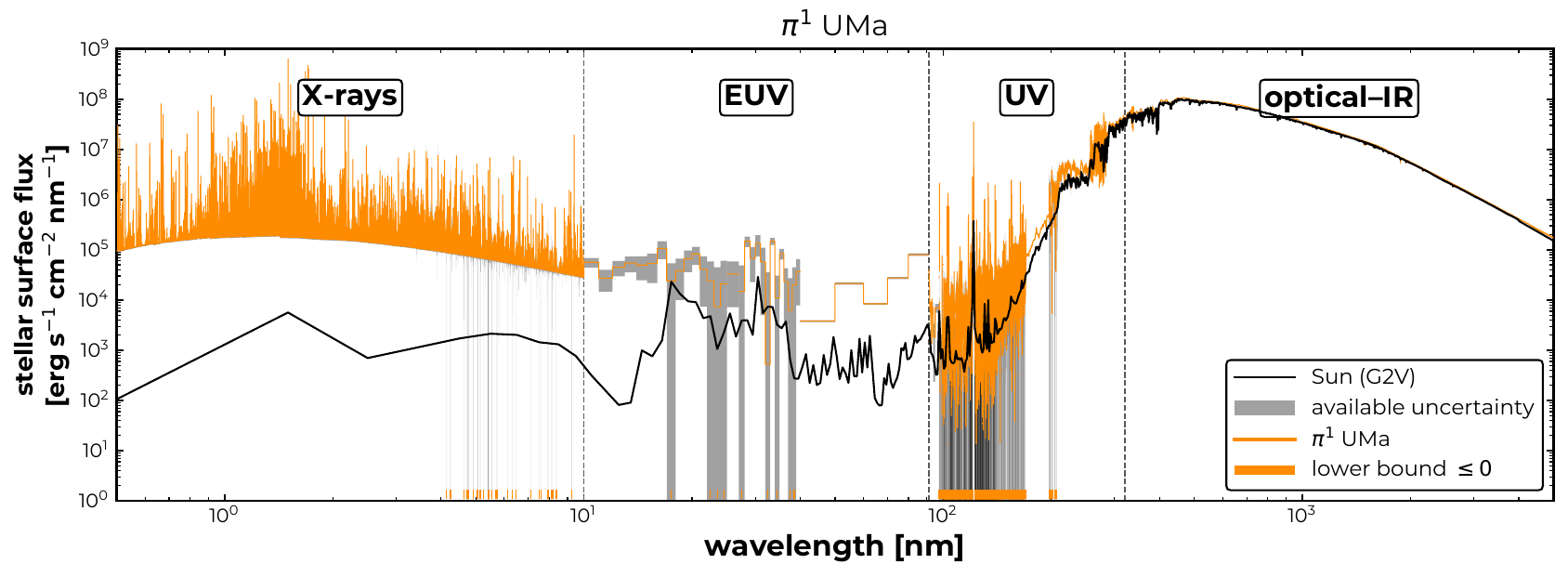}
\end{center}
\captionof{figure}{$\pi^1$ Ursae Majoris: final nominal surface-flux spectrum and available uncertainties. Target colours, shading, solar comparison and nonpositive-bound markers follow Section~\ref{sec:display}. The component origins and uncertainty scope are documented above.}\label{fig:sed:pi1_UMa}

%% file: sections/Xi_Boo_A.tex
\clearpage
\section{\texorpdfstring{$\xi$ Bootis A}{xi Bootis A}}\label{star:Xi_Boo_A}
This section establishes the identity of $\xi$ Boo A in the ultraviolet
and X-ray observations before describing its coronal spectrum, Ly$\alpha$
reconstruction, and spectral joins. Resolved component-A measurements are
distinguished from blended binary EUV evidence, and the resulting uncertainty
representation is described alongside the final spectrum.

\subsection{Identity, geometry and photospheric parameters}
The legacy local Hubble directory contained COS spectra whose native target
headers identify HD\,131156B. These were excluded from the primary. A new MAST
query recovered resolved COS observations of A from 2020 and 2026, which provide
independent spectral checks and fill selected echelle gaps. The July 1998
STIS visit has almost absent stellar signal and is excluded. The usable STIS
reference is the December 1998 E140M/E230M visit with the $0.2\times 0.06$\,arcsec
aperture. Archive identifiers, header selections and checksums are preserved
in the working notebook.

We adopt the interferometric angular diameter $\theta=1.124\pm 0.009$\,mas,
$F_{\rm bol}=(39.836\pm 2.568)\times 10^{-8}$\,erg\,s$^{-1}$\,cm$^{-2}$,
$T_{\rm eff}=5545\pm 92$\,K, $\log g=4.561\pm 0.017$ and
$[\mathrm{Fe/H}]=-0.10\pm 0.04$ from \citet{Karovicova.et.al.2022A&A...658A..47K}.
Gaia DR 3 source 1237090738916392704 identifies A and gives
$d=6.753585$\,pc. Combining this distance with $\theta$ yields
$R=0.81616\,R_\odot$. The surface conversion is
\begin{equation}
 F_{\lambda,\rm surf}=\frac{4}{\theta_{\rm rad}^{2}}F_{\lambda,\oplus}.
\end{equation}
This formulation retains the dependence between radius and distance rather
than adding them as independent errors. The common angular-diameter mode is
also retained when the same parameter controls a photospheric or EUV estimate.

Neither component has a Gaia XP spectrum. No exact match was found in the
supplied LOWLIB list. The optical/IR segment is therefore explicitly a
PHOENIX NewEra atmosphere \citep{Hauschildt.et.al.2025A&A...698A..47H}, interpolated in temperature, gravity
and metallicity and normalized by the measured bolometric flux. Separate
Gaia photometry checks the normalization: model-to-observed flux ratios for A
are approximately 1.035, 1.066 and 1.029 in G, BP and RP. These differences
are comparable to the bolometric normalization uncertainty. The B
contamination template is instead normalized to its resolved Gaia G flux;
its assumed literature radius is not treated as a measured constraint.

\subsection{Resolved Chandra spectrum and X-ray uncertainty}
We use the native source and background products of Chandra ObsID\,8899
(2008 May 2). The source centroid identifies A. A spatial decomposition of
the cross-dispersion profile indicates a B contribution below about 0.12\%
inside A's extraction aperture in the intervals used for the principal
$0.2$--$2.4$\,keV fit. Contamination increases at longer wavelengths and
prevents treating the full long-wavelength extraction as a clean spectrum
of A.

Both dispersion arms are fitted jointly with the Poisson W statistic. The
two sky strips have a combined area ten times the source area, as verified
from the native extraction metadata. Because HRC-S cannot sort overlapping
diffraction orders, the model includes the response matrices and effective
areas for orders 1--8\footnote{Chandra X-ray Center, \emph{Fitting Multiple Orders of HRC-S/LETG Data}, updated 3 December 2025: \url{https://cxc.cfa.harvard.edu/sherpa/threads/grating_hrcsletg/}.}. The thermal model is TBabs times four
vapec components, using APEC 3.1.3, XSPEC 12.14.0k and the Wilms abundance
reference. C, N, O, Ne, Mg, Si, S and Fe are fitted, tied across temperatures;
the remaining specified metals track Fe. The selected temperatures are
approximately 0.169, 0.297, 0.630 and 1.081\,keV. A fifth component introduces
strong degeneracy and is not adopted.

The unabsorbed flux on the exact $0.5$--$10$\,nm interval is
\begin{equation}
 F_X=(1.7266\pm 0.0175)\times 10^{-11}\ {
 \rm erg\,s^{-1}\,cm^{-2}}.
\end{equation}
The quoted error is the local propagation of the full fitted parameter
covariance. Alternative thermal/abundance descriptions give a flux range
of approximately $(1.714$--$1.734)\times 10^{-11}$ in the same band.
An independent response-based band-flux calculation over $0.5$--$6.2$\,nm
gives $(1.554\pm 0.020)\times 10^{-11}$, compared with $1.533\times 10^{-11}$
from the selected thermal model. Thus the integrated flux is robust at this
level, although some resolved lines retain significant model residuals.
The statistical covariance is not presented as a complete uncertainty for
individual atomic line predictions. Small detector shifts and additional
broadening were also tested and do not remove all residuals.

\citet{Wood.and.Linsky.2010ApJ...717.1279W} analysed the same Chandra observation using a
different extraction and emission-measure analysis. Their wavelength-dependent
binary contributions support the need for component separation. Their Table 1
luminosities are based on ROSAT and are not misidentified here as independent
Chandra flux measurements.

\subsection{Ly$\alpha$, airglow and independent-line checks}
The intrinsic Ly$\alpha$ profile is inferred from native STIS order 122.
The model combines a Voigt emission profile, a freely varying self-reversal,
H I and D I interstellar absorption, and additional broad H I absorption
representing the combined heliospheric/astrospheric contribution. It is
convolved with the tabulated line-spread function for the actual slit and
integrated over detector pixels. Native ERROR and serious data-quality bits
enter the likelihood. Stellar and MgII interstellar velocities constrain
velocity offsets; neither the published Ly$\alpha$ flux nor H I column is
used as a normalization prior. The central geocoronal window is excluded at
$-V_{\rm HELIO}=+19.556$\,km\,s$^{-1}$, with a nominal half-width of 20\,km\,s$^{-1}$.

Two independent ensembles sample the conditional model posterior. The
reported integral over $120.8$--$122.4$\,nm is
\begin{equation}
 F_{\mathrm{Ly}\alpha}=(2.1065\pm 0.0907)\times 10^{-11}\ {
 \rm erg\,s^{-1}\,cm^{-2}}.
\end{equation}
The released integrated flux and six intrinsic-profile diagnostics pass the autocorrelation and independent-ensemble tests, with more than 160 estimated correlation times retained. The full nuisance-parameter criterion does not pass for the slowly mixing interstellar-width/extra-absorber partition. That decomposition is not released as a physical measurement; the flux/profile convergence and the conditional scope are recorded separately.
Airglow masks of 12 and 30\,km\,s$^{-1}$, alternative D/H, turbulent broadening,
the LSF, an omitted extra absorber and mild fixed reversal are separately
tested. Their flux range is approximately $(1.95$--$2.30)\times 10^{-11}$;
this is method sensitivity, not a confidence interval.

The $1.85\times 10^{-11}$ flux of \citet{Wood.et.al.2005ApJS..159..118W} was reconstructed
from the same STIS observation. The present value is about 14\% higher, so
this is a comparison of reconstruction methods rather than independent
observational validation. CII 1335.7 alone and the CIV doublet give
Ly$\alpha$ predictions of approximately $1.76$ and $2.08\times 10^{-11}$
using \citet{Linsky.et.al.2013ApJ...766...69L}. The present reconstruction is consistent with their
approximately 22\% empirical scatter. The CII integration has 98\% supported
bandwidth, which is retained in its audit. An exploratory constant-continuum
MgII profile fit had unacceptable residuals and was rejected; none of its
derived fluxes or reversal depths enters the adopted reconstruction.

Trace-relative STIS source and sky profiles are measured around N I and O I
and compared with stellar control lines. The narrow-slit O I profiles do not
require an independently measurable extended-airglow subtraction. Their
measured fluxes and errors remain. The COS N I 113.3--113.6\,nm region is
explicitly flagged as a coarse estimate rather than claimed as a recovered
intrinsic line. Where no component-A observation is available below 106.55\,nm,
WHI/SIRS \citep{Woods.et.al.2009GeoRL..36.1101W} supplies only a labelled shape template, with the
\citet{Linsky.et.al.2014ApJ...780...61L} total FUV and Lyman-series relations setting
the band partition. The total FUV relation already includes the Lyman series;
these lines are not counted twice.

\subsection{EUV measurements, binary blending and model selection}
The 1993 and 1997 EUVE night images are independently extracted with source
and sky strips. Poisson covariance is propagated through higher-order
removal, H/He attenuation correction and 1\,nm rebinning. The nominal
foreground is $\log N_{\rm HI}=17.92$, with neutral He/H=0.1. Column and
extraction alternatives are retained. The possible contribution from
unmeasured incident photons below 7\,nm is checked using Chandra and changes
the $10$--$40$\,nm integral by only $1.38\times 10^{-14}$ for the A-only driver.

The measured binary fluxes are $(9.99\pm 0.65)\times 10^{-12}$ in 1993 and
$(10.61\pm 0.25)\times 10^{-12}$ in 1997, in $10$--$40$\,nm. These measure
\emph{A+B}. They are released separately and no constant Chandra count
fraction is imposed on their EUV spectrum. The A-specific SED uses the
eight broad bands of \citet{Linsky.et.al.2014ApJ...780...61L}, conditioned on the reconstructed
Ly$\alpha$ flux and stellar geometry. In $10$--$40$\,nm their sum is near
$1.04\times 10^{-11}$, compatible with the unresolved observations at the
precision of the proxy. The empirical RMS values 29.0\%,56.6\% and 41.0\%
are included for these three bands. A common residual mode is an explicit
covariance convention, not a correlation measured by the calibration paper.
The long-EUV bands remain conditional on the adopted model. The binary's
EUVE data were part of the calibration literature, so this comparison is
not independent validation of that empirical relation.

We also evaluate the recent relation of \citet{Sanz-Forcada.et.al.2025A&A...693A.285S} (Eq.~\ref{eq:sf25}),
with exactly $0.5$--$10$\,nm for X and $10$--$92$\,nm for EUV. It predicts
$4.815\times 10^{-11}$ at Earth; the adopted SED gives approximately
$2.31\times 10^{-11}$ in the same EUV band, within the published 0.406\,dex
dispersion. This comparison does not force an observed EUVE normalization.

For the magnetic alternative, \citep{Hahlin.et.al.2025A&A...696A...4H} gives a total intensity-field
measurement $\langle B\rangle=591\pm 60$\,G at 2008.08, close to the Chandra
epoch. With the revised radius, $\Phi=2\pi R^2\langle B\rangle=
1.197\times 10^{25}$\,Mx. The \citep{Namekata.et.al.2023ApJ...945..147N} prediction with this field
strongly exceeds the measured binary short-EUV flux. It is not selected
as the primary normalization. It remains a documented comparison; no
universal $10^{24}$\,Mx fallback is used.

\subsection{UV joins and the public uncertainty representation}
The IUE SWP continuum exceeds the resolved HST continuum. We test an
empirical additive alignment of $1.063\times 10^{-13}$\,erg\,s$^{-1}$\,cm$^{-2}$\,nm$^{-1}$,
estimated on matched continuum windows; an uncorrected spectrum and a
multiplicative alternative are retained. The IUE documentation describes
long-wavelength scattered light in cool-star SWP spectra \citep{Grady.and.Imhoff.1985IUE}.
The empirical alignment does not uniquely distinguish that effect from
aperture or epoch differences. The long 1979 LWR exposure supplies the blue
NUV; short 1984 LWP exposures anchor its useful overlap and the red NUV.
Pixels with serious quality flags are rejected. IUE beyond 320\,nm is not
adopted because of the archive's calibration guidance.

IUE portions remain identified as A-dominated estimates with possible
binary contribution. A Gaia-scaled B photosphere provides an explicit
contamination alternative. The transition to NewEra at 320\,nm is not
artificially smoothed: a localized observed/model continuum mismatch
remains visible in the notebook. This does not imply that the model's
formal normalization uncertainty accounts for all UV opacity errors.

The FITS product retains a sparse base covariance, shared calibration and
prediction modes, the X-parameter Jacobian/covariance and joint Ly$\alpha$
profile samples. An integrated flux uses actual bin widths $\boldsymbol w$,
with variance $\boldsymbol w^T C\boldsymbol w$. Method scenarios are named
spectra, not independent pixel noise. Surface uncertainties include the
angular-diameter dependence and its cross-correlation with model components.
No single column is labelled as a universally complete total error.

The nonnegative companion pools adjacent bins while conserving their signed
integrated energy; it does not clip negative measurements. Its covariance
is propagated through the same frozen pooling operator. This is a practical
irradiation input, not a positivity-constrained posterior. The portable
notebook shows the full SED with and without uncertainty, line regions and
linear views of weak measurements. Figures are not saved unless the user
sets \texttt{SALVAR\_PDF=1}.

\subsection{Diagnostic figure}
\begin{center}
\includegraphics[width=\linewidth]{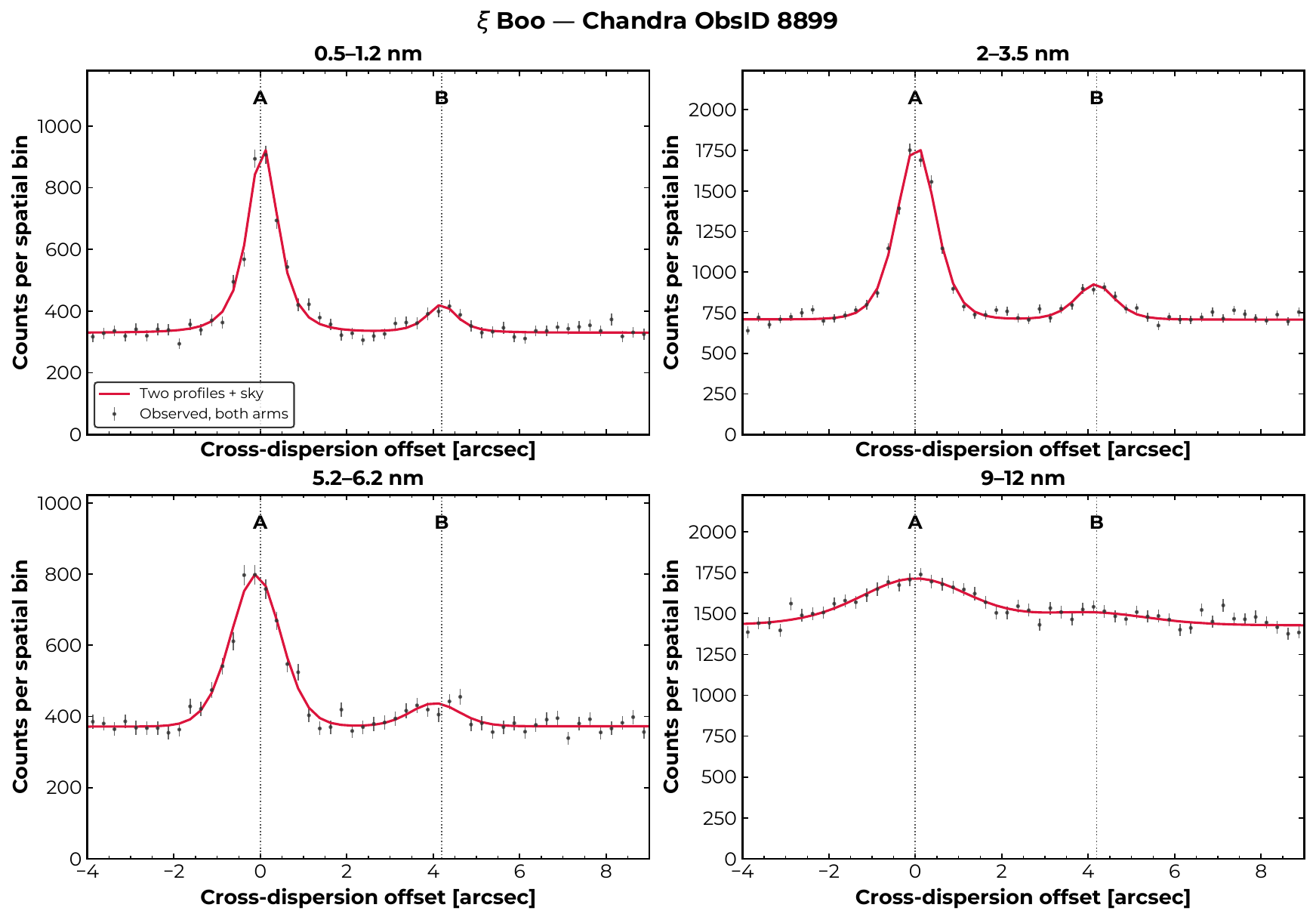}
\end{center}
\captionof{figure}{Chandra LETG/HRC-S cross-dispersion diagnostics for $\xi$ Boo. Points sum the two archived dispersed arms in the stated wavelength bands; curves show the recorded two-source spatial fits. The B trace is offset from the A extraction. This supports a component-aware X-ray extraction, but does not resolve the separately observed EUVE A+B spectrum.}\label{fig:diagnostic:Xi_Boo_A}

\newpage

\subsection{Final spectrum with available uncertainties}
Figure~\ref{fig:sed:Xi_Boo_A} shows the adopted full SED and its available uncertainty bounds.

\begin{center}
\includegraphics[width=\linewidth]{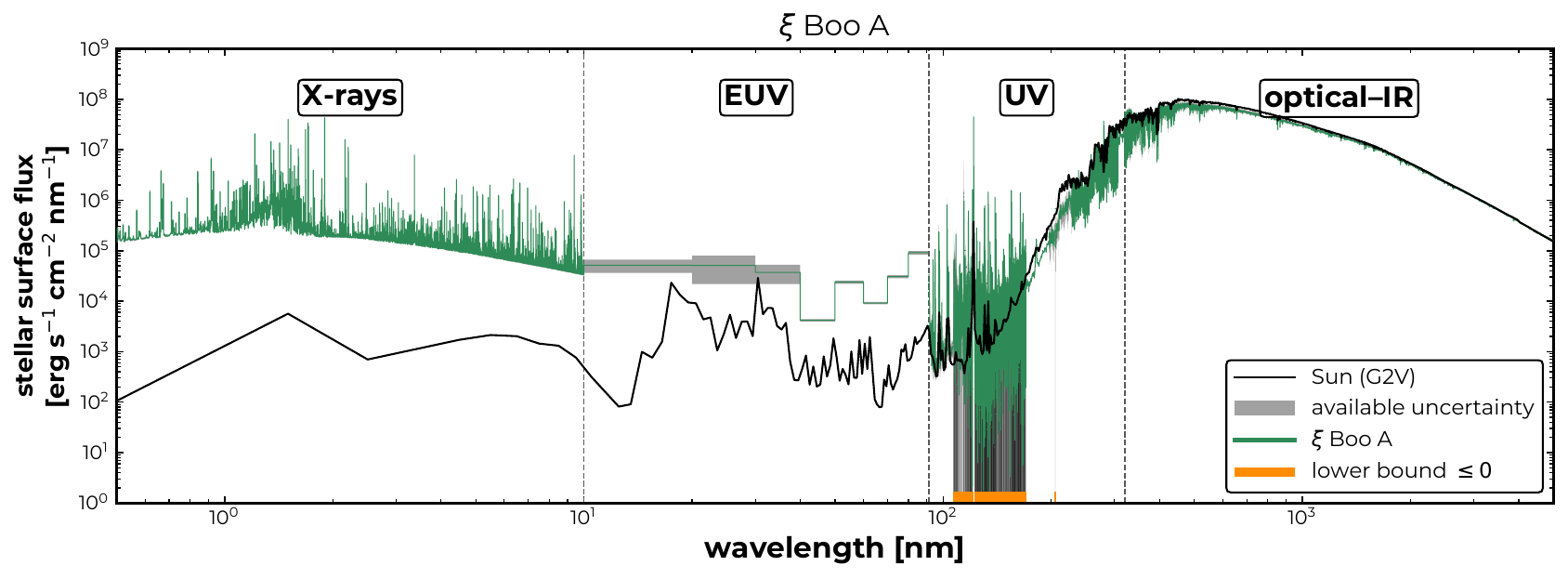}
\end{center}
\captionof{figure}{$\xi$ Bootis A: final nominal surface-flux spectrum and available uncertainties. Target colours, shading, solar comparison and nonpositive-bound markers follow Section~\ref{sec:display}. The component origins and uncertainty scope are documented above.}\label{fig:sed:Xi_Boo_A}

%% file: sections/10_Tau.tex
\clearpage
\section{\texorpdfstring{10 Tauri}{10 Tauri}}\label{star:10_Tau}
This section explains the source-identification checks and low-count
X-ray inference for 10 Tau, the empirical treatment of Ly$\alpha$ and EUV
emission, and the optical/infrared completion. The final part specifies how
to use the adopted nominal spectrum and its conditional intervals while
retaining model sensitivity as a separate diagnostic.

\subsection{Identity and stellar parameters}
The native header of legacy IUE exposure SWP40060 identifies HD\,1581.
A fresh MAST search recovered SWP42341 (1991 August 28, large aperture,
7199.8\,s), which identifies HD\,22484, together with the high-dispersion
long-wavelength exposures LWR12651, LWP02790 and LWP20882. The Hubble
inventory contains STIS CCD spectroscopy, represented by the exact-target
LOWLIB product, but no suitable high-resolution Ly$\alpha$ observation.
The available Copernicus measurements sample isolated wavelengths rather
than a resolved Ly$\alpha$ profile. Input identifiers, checksums and
selection decisions are retained in the working notebook.

We adopted the interferometric and bolometric measurements of \citet{Boyajian.et.al.2012ApJ...746..101B}, listed in Table~\ref{10_Tau:tab:par}. The Hipparcos parallax is
consistent with the Gaia DR3 value $71.837\pm0.150$\,mas, but the latter has
RUWE $=3.06$. We retained the explicitly referenced Hipparcos distance
rather than treating that Gaia solution as an uncomplicated precision
constraint. We recomputed the radius from the same distance and angular
diameter; these two quantities were not subsequently treated as independent.
The geometry was
\begin{equation}
 F_{\lambda,\mathrm{surf}}=4F_{\lambda,\oplus}/\theta_{\mathrm{rad}}^2,
 \qquad 4/\theta_{\mathrm{rad}}^2=1.45633\times10^{17}.
\end{equation}
The tabulated mass and age are isochrone-based estimates. They do not
establish an analogue of a particular terrestrial geological epoch.

\begin{table}[ht]\centering
\caption{Adopted parameters. Values and measurement references are from \citet{Boyajian.et.al.2012ApJ...746..101B}; radius and gravity are recomputed consistently.}
\label{10_Tau:tab:par}
\begin{tabular}{ll}\toprule
Parameter & Value\\\midrule
Spectral type & F9 IV--V (F9 V in the interferometric compilation)\\
Parallax & $71.60\pm0.54$\,mas\\
Distance & $13.966\pm0.105$\,pc\\
Angular diameter & $1.081\pm0.014$\,mas\\
Radius & $1.623\pm0.024\,R_\odot$\\
Bolometric flux at Earth & $(5.123\pm0.069)\times10^{-7}$\,erg\,s$^{-1}$\,cm$^{-2}$\\
Effective temperature & $5997\pm44$\,K\\
$[\mathrm{Fe/H}]$; $\log g$ & $-0.09$; $4.074$\\
Mass; age & $1.139\pm0.016\,M_\odot$; $5.7\pm0.4$\,Gyr\\\bottomrule
\end{tabular}\end{table}

\subsection{X-rays: low-count likelihood and physical sensitivity}
We used the existing XMM--Newton reduction of ObsID\,0134540601 and
jointly fitted the PN and MOS1 source and background spectra. MOS2 did
not provide a useful net signal and was not added as an independent
flux estimate. The model was \texttt{TBabs*APEC}, with
$N_{\rm H}=10^{18}$\,cm$^{-2}$, solar coronal abundance, zero redshift,
Wilms abundances and the corresponding Verner-based absorption model.
The nominal APEC version was 3.0.9, reproducing the inherited fit;
APEC 3.1.3 was tested separately. XSPEC 12.15.1 fitted 0.2--2.0\,keV using
the Poisson W statistic, with the native source/background scaling.
We did not fit background-subtracted counts with a Gaussian likelihood.

The best one-temperature fit yielded $kT=0.10984$\,keV,
normalization $9.7827\times10^{-6}$ and $W=94.613$ for 85 degrees of
freedom. A second temperature improved W by less than one and was not
adopted. We exported unabsorbed spectra on exactly 0.5--10\,nm and
separately integrated 0.1--2.4\,keV for the ROSAT-band Ly$\alpha$ proxy.
These two bands are not interchangeable.

The adopted nominal X spectrum is the direct APEC evaluation at this
minimum-W solution, not an average of models at different temperatures.
Its 0.5--10\,nm flux is $1.67868\times10^{-14}$\,erg\,s$^{-1}$\,cm$^{-2}$;
its separate 0.1--2.4\,keV flux is $1.88729\times10^{-14}$ in the same units.
EUV and line-proxy normalizations use the central empirical relations at
these best-fit fluxes and the adopted distance. Thus one nominal spectrum
is recommended for modelling, with its uncertainty distribution provided separately.

We profiled the Poisson background analytically and evaluated a
response-folded temperature--normalization grid. Its statistic agrees
with direct XSPEC evaluations to better than $10^{-3}$. The conditional
sampling measure was uniform in $\log kT$ over 0.05--1.5\,keV and in
$\log$ normalization over $10^{-9}$--$10^{-3}$. We evaluated the spectral means and joint covariance directly from all
posterior grid weights. A set of 131072 low-discrepancy Sobol latent samples
supplies mixed-component and surface-flux marginal quantiles; pure X and
proxy Earth-flux quantiles are evaluated from their weighted distributions. The mean 0.5--10\,nm flux was
$1.971\times10^{-14}$\,erg\,s$^{-1}$\,cm$^{-2}$, with 16th, 50th and
84th percentiles $(0.940,1.594,2.826)\times10^{-14}$.
The mean differs from the median because the distribution is asymmetric.
Deterministic moments avoid Monte Carlo noise in the long hard-X tail.
The 16--84\% interval is not a minimum/maximum envelope and need not include
any chosen point estimate. The physical best fit is adopted without changing
the distribution,
its covariance and its quantiles. Main plots show one solid nominal line and
transparent 16--84\% shading. Means and medians appear only in diagnostic panels;
no symmetry or enclosure of the nominal is imposed. Wavelengths below
0.620\,nm and above 6.199\,nm lie outside the fitted 0.2--2\,keV band.

The source is sufficiently weak that the assumed measure matters:
a prior uniform in linear normalization gives a mean
$3.679\times10^{-14}$ in the same band. Raising the fitted low-energy
threshold to 0.25\,keV gives $1.257\times10^{-14}$. These are method
sensitivities, not extra independent Gaussian errors. Tests of
$N_{\rm H}=10^{17}$--$3\times10^{18}$\,cm$^{-2}$, abundance 0.5--2
solar and APEC 3.1.3 are also supplied. A targeted background-only
simulation gave one exceedance of the observed $\Delta W=20.47$ in
500 trials, or the finite-simulation estimate $p=2/501$. This is conditional
on the adopted extraction/background model and is not an archive-wide
source-detection significance.

\subsection{Ly$\alpha$, airglow and empirical EUV}
The observed low-resolution IUE Ly$\alpha$ feature cannot be interpreted
as an intrinsic stellar line. No resolved profile reconstruction is
claimed. C IV is not significantly detected
($(0.92\pm2.79)\times10^{-14}$\,erg\,s$^{-1}$\,cm$^{-2}$), and the
measured C II blend, $(6.76\pm1.50)\times10^{-14}$, cannot be substituted
for the isolated 1335.7\,\AA\ component used by the calibration.
The high-dispersion Mg II data do not supply a sufficiently constrained,
ISM-corrected intrinsic flux.

We therefore adopted the F5--G9 X-ray proxy of \citet{Linsky.et.al.2013ApJ...766...69L},
\begin{equation}
 \log f_{\mathrm{Ly}\alpha}(1\,\mathrm{au})
 =1.156+0.316\log f_X(1\,\mathrm{au}),
\end{equation}
where $f_X$ uses 0.1--2.4\,keV and fluxes are in cgs units.
We propagated the joint X-ray samples and distance and represented the
published 42.3\% fractional RMS by a positive, mean-one lognormal
multiplier with the same fractional standard deviation. This is our
sampling convention, not a published probability distribution.
The nominal intrinsic line flux, from the central relation at the best-fit
ROSAT-band flux, is $9.606\times10^{-13}$\,erg\,s$^{-1}$\,cm$^{-2}$.
The unchanged distribution has mean
$9.75\times10^{-13}$\,erg\,s$^{-1}$\,cm$^{-2}$; the 16th, median and
84th percentiles are $(5.58,8.77,13.82)\times10^{-13}$.
About 36\% of the input X-ray samples lie below the calibration's
$0.12$\,erg\,s$^{-1}$\,cm$^{-2}$ lower bound at 1\,au. Thus part of
this estimate is an extrapolation, and the dwarf calibration's
applicability to the F9 IV--V classification is conditional.
Neither a precise stellar line shape nor independent validation of this
proxy is established by these data.

The 119.6--120.6, 120.6--122.6 and 129.7--131.1\,nm windows were replaced
with explicitly labelled N I, Ly$\alpha$ and O I estimates. Quiet-solar
line excesses were scaled by the inferred Ly$\alpha$ strength and added
to a continuum measured in adjacent IUE sidebands. The continuum
covariance was propagated. Original contaminated values remain in the
input table. These replacements are coarse irradiation estimates, not
observational recoveries of resolved stellar N I or O I profiles.

For the EUV we used the Flare Irradiance Spectral Model, version 2
(FISM2; \citep{Chamberlin.et.al.2020SpWea..1802588C}), solely as a spectral-shape template. We recomputed
the mean of 24 supplied daily spectra sampling 2019--2020, retained its
native 0.1\,nm bins, and normalized its integral over exactly 10--92\,nm
with the updated relation of \citet{Sanz-Forcada.et.al.2025A&A...693A.285S} (Eq.~\ref{eq:sf25}).
Here $L_X$ is integrated over 0.5--10\,nm. We used the published
0.406\,dex residual dispersion as one common predictive factor per
spectrum, without tapering it toward an observed boundary. The adopted nominal
EUV flux is $1.24902\times10^{-13}$\,erg\,s$^{-1}$\,cm$^{-2}$, from the central
relation at the best-fit X flux; it is not multiplied by the lognormal
mean correction. The conditional distribution has mean EUV flux $2.140\times10^{-13}$\,erg\,s$^{-1}$\,cm$^{-2}$;
the 16th, median and 84th percentiles are
$(0.424,1.213,3.474)\times10^{-13}$. Relation coefficients are fixed
at their published central values. The detailed solar-to-stellar shape
is a model assumption, motivated by the near-solar temperature and
soft corona, not a measured EUV line spectrum.

The 92--115\,nm gap uses the solar template normalized by the
\citep{Linsky.et.al.2014ApJ...780...61L} 91.2--117\,nm ratio, $10^{-1.025}$ times Ly$\alpha$,
with its 35\% common predictive RMS. Only the required subinterval is
inserted, so no wavelength bin is counted twice. Alternative broad EUV
shapes from that work are provided at the same 10--92\,nm energy.
Because Ly$\alpha$ is X-derived here, this comparison is not independent
validation of the X-based EUV normalization. No arbitrary unsigned
magnetic flux, including $10^{24}$\,Mx, is used.

The EUVE RAP catalogue \citep{McDonald.et.al.1994AJ....108.1843M} lists HR\,1101/HD\,22484 with
Lexan/B, Al/Ti/C and deep-survey Lexan rates of 3.1, 3.2 and
3.7\,counts\,ks$^{-1}$, respectively, and a probable identification.
We checked these entries against the printed Table 2. The archive search
recovered survey photometry rather than a calibrated stellar spectrum.
These counts are recorded as ancillary evidence; they are not converted
into flux points without the corresponding response and error treatment.

\subsection{UV quality, LOWLIB and the infrared tail}
NEWSIPS quality flag $-1024$ denotes saturation \citep{Garhart.et.al.1997IUENN..57....1G}\footnote{NEWSIPS data-quality flags: \url{https://archive.stsci.edu/iue/manual/newsips/node20.html}.}.
The correct SWP observation becomes saturated at its red end.
We used its selected 115--180\,nm portion, retaining its reported
uncertainties. A single rejected reseau bin was interpolated using adjacent
valid bins with the resulting covariance. Three IUE longwave exposures
were also audited using calibrated noise and central-order selection;
their 185--210\,nm continua did not provide a reliable replacement anchor.
The shortest LOWLIB wavelengths show a large excess relative to the
unsaturated IUE continuum and were not used to rescale IUE.

LOWLIB \citep{Pal.et.al.2023ApJS..266...41P} replaced NGSL from 200 to 1000\,nm. We selected
the scattered-light and slit-offcenter corrected flux with its matching
reported errors, without imposing an additional dust correction.
The 180--200\,nm interval was filled with the NewEra atmosphere
\citep{Hauschildt.et.al.2025A&A...698A..47H}, using a log-linear change in its multiplicative scale
between constraints from 175--178\,nm IUE and 200--205\,nm LOWLIB.
Relative to the bolometric atmosphere, those scales are 0.664 and 0.818.
This is a model-filled gap, not a correction applied globally to the
observations. Detailed absorption structure in that interval is therefore
model dependent.

NewEra was interpolated in temperature, gravity and metallicity.
The 1000--5000\,nm photospheric tail was anchored to 850--950\,nm LOWLIB,
with scale 0.982 relative to the bolometric model. We propagated the
three measured normalization anchors, their mutual covariance and
their cross-covariance with the retained observed bins. Temperature
changes induced by bolometric-flux and angular-diameter uncertainties
were propagated together with the surface conversion. Alternative
temperature, metallicity and gravity choices remain separate method
scenarios. The atmosphere comparison of \citet{Caffau.et.al.2016A&A...585A..16C} uses its
temperature and metallicity at the adopted gravity; it is not an
unacknowledged extrapolation outside the downloaded grid.
The released IR segment describes the stellar photosphere; no arbitrary
circumstellar component was added to mimic the near-IR excess discussed
by \citet{Absil.et.al.2013A&A...555A.104A}.

\subsection{Public product and uncertainty use}
The FITS product contains wavelength edges, nominal flux densities at Earth,
at the stellar surface and at 1\,au, source and quality labels, native
and propagated errors, conditional bounds and separate method scenarios.
It retains the original input values and the sparse observation mapping.
The columns \texttt{F\_EARTH\_NM}, \texttt{F\_SURFACE\_NM} and
\texttt{F\_1AU\_NM} are the recommended nominal SED. Separate
\texttt{F\_MEAN\_*} and \texttt{F\_MEDIAN\_*} columns describe the
distribution for diagnostic purposes. \texttt{NOMINAL\_INFO} records the
adopted fit parameters and proxy amplitudes. Conditional standard deviations
are not distances from the nominal to its percentile bounds, and should not
be used to manufacture a symmetric interval about the best fit.
For the Earth-flux vector the available covariance is
\begin{equation}
 C=C_{\rm base}+UKU^{\mathsf T}
   +B_{\rm model}C_{\rm model}B_{\rm model}^{\mathsf T}.
\end{equation}
The low-rank term includes cross-covariances of measured anchors, so
$K$ alone need not be positive definite. The full sum has the covariance
of the stated construction. The model basis and its analytic coefficient covariance preserve the
shared X/EUV/Ly$\alpha$ driver and its skewed distributions. To integrate a band, the notebook computes $w^{\mathsf T}Cw$
with wavelength-overlap weights, rather than summing pixel errors as
if every bin were independent. Surface conversion retains the common
angular-diameter mode; positive model flux intervals use a positive
angular-diameter transformation rather than a symmetric lower error
that could cross zero solely because of linearized geometry.
Because the distribution is unchanged, the surface covariance transform
uses the posterior mean Earth-flux vector. Coherent model realizations are
constructed as $F_{\rm mean}-B_{\rm model}\mu_{\rm model}
+B_{\rm model}a$, with the retained latent coefficients $a$; they must not
be shifted to the newly adopted nominal. Observational and anchor covariance
is additional to these model realizations. Method scenarios remain diagnostic
posterior means under alternative assumptions, not alternative nominal SEDs
or confidence bounds.

All 9621 central SED bins were already nonnegative. No negative observed
value was clipped to create the adopted spectrum; signed original input
values remain accessible. Full-spectrum plots, with and without shading,
and selected linear/logarithmic panels are included in an executed,
portable notebook. PDF figure export is off by default.

The product addresses the source-specific referee requests for intrinsic
Ly$\alpha$ estimation, contamination handling, the current X/EUV scaling,
explicit shape assumptions, updated libraries and distributed uncertainties.
Its errors are the available conditional budget, not a claim that every
calibration or model discrepancy is known. In particular, this star's
weak X-ray prior dependence, extrapolated Ly$\alpha$ proxy, detailed solar
line transfer and model-filled UV interval remain identifiable assumptions.
No unmeasured instrumental or epoch variance was invented. The complete
manuscript, catalogue-level comparisons and response letter remain separate
from this source note.

\FloatBarrier
\begin{minipage}{\linewidth}
\subsection{Final spectrum with available uncertainties}
Figure~\ref{fig:sed:10_Tau} shows the adopted full SED and its available uncertainty bounds.

\begin{center}
\includegraphics[width=\linewidth]{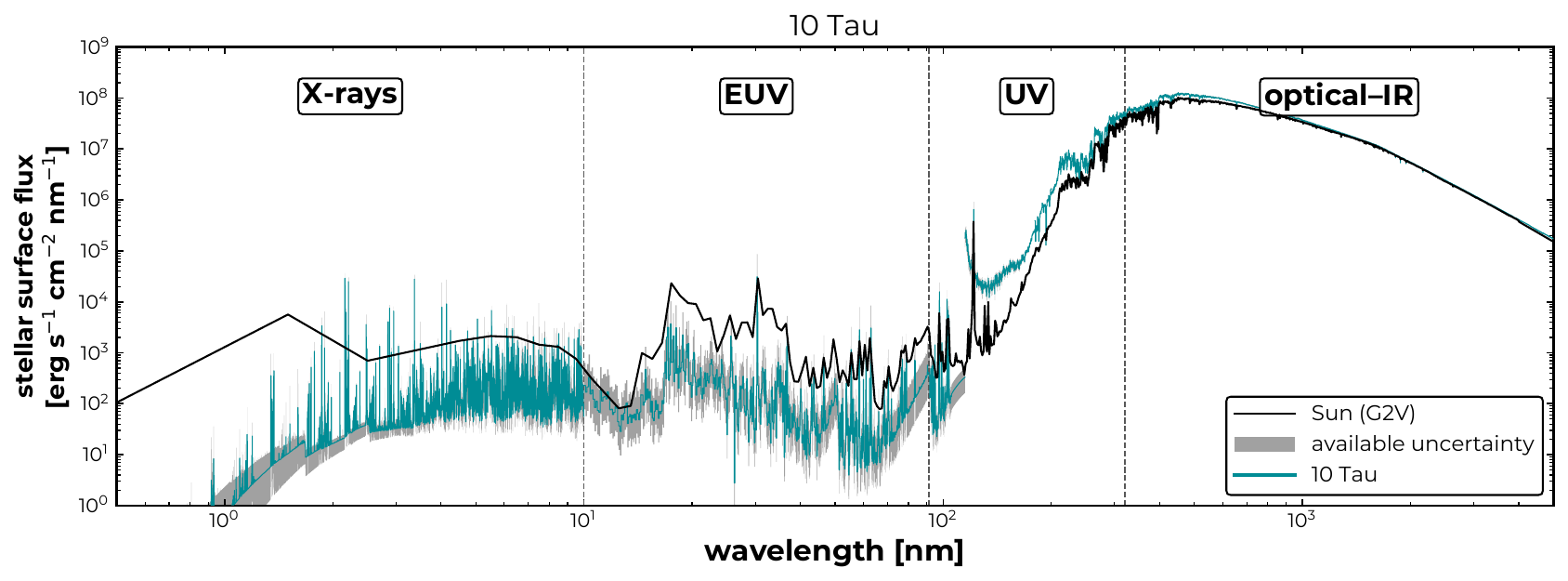}
\end{center}
\captionof{figure}{10 Tauri: final nominal surface-flux spectrum and available uncertainties. Target colours, shading, solar comparison and nonpositive-bound markers follow Section~\ref{sec:display}. The component origins and uncertainty scope are documented above.}\label{fig:sed:10_Tau}
\end{minipage}
\FloatBarrier

%% file: sections/alpha_Cen_A.tex
\clearpage
\section{\texorpdfstring{$\alpha$ Centauri A}{alpha Centauri A}}\label{star:alpha_Cen_A}
This target section accompanies the component-A
PHAROS 2.0 SED, covering 0.5--5000~nm. It records the adopted calibration,
reconstruction, source selection and uncertainty propagation. Observations,
empirical proxies, atmosphere models and the optical twin are distinguished
in the public FITS. The nominal irradiation spectrum is nonnegative, while
the signed measurements before adaptive grouping are retained separately.

\subsection{An empirical optical template with explicit provenance}
The optical spectrum in the previous assembly was not a direct spectroscopic
observation of $\alpha $ Cen A. It used Gaia DR3 XP spectroscopy of HD~30708
(source identifier 174156255225869440), multiplied by one amplitude. This distinction
is retained in both the data provenance and the plot labels. Photometric measurements
of $\alpha $ Cen A are available; it is the optical spectral shape that was borrowed.

The selection of HD~30708 is now traceable to an email from E.~Mamajek to V.~Sumida,
provided by the author during this revision. The quoted donor parameters are
$ T_{\rm eff}=5783\pm 54$~K, $\log g=4.31\pm 0.07$, and
$[\mathrm{Fe/H}]=+0.17\pm 0.04$. It was the first-ranked object in the supplied
list of analogues to a target with $ T_{\rm eff}=5804$~K, $\log g=4.32$ and
$[\mathrm{Fe/H}]=+0.22$. The ranking metric is preserved as supplied, without
interpreting it as a statistical $\chi^2$ because its definition was not included.
The attribution is a personal communication, not a published parameter table.
The private email is retained with the working files, rather than redistributed
as part of the public SED.

\subsection{Consistent geometry and bolometric normalization}
We use the parallax $750.81\pm 0.38$~mas and mass $1.0788\pm 0.0029\, M_\odot $
from \citet{Akeson.et.al.2021AJ....162...14A}, the limb-darkened angular diameter $8.502\pm 0.038$~mas
from \citet{Kervella.et.al.2017AA...597A.137K}, and the bolometric flux
$(2.716\pm 0.027)\times 10^{-5}$~erg~s $^{-1}$~cm $^{-2}$ reported by \citet{Boyajian.et.al.2013ApJ...771...40B} and quoted by \citet{Kervella.et.al.2017AA...597A.137K}. These give a consistent distance
and radius, rather than combining a radius inferred at one parallax with another
distance. The central derived values are
\[
d=1.331895\ {\rm pc}, \qquad R=1.217487\ R_\odot, \qquad
T_{\rm eff}=5794.9\ {\rm K}, \qquad \log g=4.3001.
\]
The surface-flux conversion is
\[
 F_{\lambda, \star}=\frac{4}{\theta_{\rm rad}^{2}}f_{\lambda, \oplus}.
\]
Its uncertainty depends on the angular diameter, not on independently sampled
radius and distance. The temperature is derived from the same bolometric flux
and angular diameter; their shared effects are propagated consistently.

PHOENIX NewEra spectra \citep{Hauschildt.et.al.2025A&A...698A..47H} are interpolated in temperature, gravity
and metallicity, adopting $[\mathrm{Fe/H}]=+0.22$ for the target. Flux-conserving
integration is used to construct the model bins. The 336--1000~nm fraction of the
bolometric model specifies the normalization of the Gaia template:
\[
 k=\frac{f_{\rm bol}\, b_{336:1000}(T, g, Z)}{\int_{336}^{1000}f_{\lambda, \rm donor}\, d\lambda}.
\]
The resulting candidate amplitude is $ k=507.21456$, compared with 504.23495 in the
previous assembly, a change of approximately 0.59\%. This small amplitude change
does not establish that the old uncertainty prescription or filter convention
was correct. The new normalization is explicitly conditional on a model
bolometric correction.

\subsection{Native Gaia errors and the photometric checks}
The continuous XP coefficient covariance is transformed to sampled flux using
the calibrated basis functions. The reconstructed fluxes and diagonal errors
agree with the archived sampled products to better than 0.1\%. The full
cross-wavelength covariance is retained. For sampled donor flux $\boldsymbol f $
and quadrature vector $\boldsymbol w $, normalization by the same samples gives
\[
 \boldsymbol y=k\boldsymbol f, \qquad
 J=k\left[I-\frac{\boldsymbol f\boldsymbol w^T}
 {\boldsymbol w^T\boldsymbol f}\right], \qquad
 C_{\rm Gaia}=J C_f J^T.
\]
The normalization and donor samples therefore are not treated as independent.
Shared bolometric-flux, angular-diameter, parallax and mass modes are added and
retained separately. Independence is assumed where the literature does not
provide covariance; in particular, an orbital mass--parallax covariance is not
invented. The available optical errors are conditional errors, not an estimate
of every possible template mismatch.

The cached SVO Generic/Bessell filters identify energy-counting responses.
Synthetic photometry therefore uses $\int f_\lambda T\, d\lambda/\int T\, d\lambda $
and the corresponding cached zero points. The earlier photon weighting is not
retained. The 0.015~mag errors assigned by the old code have no identified
individual measurement source and are removed. The $ BV R_C I_C $ values in
\citep{Bessell.1983PASP...95..480B} are used as checks rather than assigned artificial precision.
The printed $ V-I=0.639$ is inconsistent with $(V-R)+(R-I)=0.693$ in that table;
both interpretations are recorded. The corresponding colour residuals are
shown in the notebook rather than hidden by a wavelength-dependent rescaling.

A separate NewEra target/donor spectral-ratio test, smoothed to 10~nm to isolate
broad colours, changes the normalized template by less than 0.5\% between
400 and 950~nm. Donor-parameter variations use the uncertainties supplied by
Mamajek. These tests support retaining the unwarped empirical shape, but do not
turn the model difference into a measured statistical error. The 10~nm smoothing
is a broad-colour diagnostic, not an exact Gaia line-spread function.

\subsection{Adopted ultraviolet scale and quality treatment}
Ten native STIS spectra from program~7263 were recovered from MAST, including
E140H/1598 and both E230H/2263 integrations absent from the previous local
coadds. The FUV visits are from 1999 February and the NUV observations from
1999 April. We propagate native ERROR arrays through deterministic
order/exposure weights, serious-DQ rejection, edge trimming and flux-conserving
resampling. Native pixel covariance induced by the resampling is retained.

For the NUV, identical supported overlap intervals determine relative scales
by generalized least squares. E230H/1763, using the supported aperture, sets
the reference. The relative factors are 1.066 at 2013~\AA, 1.656 and 1.790 for
the two 2263~\AA\ exposures, 1.950 at 2513~\AA, 1.808 at 2762~\AA, and 1.679
at 3012~\AA. The latter observations used the neutral-density aperture.
Their large corrections are consistent with the pattern independently
documented in the dedicated StarCAT processing \citep{Ayres.2010ApJS..187..149A}; the tested
broad-band splice fluxes agree with that reduction at about 1--3\%.
This is a processing check using the same observations, not an independent
measurement of stellar variability. Three large-aperture IUE LWR spectra
from 1980 January 12 provide a historical comparison; their approximately 19\%
excess at 300--310~nm is not imposed as a new absolute calibration.

Native FUV amplitudes are retained. The line-dominated FUV overlaps are not
forced into a single multiplicative chain: the CII region gives a different
ratio from adjacent weak intervals. We select the interiors of the settings
and retain blue/red overlap alternatives as sensitivity scenarios. A
representative scale budget follows StarCAT section 3.7.2: a common 3\%
radiometric term and a 7\% repeatability term for each independent FUV
exposure or the shared NUV anchor (about 7.6\% in quadrature). This is an
adopted instrumental calibration budget, not a measured epoch-to-epoch
variance. Scale/data cross-covariance is propagated because an overlap-derived
scale is not independent of the input spectrum.

Two current-pipeline DQ32 regions, 123.8325--123.9200 and 159.5260--159.7285~nm,
coincide with repeller-wire/p-flat blemishes \citep{Sohn.2019stid.book....7S}. They are replaced
by valid pixels in the published dedicated StarCAT reduction of the same
observations; the NV line is not erased by linear interpolation. Reported
StarCAT errors are retained. Because its oversampling covariance is not
supplied, the errors within each short replacement are taken as fully
positively correlated: a conservative integrated-flux bound for positive
weights, not a claim of new statistical precision. Smaller DQ gaps have
explicit interpolation flags and the corresponding interpolation covariance.

Native NI and OI profiles are kept unless subtraction is supported by a
positive terrestrial residual. The proper STIS velocity convention places
the geocorona near+22.04~km~s $^{-1}$, whereas the stellar reference is near
$-23.45$~km~s $^{-1}$. NI tests give no positive excess above $3\sigma $.
The strong OI triplet is not adequately represented by a local quadratic;
negative fitted excesses therefore do not justify subtraction. Near
OI 1355.598~\AA, the red feature contains CI 1355.844~\AA\ \citep{Pagano.et.al.2004A&A...415..331P}.
A joint stellar-blend/terrestrial test yields only a $1.5\sigma $ nominal
terrestrial residual, with $ Q\simeq 42$ for 45 degrees of freedom.
These observed profiles and errors are retained. This is not a claim to
reconstruct every narrow interstellar metal absorption core.

\subsection{Resolved FUSE observations instead of a wholly synthetic far UV}
Observation P1042601 targets HD128620, component A, through the
4$\times $20~arcsec MDRS aperture on 2001 June 25 \citep{Redfield.et.al.2002ApJ...581..626R}.
The archived CalFUSE 3.2.1 exposures and all detector channels were audited.
Non-guiding SiC channels suffer severe throughput losses after exposure 002;
LiF also drops in 007/008. Consequently the automatic ALL/ANO/NVO coadds
are not adopted as calibrated stellar spectra. This behaviour is a known
MDRS failure mode discussed in the archive handbook \citep{Sonnentrucker.et.al.2009FUSE}.

Exposure 001 is entirely orbital night, with 1219~s of good time and stable
guide-channel rates. We use SiC2A over 92--98.8~nm, LiF1A over 98.8--108~nm,
SiC2B over 108--109.4~nm, and LiF1B above 109.4~nm up to the STIS join at 114.2~nm.
QUALITY and native Gaussian ERROR are propagated through 0.005~nm bins;
no correction is fitted to force agreement with a literature line flux.
The measured CIII 977, OVI 1032 and CIII 1176 integrals are respectively
$2.115\times 10^{-12}$, $9.945\times 10^{-13}$ and $8.662\times 10^{-13}$
erg~s $^{-1}$~cm $^{-2}$. They differ by about 4--7\% from the dedicated
earlier reduction in \citep{Redfield.et.al.2002ApJ...581..626R}, with the expected calibration and
extraction differences. That comparison is not independent astrophysical
evidence, since it uses the same visit. STIS 1999 CIII 1176 is approximately
11\% brighter than the selected FUSE exposure; we retain the native
calibrations rather than estimate an epoch variance from this one comparison.

Orbital night reduces airglow but does not remove it. Higher-Lyman and
prominent HI/OI/NI/NII airglow neighbourhoods are therefore explicitly
replaced by quiet-Sun FISM2 cells scaled to the reconstructed Ly $\alpha $
flux. These are proxies at the native 0.1~nm solar resolution, not measured
intrinsic stellar line profiles. A coherent $\pm 35\%$ proxy-amplitude test
is retained as an adopted sensitivity range, not a measured per-line
confidence interval. A separate $\pm 15\%$ FUSE amplitude test represents
the calibration sensitivity. The signed original nightly spectrum remains
in the working data; no contaminated pixels are silently declared intrinsic.

\subsection{Ly $\alpha $ reconstruction and its validation}
The adopted model has two positive Voigt emission components, with separate
narrow and broad Gaussian widths and a shared Lorentz width, multiplied by
a central Gaussian optical-depth reversal of independently fitted width,
depth and offset. The transmission includes interstellar HI and DI plus
heliospheric and astrospheric nuisance HI absorption. We convolve the
absorbed profile with the tabulated STIS E140H/1200~\AA\ point-source LSF
for the actual 0.2$\times $0.09 aperture \citep{Rickman.and.Brown.2025stii.book...25R}, sampled at the
observed pixel scale. The geocoronal core is excluded within 20~km~s $^{-1}$
of its predicted location;15 and 25~km~s $^{-1}$ masks are also tested.

DI/HI is fixed to $1.5\times 10^{-5}$ in the nominal fit, with
$1.3$ and $1.8\times 10^{-5}$ sensitivity trials. ISM width/velocity,
stellar velocity, profile offsets and a small baseline have explicit
regularizing constraints. Weak wall priors are centred on the components
in \citep{Linsky.and.Wood.1996ApJ...463..254L}, Table 4, model 2A: $(\log N, b, v)=(14.439, 22.39, -8)$ and
$(14.575, 24.83, -20)$, with velocity units km~s $^{-1}$.
The adopted widths of these priors (0.5~dex in column, 6~km~s $^{-1}$
in Doppler width and 6/10~km~s $^{-1}$ in velocity) are assumptions,
not errors quoted by that paper. There is no prior on the intrinsic
integrated flux. Degenerate/boundary wall parameters are not interpreted
as new heliosphere or astrosphere measurements. The adopted kinematic
box restricts heliospheric velocities to $[-18, 0]$ and astrospheric
velocities to $[-24, -18]$~km~s $^{-1}$, avoiding the redshifted ``astrospheric''
solution permitted by the initial unrestricted nuisance fit. The
astrospheric Doppler-width range is 12--45~km~s $^{-1}$, allowing the
narrower heated component also present in the alternative fits of \citet{Linsky.and.Wood.1996ApJ...463..254L}. The ISM width constraint, $9.5\pm 0.6$~km~s $^{-1}$, is
consistent with their MgII/FeII/DI-derived $ T=5400\pm 500$~K and
turbulent width 1.20~km~s $^{-1}$, which imply $ b_H=9.51$~km~s $^{-1}$.
A further check fixes both wall velocities to $-8$ and $-20$~km~s $^{-1}$;
its intrinsic flux, $8.9681\times 10^{-11}$~cgs, changes by less than 1\%.

The best fit has data $\chi^2=811.35$ for 719 unmasked pixels and 20
parameters, and a penalized objective $ Q=821.72$. It improves the structured
residuals of the earlier tied-reversal, single-component profile. The
intrinsic 1208--1224~\AA\ integral is
\[
f_{\rm Ly\alpha}=8.8949\times 10^{-11}\ {\rm erg\ s^{-1}\ cm^{-2}}.
\]
The local joint covariance gives a conditional flux standard deviation
of $8.50\times 10^{-12}$ in the same units. A separate scan fixes the
integrated flux and refits all 19 nuisance parameters from two starting
solutions at each flux. The interpolated $\Delta Q=1$ range is
$[8.42, 9.58]\times 10^{-11}$; the $\Delta Q=4$ range is
$[8.15, 10.47]\times 10^{-11}$. These are conditional profile diagnostics,
not a complete uncertainty budget for the hidden stellar core. Including
the separate 7.6\% calibration budget gives an indicative asymmetric
flux uncertainty of approximately $(-0.83, +0.97)\times 10^{-11}$.

The GHRS reconstruction compiled by \citet{Wood.et.al.2005ApJS..159..118W},
$1.01\times 10^{-10}$~erg~s $^{-1}$~cm $^{-2}$, is consistent with
the nominal STIS result (about 12\% lower) within the conditional $\Delta Q=4$
range before calibration is added. Thus there is no significant
disagreement established by these tests. The epochs differ. D/H, LSF,
mask and wall-prior alternatives remain separate model sensitivities.
The exploratory 20-dimensional MCMC mixes too slowly to establish global
convergence and is not used to quote credible intervals.

In the SED, the central 1211--1218~\AA\ interval is replaced by the
intrinsic model. Outside it, through 1208--1224~\AA, an additive
Ly $\alpha $ absorption-wing correction preserves the surrounding observed
spectrum. In particular, OV 1218.34~\AA\ and its native error are retained.

\subsection{X/EUV and correlated uncertainty}
The component-A extraction of Chandra ObsID 7432 is retained. Both LETG
arms, source/background spectra and responses for orders 1--8 enter the
0.2--2~keV W-statistic fit in XSPEC. Comparisons include APEC 3.0.9/3.1.3,
one/two temperatures, abundance and foreground-column assumptions.
Hot-start 2T trials do not improve the identifiable 1T free-abundance solution.
The adopted APEC 3.1.3 model has $ N_H=10^{17.61}$~cm $^{-2}$,
$ kT=0.090631$~keV, $ Z=0.10156$ relative to the Wilms mixture, and
$ W=9930.001$. The global coronal abundance is an effective model
parameter and is not the photospheric metallicity.

The unabsorbed 0.5--10~nm integral is
$2.7654\times 10^{-12}$~erg~s $^{-1}$~cm $^{-2}$, with local-covariance
16th--84th percentiles $[2.6576, 2.8546]\times 10^{-12}$.
Parts of that interval are model extrapolation outside the fitted band.
Joint temperature--abundance--normalization draws feed the updated
\citep{Sanz-Forcada.et.al.2025A&A...693A.285S} relation (Eq.~\ref{eq:sf25}),
for X=0.5--10~nm and EUV=10--92~nm. The central EUV flux is
$1.9140\times 10^{-11}$~erg~s $^{-1}$~cm $^{-2}$; the 0.406~dex predictive
residual is a correlated whole-spectrum amplitude. Shared X samples
and parallax dependence are propagated rather than sampled independently.

The nominal EUV shape is the mean of 24 quiet-Sun FISM2 daily spectra
from 2019--2020 \citep{Chamberlin.et.al.2020SpWea..1802588C}. A second shape uses the Ly $\alpha $-based
band relations of \citet{Linsky.et.al.2014ApJ...780...61L}, normalized to the same Sanz-Forcada
total, as a method-sensitivity test. Neither is direct component-A EUV
spectroscopy. The archival EUVE spectrum listed by \citet{Linsky.et.al.2014ApJ...780...61L}
contains A+B and is not assigned to A. No assumed $10^{24}$ Mx
magnetic flux is used. Atomic-data/abundance/column alternatives are
propagated through both the X spectrum and the EUV normalization.

\subsection{Final assembly, optical joins and use of the released errors}
The spectral sequence is X 0.5--10~nm, EUV 10--92~nm, selected FUSE/proxies
92--114.2~nm, STIS with reconstructed Ly $\alpha $114.2--314.5~nm,
a PHOENIX bridge 314.5--336~nm, the calibrated Gaia optical twin 336--1000~nm,
and PHOENIX 1000--5000~nm. The bridge is represented in 0.5~nm bins and its
normalization interpolates logarithmically between the adjacent observed
UV and optical anchors. The measured mean observed/model factors are
0.684 in 309--313~nm and 1.074 in 340--350~nm. This significant atmosphere/UV
shape mismatch is explicit: the atmosphere is not used to overwrite the
observed STIS amplitude. The IR is normalized by 1.0165 using 950--990~nm.
Anchor gradients, the full Gaia covariance, and shared bolometric-flux,
angular-diameter, parallax and mass modes are propagated jointly.
The assembled 0.5--5000~nm integral is 99.15\% of the adopted bolometric
flux; the missing red tail is not spuriously renormalized into the SED.

The recommended irradiation product groups adjacent signed bins until
their integrated flux is nonnegative. It never clips individual negative
measurements to zero. The operation conserves the integrated flux exactly,
and applies the same linear map to covariance and sensitivity spectra.
The original signed assembly remains available in a separate FITS
extension. Errors are conditional on the selected grouping; grouping does
not turn a weak flux into a statistically significant detection.

The public covariance representation is
\[
C=B+UKU^{\mathsf T}+\mathrm{Cov}(\Delta F_{X, \mathrm{EUV}}),
\]
where $ B $ is sparse native/resampling covariance, $ U, K $ represent scale,
atmosphere, Ly $\alpha $ and anchor cross terms, and the last term comes
from joint X/EUV realizations. Surface-flux propagation modifies the
shared angular-diameter mode before applying $4/\theta^2$; distance and
radius are not propagated independently. Band errors use the complete
quadratic form rather than adding all pixel errors independently.
Local parameter errors, conditional percentiles and physical-method
alternatives are explicitly distinguished. The public viewer shows one
nominal line with a transparent conditional-error band; alternative
models remain in diagnostic plots and the covariance companion file.

\FloatBarrier
\begin{minipage}{\linewidth}
\subsection{Final spectrum with available uncertainties}
Figure~\ref{fig:sed:alpha_Cen_A} shows the adopted full SED and its available uncertainty bounds.

\begin{center}
\includegraphics[width=\linewidth]{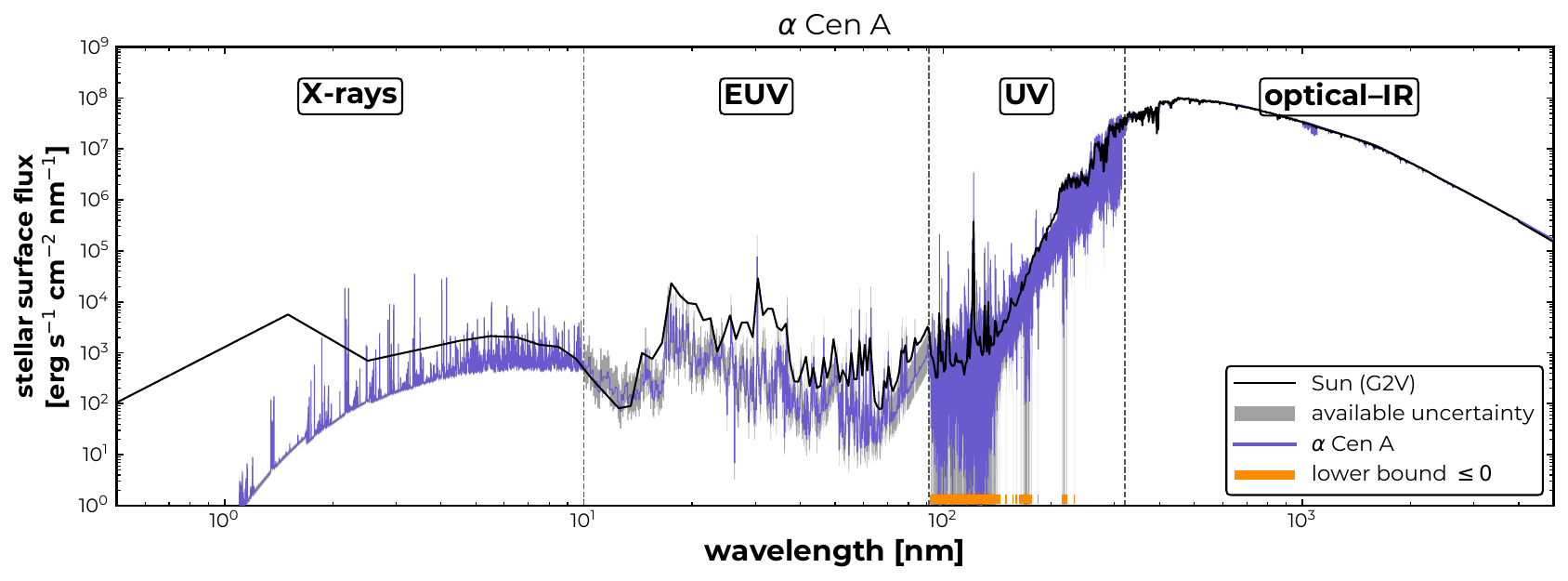}
\end{center}
\captionof{figure}{$\alpha$ Centauri A: final nominal surface-flux spectrum and available uncertainties. Target colours, shading, solar comparison and nonpositive-bound markers follow Section~\ref{sec:display}. The component origins and uncertainty scope are documented above.}\label{fig:sed:alpha_Cen_A}
\end{minipage}
\FloatBarrier

%% file: sections/beta_Com.tex
\clearpage
\section{\texorpdfstring{$\beta$ Comae Berenices}{beta Comae Berenices}}\label{star:beta_Com}
This section describes the consistent stellar geometry adopted for
$\beta$ Com, the response and background checks in the X-ray fit, and the
ultraviolet line reconstruction. It then follows the FUSE, IUE, and LOWLIB
segments into the photospheric extension and summarizes the EUV estimate
and uncertainty products.

\subsection{Stellar parameters and flux convention}
We adopt the internally consistent interferometric parameter set of \citet{Boyajian.et.al.2012ApJ...746..101B}: $\theta_{\rm LD}=1.127\pm0.011$\,mas, Hipparcos parallax $109.53\pm0.17$\,mas, and $F_{\rm bol}=(5.249\pm0.056)\times10^{-7}$\,erg\,s$^{-1}$\,cm$^{-2}$. These imply $d=9.1299\pm0.0142$\,pc and $R=1.1063\pm0.0109\,R_\odot$. Recomputing the Stefan--Boltzmann relation with modern constants gives $T_{\rm eff}=5934\pm33$\,K, consistent with the published $5936\pm33$\,K. A mass of $1.045\pm0.013\,M_\odot$ gives $\log g=4.369$. The atmospheric interpolation uses the compilation metallicity $[\mathrm{Fe/H}]=-0.06$; metallicity alternatives are kept separate from statistical errors.

The Gaia DR3 source is 1460229442589223424. Its parallax, 108.725\,mas, and RUWE of 2.46 are recorded, but are not mixed into the adopted interferometric radius/parallax pair. Age is not used to normalize any spectrum. Surface flux is obtained directly from the angular diameter,
\begin{equation}
 F_{\lambda,\rm surf}=\frac{4}{\theta_{\rm rad}^{2}}F_{\lambda,\oplus},
 \qquad \frac{4}{\theta_{\rm rad}^{2}}=1.33987\times10^{17}.
\end{equation}
The angular-diameter uncertainty is a common wavelength-correlated mode. Radius and distance errors are consequently not added as independent geometric uncertainties. All released flux densities are per nanometre; integrated fluxes quoted below are at Earth.

\subsection{X-rays: response correction and background-controlled fits}
XMM-Newton observation 0148680101, obtained on 2003 July 20, supplies EPIC-pn and RGS data. The retained pn spectrum has a 57.16\,ks exposure and selects PATTERN$\leq4$. A byte-level audit established that the legacy file named \texttt{epn\_e1\_ff20\_sdY9.rmf} was actually identical to the PPS doubles-only response. This was inappropriate for a singles-plus-doubles source spectrum. The revised fit uses the source-specific \texttt{SP4RMF0001} response. An independently downloaded version-21 singles-plus-doubles response gives consistent results. The erroneous-response results are retained only as invalid diagnostics, never as scientific alternatives.

The RGS source-excluded CCD9 light curves contain strong background flares. We select 100\,s bins with at least 90\% native frame coverage and a background rate no greater than 0.1\,count\,s$^{-1}$. The resulting useful exposures are 28.13\,ks for RGS1 and 25.07\,ks for RGS2. The source is entry~3 in the native source lists. Source and background spectra and matching responses are regenerated with SAS, preserving the original event rejection mask. Thresholds of 0.05 and 0.2\,count\,s$^{-1}$ provide sensitivity tests. Grouping requires adequate source and background counts and does not bridge detector gaps.

We fit pn over 0.32--2\,keV and RGS over 0.33--1.5\,keV with a Poisson source-plus-background likelihood (XSPEC W-statistic). The nominal model is an absorbed, three-temperature APEC~3.1.3 plasma, with Wilms abundances and $N_{\rm H}=10^{17.94}$\,cm$^{-2}$. This column is a weak local-ISM proxy, not an X-ray measurement. O, Ne, and Fe abundance groups are free and shared between temperatures; C and N follow O, and the remaining metals follow Fe. RGS cross-normalizations are fitted nuisance parameters. The adopted temperatures are approximately 0.158, 0.271, and 0.460\,keV, with O, Ne, and Fe abundances of 0.365, 0.499, and 0.972 relative to the adopted solar table. The RGS normalizations relative to pn are 0.877 and 0.961.

The three-temperature fit has $W=250.95$ for 197 degrees of freedom. Its AICc advantage over the two-temperature fit is only about one; the thermal description is therefore not uniquely determined. Freeing N provides little additional evidence. No conspicuous residual above four plotted standard errors requires an isolated line to be discarded. Fixed-model simulations are retained as diagnostics, not as calibrated goodness probabilities: they do not repeat the fitting and data-dependent grouping. The analysis is related to earlier work on this same observing campaign~\citep{Telleschi.et.al.2005ApJ...622..653T}; such a comparison is an independent analysis of shared photons, not an independent stellar epoch.

The nominal unabsorbed flux in the exact 0.5--10\,nm band required by the EUV relation is
\begin{equation}
 F_X=1.862\times10^{-12}\ \mathrm{erg\,s^{-1}\,cm^{-2}},
 \qquad \log L_X=28.269.
\end{equation}
Positive multivariate draws from the full local covariance, marginalized over the RGS normalizations, give a conditional 16--84\% integrated range of approximately $(1.73,1.99)\times10^{-12}$. These are local-covariance draws, not a global MCMC posterior. The two-temperature, free-N, and background-threshold alternatives span approximately $(1.75,1.96)\times10^{-12}$ and remain separate sensitivity spectra. Extrapolation outside the fitted energy interval is explicitly part of the plasma model.

\subsection{Ly$\alpha$, Mg\,II, and airglow}
The newly available STIS observation \texttt{ofnu03010} provides E140M spectroscopy through the $0.2\times0.2$\,arcsec aperture, with a 2055\,s exposure in 2026. Native JIT metadata identify the entire integration as orbital night, using the definition recorded in that file. We fit a Voigt emission profile multiplied by an independently variable Gaussian self-reversal, then apply H\,I and D\,I absorption, the tabulated STIS line-spread function, and native pixel integration. A $\pm35$\,km\,s$^{-1}$ interval around geocoronal Ly$\alpha$ is excluded in the correct heliocentric frame.

The metal-line cloud velocity is constrained by \citet{Malamut.et.al.2014ApJ...787...75M}. We use weak, explicitly assumed regularization around the metal-based column proxy $\log N_{\rm H}=17.94$, together with stellar/cloud velocity and background constraints. D/H is fixed at $1.5\times10^{-5}$. These assumptions are recorded with the model bounds.

An unrestricted reversal can hide considerable intrinsic emission inside the absorbed core. To restrict this degeneracy, we fit the Mg\,II doublet in the 2009 E230H observation \texttt{ob4509010}. Its core trough/peak ratios are approximately 0.72--0.76, and its reversal strength agrees with the independent Mg\,II characterization of \citet{Taylor.et.al.2024ApJ...964...80T}. Following the physical motivation discussed by \citet{Youngblood.et.al.2022ApJ...926..129Y}, we require the intrinsic Ly$\alpha$ trough/peak ratio to be at least 0.65, allowing margin for the different line-formation physics and epochs. Mg\,II is not copied as a literal Ly$\alpha$ template. Bounds of 0.60 and 0.70, a different airglow mask, and removal of the column regularizer are tested separately.

The resulting nominal intrinsic Ly$\alpha$ integral is
\begin{equation}
 F_{\mathrm{Ly}\alpha}=6.3575\times10^{-12}\ \mathrm{erg\,s^{-1}\,cm^{-2}}.
\end{equation}
A 48-walker, 12000-step conditional MCMC calculation gives 16th, 50th, and 84th percentiles of $(6.17,6.50,6.90)\times10^{-12}$. The chain length exceeds 60 times its largest estimated autocorrelation time; the minimum effective sample estimate exceeds 2100. Post-burn-in temporal halves give consistent flux distributions. This establishes useful conditional sampling within the chosen family, not uniqueness of the hidden core. The unconstrained deep-reversal fit gives approximately $4.79\times10^{-12}$; a different same-width reversal family gives $7.87\times10^{-12}$ with a poorer fit. These are model-dependence diagnostics, not endpoints of a confidence interval.

The empirical C\,II and C\,IV relations of \citet{Linsky.et.al.2013ApJ...766...69L} predict $6.66\times10^{-12}$ and $7.08\times10^{-12}$, respectively, each with about 22\% published scatter. The Mg\,II prediction is $8.54\times10^{-12}$ with 31\% scatter and a different epoch. The adopted reconstruction is consistent with these checks. C\,II/C\,IV share the STIS radiometric calibration, so the comparisons are not fully independent observations.

N\,I and O\,I neighborhoods are tested for a narrow positive component at the terrestrial velocity. No test establishes a robust excess above three standard deviations. The night data are consequently retained without an unsupported airglow subtraction, preserving their native errors and any unresolved interstellar absorption. The observed O\,V~1218.34\,\AA\ line is preserved outside the central Ly$\alpha$ replacement; the Ly$\alpha$ absorption correction in the outer wings is additive.

\subsection{FUSE, IUE, LOWLIB, and the photospheric extension}
FUSE observation A0830401 supplies the 2001 far-UV spectrum. Exposure and channel checks favor the orbital-night coadd and the sequence SiC2A, LiF1A, SiC1A, and LiF2A over 92--114.3\,nm. SiC2B is rejected because of low throughput; LiF2A avoids the red LiF1B worm-sensitive region. Native errors are propagated through a flux-conserving overlap operator. The measured C\,III~1176 flux, $(5.60\pm0.24)\times10^{-14}$, agrees with the 2026 STIS value $(5.57\pm0.66)\times10^{-14}$. This agreement is a check, not a reason to invent a general inter-epoch variability error.

Higher-Lyman and selected airglow-contaminated FUSE cells are replaced by explicitly labeled solar-template proxies normalized to the reconstructed Ly$\alpha$ integral. Their uncertainty shares the Ly$\alpha$ draw block. Original measurements remain available in the working reduction. The remaining N\,I/O\,I absorption is not asserted to be intrinsic stellar emission.

The same-star LOWLIB spectrum~\citep{Pal.et.al.2023ApJS..266...41P} already includes scattered-light and slit-offset corrections. We use the supplied corrected flux and error arrays without duplicating those corrections or imposing the catalog dust correction on this nearby star. Its shortest-wavelength region nevertheless shows a large excess relative to IUE and the photospheric trend, consistent with an unreliable scattered-light-sensitive blue edge. We exclude it below 210\,nm rather than applying a second empirical subtraction.

Long low-dispersion SWP observations saturate toward the red. The recovered 24.9\,ks, large-aperture, high-dispersion IUE exposure SWP32905 instead supplies 170.9--195\,nm. Its errors are obtained from the documented NEWSIPS conversion of native noise to absolutely calibrated flux~\citep{Garhart.et.al.1997IUENN..57....1G}\footnote{NEWSIPS flux calibration and noise propagation: \url{https://archive.stsci.edu/iue/manual/newsips/node120.html}.}. The noisy STIS overlap does not require a SWP normalization. Two short large-aperture LWR observations, LWR04834 and LWR04835, provide the next interval, combining only accepted pixels with exposure-based weights. Their common factor of $1.1122\pm0.0064$ is measured against LOWLIB over 230--260\,nm; the quoted error is conditional native-error propagation. Alternative anchor windows and the unscaled IUE solution remain method tests.

Wavelength-dependent blends join SWP to LWR over 195--198\,nm and LWR to LOWLIB over 210--220\,nm. E230H replaces LOWLIB over 257--284.5\,nm and agrees with it at approximately 1\%, without an imposed scale. Larger STIS gaps use IUE observations. Small rejected intervals use explicitly flagged linear interpolation with transformed covariance, not fabricated detections.

LOWLIB supplies the optical spectrum to 1000\,nm. Same-star Gaia XP, with its reconstructed full coefficient covariance, agrees near 500--800\,nm and shows modest color-dependent offsets elsewhere. A Gaia-based optical/IR-anchor alternative is retained separately. A PHOENIX NewEra V3 interpolation supplies 1000--5000\,nm, normalized to the LOWLIB 950--990\,nm anchor. Temperature and gravity effects are propagated through the shared stellar parameters; metallicity choices are sensitivity curves. The 0.5--5000\,nm integral is $5.343\times10^{-7}$, approximately 1.8\% above the adopted bolometric reference, before any unrepresented flux beyond 5000\,nm. No final global scaling is imposed to conceal that comparison.

\subsection{EUV estimation and uncertainty products}
No pointed EUVE spectrum of $\beta$ Com was identified in the archive search. The nearby HZ~43 spectrum is not assigned to this target, and a survey detection is not treated as a resolved spectrum. The EUV integral uses the updated relation of \citet{Sanz-Forcada.et.al.2025A&A...693A.285S} (Eq.~\ref{eq:sf25}),
giving $F_{\rm EUV}=6.943\times10^{-12}$. The 0.406\,dex predictive residual is drawn once per complete spectrum, together with the shared X-ray uncertainty. It produces an asymmetric conditional interval of approximately $(2.68,17.5)\times10^{-12}$. A standard deviation should not be interpreted as a symmetric lower/upper bound for this distribution.

The nominal spectral shape is the average of 24 monthly FISM2 daily spectra from 2014--2015~\citep{Chamberlin.et.al.2020SpWea..1802588C}. This active-solar template is an assumption appropriate for testing a more active solar-type star, not a measurement of its EUV emission. When normalized by Ly$\alpha$, it reproduces the observed FUSE C\,III~977, O\,VI~1032, and C\,III~1176 integrals to approximately 1\%, 17\%, and 2\%, respectively. These checks support its use as a template without proving its EUV shape. A quiet-solar template and the Linsky et al.\ band-ratio prescription~\citep{Linsky.et.al.2014ApJ...780...61L}, each normalized to the same EUV total, are retained as alternatives. We do not assume $10^{24}$\,Mx or equate a large-scale ZDI field to the unsigned magnetic flux required by a magnetic scaling relation.

The released covariance has the form
\begin{equation}
 C=B+UKU^{\mathsf T}+C_{X,\mathrm{EUV}}+C_{\mathrm{Ly}\alpha,\mathrm{FUSE}},
\end{equation}
where $B$ contains native-error propagation and resampling covariance; $U,K$ preserve shared normalization, stellar-parameter, and anchor cross-terms; and the two independent draw blocks preserve the respective model correlations. The angular-diameter mode converts this covariance to surface-flux units. Native calibration uncertainties are used to the extent supplied; unknown correlations and epoch scatter are not assigned invented values. Atmospheric and proxy shape alternatives remain named curves with no assigned probability.

The public \texttt{SED} extension contains one nominal irradiation spectrum. Adjacent grouping removes negative integrated bins while exactly preserving total flux; it does not clip individual measurements to zero. \texttt{SED\_SIGNED} retains the signed estimates before this operation. Bin widths and grouping flags are explicit. Pointwise uncertainty intervals may cross zero even when the nominal irradiation spectrum is nonnegative. Nonpositive interval endpoints are displayed using the common conventions in Section~\ref{sec:display}; the signed data remain available for linear diagnostics. Band integration uses bin widths and the full covariance, not independent pixel-error quadrature. Checks cover finite/nonnegative nominal values, contiguous boundaries, flux conservation, covariance diagonals and quadratic forms, FITS checksums, and serialized round trips.

\FloatBarrier
\begin{minipage}{\linewidth}
\subsection{Final spectrum with available uncertainties}
Figure~\ref{fig:sed:beta_Com} shows the adopted full SED and its available uncertainty bounds.

\begin{center}
\includegraphics[width=\linewidth]{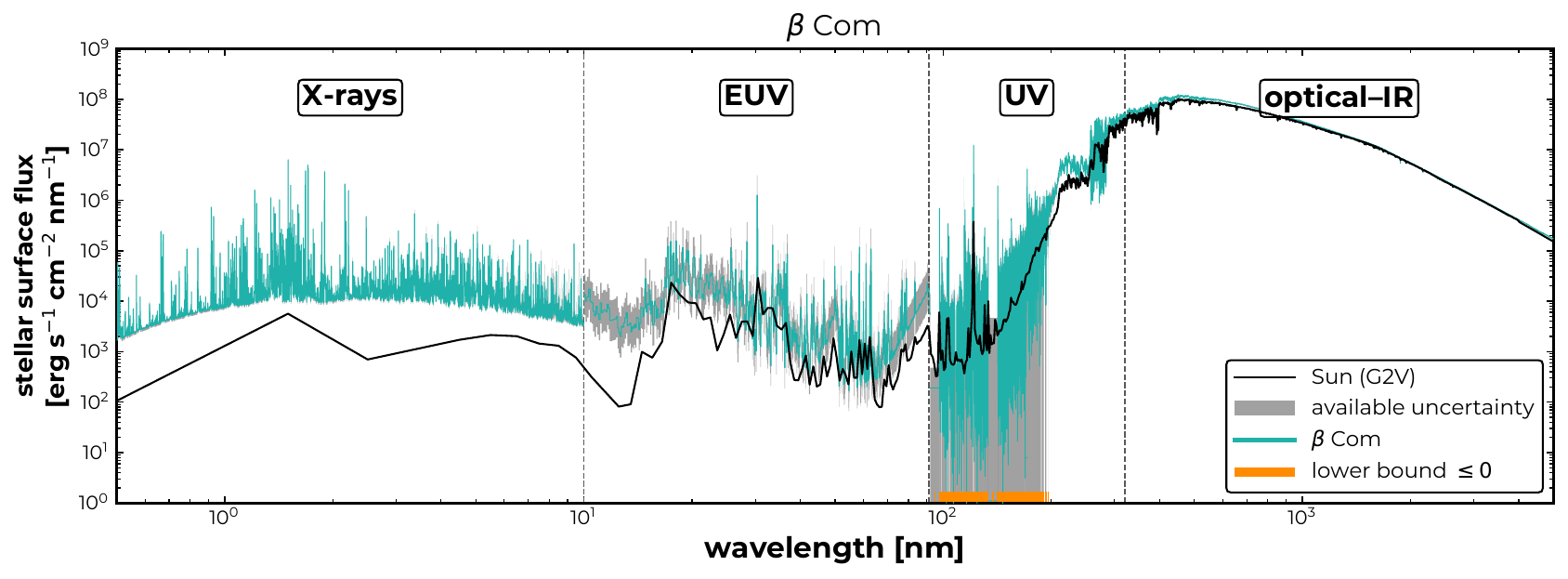}
\end{center}
\captionof{figure}{$\beta$ Comae Berenices: final nominal surface-flux spectrum and available uncertainties. Target colours, shading, solar comparison and nonpositive-bound markers follow Section~\ref{sec:display}. The component origins and uncertainty scope are documented above.}\label{fig:sed:beta_Com}
\end{minipage}
\FloatBarrier

%% file: sections/upsilon_And.tex
\clearpage
\section{\texorpdfstring{$\upsilon$ Andromedae A}{upsilon Andromedae A}}\label{star:upsilon_And}
\noindent This record accompanies the released 0.5--5000 nm spectrum of
$\upsilon$ Andromedae A. The nominal spectrum uses the selected best fits;
conditional uncertainties and named method sensitivities are provided separately.
The executed notebook retains the data-selection and validation diagnostics.

\subsection{Data identity and consistent stellar parameters}
The revision separates HD 9826 A from its wide companion. The previously extracted
quiet EPIC-pn spectrum of observation 0722030101 has an exposure of approximately
5280 s, a 40 arcsec source aperture, and a measured local background. We additionally
extracted a 25 arcsec spectrum using the same event selections and good-time
intervals, with its own aperture-dependent ancillary response. The existing image
audit places the companion approximately 55.36 arcsec from A; its aperture count
rate is much lower. Aperture comparisons, rather than a sharp-PSF assumption,
test whether contamination materially changes the inferred spectrum.

The adopted angular diameter is $1.114\pm0.009$ mas from \citet{Baines.et.al.2008ApJ...680..728B}. A parallax of $74.19\pm0.21$ mas gives $d=13.479\pm0.038$ pc and a
recomputed radius of $1.614\pm0.014\,R_\odot$. Radius values derived with different
parallaxes are not mixed. The bolometric flux
$(6.015\pm0.0126)\times10^{-7}$ erg s$^{-1}$ cm$^{-2}$ from \citet{Boyajian.et.al.2013ApJ...771...40B}, combined with the diameter, gives $T_{\rm eff}=6176\pm25$ K.
Surface fluxes use $F_{\lambda,\star}=4F_{\lambda,\oplus}/\theta^2$, with
$\theta$ in radians; the diameter uncertainty is a common multiplicative mode.
Independent radius and distance errors must not be added a second time. The NPOI
diameter $1.083\pm0.018$ mas and the Ligi et al. diameter $1.161\pm0.027$ mas are
retained as comparison measurements. Atmospheric-grid comparisons use
$[\mathrm{Fe/H}]=+0.08$ from the Boyajian compilation.

\subsection{X-ray fitting and EUV propagation}
We refitted the source and measured background with the Poisson W statistic in
0.2--2.4 keV, using APEC/AtomDB 3.1.3 and Wilms solar abundances. The absorption
column for the baseline tests is the sum of the three H I components reported by
\citet{Edelman.et.al.2019ApJ...880..117E}, $N_{\rm H}=1.176\times10^{18}$ cm$^{-2}$. Half and twice
this value are separate sensitivity calculations. Source-count and background-count
grouping choices, a 0.3--2 keV fitting band, aperture size, and abundance choices
are also retained separately.

For identical ten-source-count grouping, the one-temperature solar-abundance fit
has $W=102.08$ for 68 degrees of freedom, whereas the two-temperature fit has
$W=75.42$ for 66 degrees of freedom. The latter gives
$kT_1=0.0892$ keV and $kT_2=0.3101$ keV. A freely fitted common abundance yields
a slightly lower statistic but is poorly identified; its large best-fit value is
not interpreted as a measured coronal enrichment. The baseline two-temperature
unabsorbed flux in exactly 0.5--10 nm is $8.03\times10^{-13}$ erg s$^{-1}$ cm$^{-2}$.
The cool component is sensitive to extrapolation below the directly fitted band.
We replaced the initial short XSPEC chains with six independent ensemble runs:
two for each of three prior choices. Each run contains 48 walkers, 2000 burn-in
steps and 16000 retained steps per walker. A response-folded APEC temperature
grid and an analytic profiled Poisson-background likelihood accelerate sampling.
The exact likelihood agrees with XSPEC to $8\times10^{-12}$ in $W$ for 31 tests;
the maximum grid-interpolation discrepancy in these tests is 0.029 in $W$.

The adopted conditional distribution is uniform in log temperature over
0.02--5 keV and in log positive APEC normalization over $10^{-10}$--1, with
ordered temperatures. This is an explicit assumption that gives equal weight
to decades, not a uniquely measured prior. The lower temperature floor was
broadened from 0.04 to 0.02 keV to avoid truncating the weak cool tail. Independent
runs agree; the retained length exceeds 150 integrated-flux autocorrelation
times per walker, with more than 7500 effective integrated-flux samples per run.
The nominal curve is still the exact maximum-likelihood two-temperature fit,
not a posterior mean or median. The integrated conditional 16th, 50th and 84th
percentiles are $[0.681,0.945,2.236]\times10^{-12}$ erg s$^{-1}$ cm$^{-2}$.
We save 16384 correlated spectral realizations, preserving a common parameter
draw across wavelengths rather than randomizing bins independently.

The 0.04-keV floor and a prior uniform in physical temperatures and normalizations
are retained separately. The latter gives a substantially larger upper tail;
it is not pooled with the adopted samples into an arbitrary probability mixture.
Profile likelihood gives an additional check without parameter-volume weighting:
the $\Delta W=1$ interval is approximately $[0.644,1.523]\times10^{-12}$ and the
$\Delta W=2.71$ interval is $[0.586,3.998]\times10^{-12}$ in the same units,
nearly unchanged by the two temperature floors. These are conditional likelihood
ratio intervals with asymptotic calibration, not a total systematic error.
Abundance, aperture, band, and grouping alternatives remain distinct sensitivity
tests. No symmetry is imposed on the adopted uncertainty band.

For EUV normalization we use \citet{Sanz-Forcada.et.al.2025A&A...693A.285S} (Eq.~\ref{eq:sf25}),
where $L_X$ covers 0.5--10 nm and $L_{\rm EUV}$ covers 10--92 nm.
The baseline value at Earth is $3.03\times10^{-12}$ erg s$^{-1}$ cm$^{-2}$.
The 0.406 dex predictive residual is drawn once for each complete spectrum,
along with the X-ray realization and parallax. The X/EUV association is saved
explicitly. With the adopted X samples, the EUV 16th, 50th and 84th percentiles
are $[1.43,4.02,12.26]\times10^{-12}$ erg s$^{-1}$ cm$^{-2}$. These quantiles
include the common relation residual and are not centered by construction on
the nominal value. The alternative X priors are also propagated separately.
Active and quiet FISM2 solar spectra \citep{Chamberlin.et.al.2020SpWea..1802588C} are compared at fixed integrated
EUV flux. These are shape templates, not observed stellar EUV spectra.
No arbitrary magnetic flux of $10^{24}$ Mx is supplied. The HEASARC pointed-EUVE
query returned no observations in the target search region; the recovered FUSE
observation Q3010101 measures far-UV wavelengths instead.

\subsection{Ly$\alpha$, Mg II, and airglow}
The 2011 STIS E140M exposures obqr04020 and obqr04030 are fitted jointly in their
native wavelength samples. The forward model includes an intrinsic Voigt emission
profile with a free self-reversal, three H I/D I absorbing components, the atomic
fine-structure doublets, the STIS line-spread function, and pixel integration.
The intrinsic stellar profile and ISM are shared; additive backgrounds are fitted
separately for the exposures. The narrow geocoronal core is masked, with 15,
25, and 40 km s$^{-1}$ mask tests. Night flags do not imply zero geocoronal emission.

\citet{Edelman.et.al.2019ApJ...880..117E} use the same HST photons. Their three-cloud structure and
relative column/width pattern initialize the conditional model; their fitted H I
column is not counted as an independent likelihood. Total column, common width
scale, and velocity offset are refitted. Equal cloud fractions and alternative
D/H ratios test this conditional description. The E230H Mg II h and k lines
are fitted with a self-reversed profile and three foreground components to check
plausible core contrast. Mg II shape information is not equated with a direct
measurement of the hidden Ly$\alpha$ core.

The adopted Ly$\alpha$ reconstruction integrates to $3.36\times10^{-12}$ erg s$^{-1}$
cm$^{-2}$, with conditional 16th--84th percentiles of
$[3.11,3.72]\times10^{-12}$ from bounded correlated local-covariance draws. These quoted line integrals use the intrinsic profile over $\pm800$ km s$^{-1}$ about the Ly$\alpha$ rest wavelength.
The physical and mask alternatives remain separate method tests. Removing the
self-reversal worsens the fit substantially. \citet{Linsky.et.al.2013ApJ...766...69L} C II, C IV,
and Mg II proxies predict approximately $4.4$--$4.8\times10^{-12}$ in the same
units; the difference is about 1--1.5 of their empirical RMS scatters rather than
an independent contradiction. Current MAST E140M files are byte-identical to the
local files and specify flux per Angstrom. No numerical flux was digitized from
the Edelman figure, and the paper supplies no tabulated integrated intrinsic flux.

N I and O I narrow-excess tests are recorded for both exposures and several assumed
widths. None exceeds approximately $2.7\sigma$ in these local tests. Because
stellar and interstellar structure can mimic a narrow excess, no unvalidated
subtraction is applied. These diagnostics do not recover flux hidden by the ISM.

\subsection{UV calibration and optical comparisons}
Same-star LOWLIB \citep{Pal.et.al.2023ApJS..266...41P}, using its corrected flux and supplied
errors, supplies the selected near-UV and visual regions. Its blue G230LB end
still exceeds E140M/IUE. An empirical residual uses the wavelength dependence of
the library's scatter correction \citep{Worthey.et.al.2022arXiv221105043W}. On identical supported
165--170 nm intervals, SWP18651 is multiplied by $0.77540\pm0.01121$ to match
E140M. A fit over 171.5--180 nm gives an additional scatter-template amplitude
$0.96759\pm0.02434$, with $\chi^2=13.55$ for 16 degrees of freedom. In the
held-out 180--185 nm region, the fully supported common width is 1.9 nm and the
integrated residual is only $0.76\sigma$. The covariance includes the common
SWP scale and its correlation with the residual amplitude. The template shape
is fixed; this empirical residual is not an automatic repeat of the library
pipeline correction.

The selected E140M segment spans 115--170.9 nm. Larger DQ gaps use the scaled
SWP spectrum; gaps no wider than 0.05 nm are flagged interpolations. SWP supplies
170.9--179 nm and is blended with corrected LOWLIB across 179--181 nm, before
the rejected SWP red pixels. Corrected LOWLIB supplies 181--330 nm except for
the 257--284 nm wide-aperture 2013 E230H spectrum. The latter receives a
$0.969932\pm0.000475$ scale from the corrected LOWLIB 258--275 nm integral.
The fainter narrow-aperture 2011 E230H spectrum is not the continuum donor.
The quoted scale errors are propagated statistical errors, not total absolute
calibration precisions. A common 4 percent STIS echelle mode represents the
handbook's 8 percent two-sigma accuracy for the $0.2\times0.2$ arcsec aperture
\citep{Rickman.and.Brown.2025stii.book...25R}; its transfer through the normalization fits is retained.

\subsubsection{FUSE selection and native covariance}
We retrieved the six Q3010101 time-tag exposures and all calibrated channels.
The native QUALITY field denotes the percentage of usable aperture, rather
than a bit mask; we require at least 90 percent. Exposures 003 and 005 are
fully orbital-night, totaling 696 s. Below 100 nm the selected donor is night
SiC1B, with night SiC2A only where full support is missing. Between 100 and
108.2 nm we use all-time LiF1A, with LiF2B for support gaps. Above 108.2 nm we
use LiF2A, with the documented overlapping LiF fallbacks. The channel choices
follow the instrument guidance \citep{Sonnentrucker.et.al.2009FUSE} and observed cross-channel checks. SiC2B shows
excess O VI relative to LiF; LiF1B has an inconsistent 110--112 nm background
and is vulnerable to the documented worm. Neither is a nominal donor.

Coaddition uses exposure times and geometric bin overlaps, not measured flux
or noisy inverse variance as weights. The selected output stores a native-to-output
operator, native variances, and the full sparse overlap covariance. There are
2730 output bins of width 0.01 nm, of which 2636 have complete measurement support.
No scale to HST is applied. Native signed fluxes are retained.

Day-only and night-only exposures form independent comparison samples. In the
selected LiF1A channel, O VI 103.2 and 103.8 nm differ by 0.89 and 0.83 standard
deviations respectively. C III 117.6 nm in LiF2A differs by 1.24 standard
deviations. The SiC2A C III 97.7 nm difference is 3.56 standard deviations,
supporting night selection in this part of the spectrum. These are individual
diagnostics, not a multiple-testing discovery claim. All-time and night-only
coadds share photons and are not treated as independent. Ly$\beta$ airglow is
strong even at night. Explicit H I/neutral masks and support gaps flag 565 bins
for intrinsic completion in the full diagnostic range. The final FUSE segment
ends at 115 nm, where E140M becomes the donor. Over 117.4--117.75 nm,
FUSE and STIS C III integrals are $(4.21\pm0.39)$ and $(4.95\pm0.29)$
$\times10^{-14}$ erg s$^{-1}$ cm$^{-2}$, respectively: a $1.53\sigma$
difference. No cross-epoch scaling is imposed from this line.

The final continuum audit further restricts these channel candidates. LiF1A
100--101 nm is negative at $4.7\sigma$, and 106--107 nm at $3.8\sigma$;
overlapping LiF2B backgrounds disagree. The short-channel continuum is not
detected. A whole-range positivity grouping would therefore erase real emission
lines by combining them with an unreliable baseline. We instead admit only the
detected 103.1--103.9 nm complex and 108.7--115 nm red continuum, excluding the
original airglow/support masks. The remaining intervals are explicit proxies.
All selected signed FUSE measurements, including excluded regions, are supplied
in the separate FUSE\_NATIVE extension. No ad hoc background subtraction is
applied. Grouping is confined to contiguous admitted measurement intervals and
cannot consume neighbouring proxy cells or erase the O VI complex.

Unusable or undetected FUSE windows receive explicitly labelled 0.1 nm solar-template cells,
with uncertainty propagated from the reconstructed Ly$\alpha$ flux. Higher
Lyman lines use the total-series relation of \citet{Linsky.et.al.2014ApJ...780...61L},
$F_{\rm series}/F_\alpha=10^{-1.798}f_\alpha^{0.351}$, where $f_\alpha$
is the 1 AU Ly$\alpha$ flux divided by $(R_\star/R_\odot)^2$.
Here $f_\alpha=9.95$ and the complete-series prediction is
$1.197\times10^{-13}$ erg s$^{-1}$ cm$^{-2}$. FISM2 partitions this total
among the masked hydrogen windows before clipping the adopted segment to
92--115 nm; those retained windows contain $9.865\times10^{-14}$.
Other airglow/support gaps use a Ly$\alpha$-scaled solar template. These are
coarse proxy cells, not recovered individual stellar line profiles.
The active solar shape has a marginally smaller descriptive C III/O VI
log-ratio discrepancy than the quiet shape, but the comparison does not
establish a significant preference or validate the unobserved EUV. We retain
the active template as nominal and the quiet template as a method alternative.
Direct solar scaling of the hydrogen windows is a second alternative.
The published 35 percent scatter for the entire 91.2--117 nm band is not
misassigned as an error on every missing hydrogen line.

\subsubsection{Independent red-color validation and selected optical/IR scale}

Gaia XP \citep{De.Angeli.et.al.2023A&A...674A...2D} was independently reconstructed from its continuous coefficients, preserving
the full statistical wavelength covariance. LOWLIB/Gaia ratios vary from about
0.96--0.98 in the blue/visual to 0.85--0.88 in the red, so a single scale factor
does not resolve the comparison. Checksum-verified PHOENIX NewEra models \citep{Hauschildt.et.al.2025A&A...698A..47H} spanning
the relevant temperature, gravity, and metallicity were interpolated and normalized
to the adopted bolometric flux. LOWLIB/model ratios are 0.988 in 400--450 nm,
0.984 in 500--600 nm, and 0.886 in 850--950 nm. Thus the independent photospheric
comparison also reveals a red calibration difference. We retrieved the original
Johnson photometry from Morel \& Magnenat's catalogue, row HD 9826, reference
8029 \citep{Johnson.et.al.1966CoLPL...4...99J,Morel.and.Magnenat.1978A&AS...34..477M}: $B=4.64$, $V=4.10$, $R=3.64$, $I=3.35$ mag.
These are original Johnson filters, not Cousins $R,I$. Synthetic magnitudes use
the energy-response SVO Johnson passbands \citep{Rodrigo.et.al.2012ivoa.rept.1015R}\footnote{The adopted Johnson $I$ profile is available at \url{https://svo2.cab.inta-csic.es/theory/fps/fps.php?ID=Generic/Johnson.I}.} and CALSPEC Vega \citep{Bohlin.et.al.2020AJ....160...21B}\footnote{CALSPEC reference file \texttt{alpha\_lyr\_stis\_011.fits}: \url{https://ssb.stsci.edu/cdbs/calspec/alpha_lyr_stis_011.fits}.} assigned zero magnitude.
The observed $R-I=0.290$ mag compares with 0.290 for Gaia, 0.277 for the
bolometric-normalized NewEra model, and 0.219 for LOWLIB. A common Vega magnitude
offset cancels in this color. Approximately six percent of the I-band integral
uses the model tail beyond 1000 nm. The compilation gives no per-target formal
errors, so the color is an independent qualitative check, not a precision
chi-square calibration. It favors the Gaia red shape; it does not identify the
physical cause of the LOWLIB discrepancy.

We retain LOWLIB through 560 nm and match Gaia to the LOWLIB integral over
500--560 nm. The multiplicative factor is $0.97892\pm0.00167$, where the
uncertainty includes the supplied statistical errors and full Gaia covariance.
Gaia supplies 560--980 nm. A labelled 980--1000 nm transition combines Gaia and
NewEra with a linearly increasing model fraction, avoiding a few-percent edge
step without claiming that this mixed region is a pure observation. NewEra
supplies the longer wavelengths, normalized to the selected Gaia 950--990 nm
integral; its factor relative to the bolometric model is 1.00932, rather than
the rejected LOWLIB-red anchor of 0.88148. No polynomial correction is applied
to the original library spectrum.

The normalization denominators contain the same measurements as the output
spectrum. Their full Jacobians therefore propagate both positive and negative
cross-covariances; the visual scale and the IR anchor are not independent
per-pixel errors. The optical covariance and the optical--IR anchor cross-mode
are saved explicitly. An independent nonlinear Monte Carlo check with 8192
draws reproduces representative band standard deviations to within one percent.
The temperature derivative is evaluated on the bracketing grid; gravity and
metallicity alternatives remain method sensitivities. In the final surface-flux
assembly, temperature must retain its dependence on bolometric flux and angular
diameter, and the diameter also sets the shared $4/\theta^2$ surface conversion.
Repeating the same stellar-parameter error as an independent temperature error
would double-count information. Bright-star Gaia systematics are not declared
removed by this relative normalization, and no guessed epoch scatter is added.

\subsection{Final assembly, covariance, and validation}
The assembled signed spectrum has 65325 bins spanning 0.5--5000 nm without
gaps or overlaps. Its 3774 negative nominal bins are retained in the signed
extension. The recommended irradiation extension groups adjacent bins until
each integrated group is nonnegative, producing 59396 bins. No negative
measurement is clipped to zero. The grouping operator propagates all covariance
and sample blocks and conserves the X, EUV, FUV, NUV, optical, IR, and total
integrals of the adopted assembly (not the rejected native continuum). Uncertainties are conditional on the resulting grouping.
The covered bolometric integral is 0.98014 of the adopted bolometric flux;
the finite 5000 nm endpoint omits the longer-wavelength tail.

Ly$\alpha$ replaces 121.35--121.80 nm. Outside that interval, the modelled
absorbed wing contribution is added over 120.8--122.4 nm, preserving the
observed O V line at 121.834 nm. Bin quadrature is used instead of evaluating
only line centers. The reconstruction remains conditional on the stated
profile and cloud constraints. Neutral-metal interstellar absorption is not
claimed to have been fully reconstructed by the airglow tests.

The public covariance representation is
\begin{equation}
C=B+UKU^T+\operatorname{Cov}(\Delta_X)
 +\operatorname{Cov}(\Delta_\alpha).
\end{equation}
$B$ retains native statistical and overlap covariance; $U,K$ retain common
normalization, stellar-parameter, and calibration modes. The metric $K$ can
contain negative cross-terms; only the complete covariance is meaningful.
The X/EUV block uses 4096 paired realizations from a complete power-of-two
Sobol prefix, preserving the Sanz-Forcada residual across wavelengths.
It is independent of the 4096-realization Ly$\alpha$/FUV block.

The Ly$\alpha$ fit uses some of the same native pixels as the retained wings.
We recover their parameter--data cross-covariance from the fit Jacobian and
native variances. The conditional residual covariance subtracts the shared
parameter contribution, while the joint line/proxy sample block restores
its correlated effect on the wings. This avoids treating the fitted line and
the pixels constraining it as independent information. It remains a local
Gaussian approximation with bounded parameter draws, not a global posterior.
Optical/IR anchor covariance is retained similarly. The angular diameter
jointly controls surface conversion, photospheric temperature and the
higher-Lyman proxy; bolometric flux and parallax have separate shared modes.

The published nominal curve is the best-fit spectrum, not a sample mean.
The 16th--84th percentile bands are asymmetric where the inference requires it;
they need not be centered on the best fit. Negative lower limits remain in
the data; logarithmic plots extend the supported shading to the displayed floor and mark nonpositive lower endpoints (Section~\ref{sec:display}).
Method alternatives are not labelled confidence limits, and the available
uncertainty is not promoted to a measured universal systematic error.
All supplied errors and the reproducible parameter contributions described
above are propagated; no arbitrary epoch-variability percentage is introduced.

The release includes the main FITS, a covariance/sensitivity FITS, component
and band tables, input hashes, a portable executed plot notebook, and this
English source/PDF. Validation checks contiguous support, positive irradiation
flux, conserved integrals, covariance diagonals and positive quadratic forms,
sample linkage, FITS checksums, and notebook execution. Figures display inline
by default; a single \texttt{SALVAR\_PDF=1} switch enables PDF export.

\subsection{Final spectrum with available uncertainties}
Figure~\ref{fig:sed:upsilon_And} shows the adopted full SED and its available uncertainty bounds.

\begin{center}
\includegraphics[width=\linewidth]{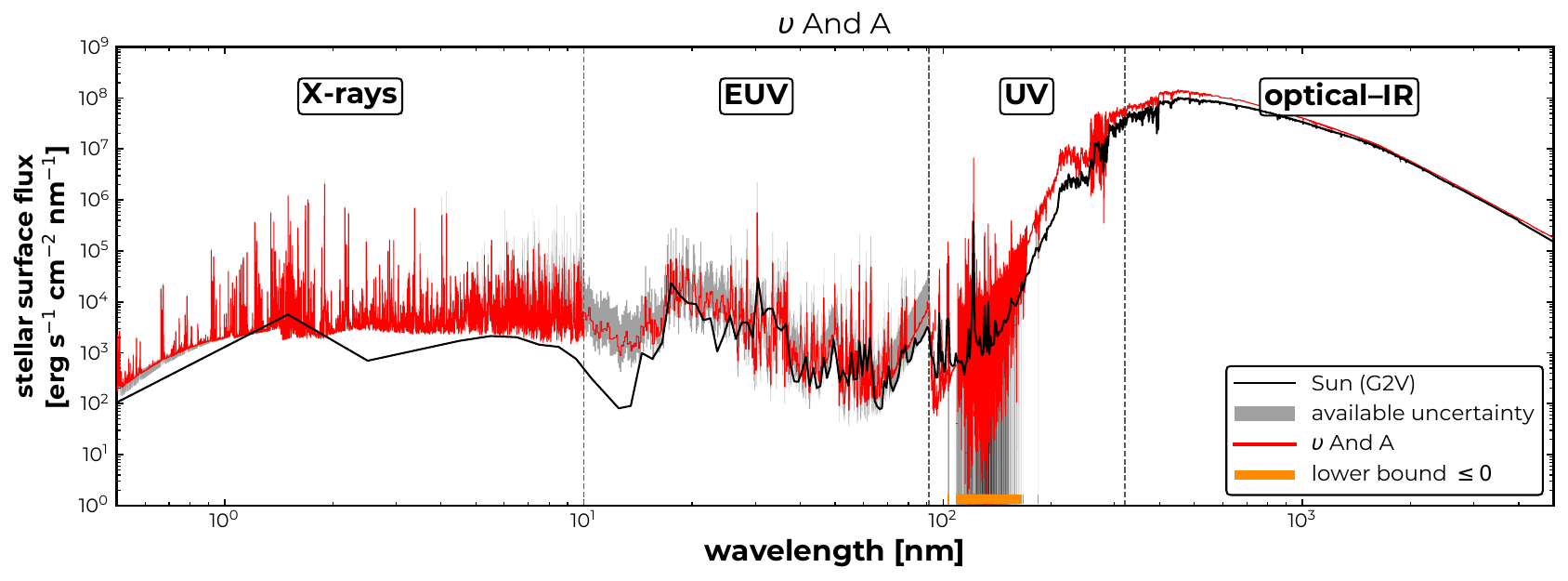}
\end{center}
\captionof{figure}{$\upsilon$ Andromedae A: final nominal surface-flux spectrum and available uncertainties. Target colours, shading, solar comparison and nonpositive-bound markers follow Section~\ref{sec:display}. The component origins and uncertainty scope are documented above.}\label{fig:sed:upsilon_And}

%% file: sections/47_UMa.tex
\clearpage
\section{\texorpdfstring{47 Ursae Majoris}{47 Ursae Majoris}}\label{star:47_UMa}
This section explains how source confusion in the X-ray archive and
the IUE/COS discrepancy were evaluated for 47 UMa. It presents the adopted
high-energy solution, the treatment of Ly$\alpha$ and contaminated ultraviolet
intervals, and the optical/infrared scale used in the released composite.

\subsection{X-ray identity and the adopted X/EUV solution}
Gaia DR3 astrometry propagated to the 2006 June 11 XMM observation
(0304203401) places the star at $\alpha=164.8658084^\circ$,
$\delta=40.4303544^\circ$. The nearest existing PPS extraction,
source 0002, is centred 22.42 arcsec away with an approximately
23 arcsec extraction radius. Pointing coordinates do not establish
the identity of an extracted source. The confusion is consistent
with the case discussed by \citet{Sanz-Forcada.et.al.2010IAUS..264..478S}; this PPS spectrum is
rejected as the stellar anchor.

New SAS 22.1 spectra and responses were generated at the stellar
position: PN apertures of 10, 12, and 15 arcsec, and MOS1/2 apertures
of 12 arcsec. PN uses \texttt{PATTERN}$\leq4$, MOS uses
\texttt{PATTERN}$\leq12$, and both use \texttt{FLAG}=0. Background
regions lie on the same CCD and exclude the neighbouring source.
A 10--12 keV, single-event background audit in 100 s bins found
median/maximal rates of 0.11/0.20, 0.05/0.15, and 0.06/0.13
counts s$^{-1}$ for PN, MOS1, and MOS2, respectively. Bin exposures
are computed from the union of event-file GTIs. No isolated strong
high-energy flare justifies a new time cut, so the existing GTIs
are retained. This result does not identify the origin of the
low-energy excess.

Fits use Poisson source/background statistics (W-stat). The
published plasma prescription is reproduced with APEC 3.0.9,
Anders--Grevesse abundances, iron at $-0.30$ dex relative to that
scale, $\log T=6.25$, $\log EM=48.32$, and
$\log N_{\rm H}=18.3$ \citep{Sanz-Forcada.et.al.2025A&A...693A.285S}. It differs from the initial
diagnostic model that scaled all metals together. With temperature
and abundances fixed to the published values, the PN normalization
fit yields $\log L_X=26.400$ when fitting 0.2--2 keV and
$25.743$ when fitting 0.4--2 keV. Both luminosities are integrated
over the same 0.5--10 nm band. The latter is close to the published
$25.75$, but agreement does not prove that this energy cut is
uniquely correct; the photons are not an independent data set.
The soft excess and temperature--normalization degeneracy remain
diagnostic limitations, not evidence for a securely measured hotter
or more luminous stellar component.

The adopted nominal uses the published target plasma rather than
the confused PPS source or an unconstrained new extrapolation.
Re-evaluation of that plasma gives
$F_X=2.489\times10^{-15}$ erg s$^{-1}$ cm$^{-2}$ at Earth and
$\log L_X=25.759$ in 0.5--10 nm. The EUV is taken from column F4
of the machine-readable \citet{Sanz-Forcada.et.al.2025A&A...693A.285S} spectrum, with its 1 au
normalization converted explicitly. The resulting 10--92 and
10--50.4 nm luminosities reproduce the published 26.62 and 26.26
within 0.012 dex. The 10--92 nm flux at Earth is
$1.786\times10^{-14}$ erg s$^{-1}$ cm$^{-2}$.
This is a coronal/transition-region reconstruction, not a direct
EUV observation. No arbitrary magnetic flux, no Namekata
normalization, and no second X-to-EUV regression is applied.

\subsubsection{Meaning of the X/EUV bounds}
The published marginal bounds are $\log T=6.00$--6.39,
$\log EM=47.85$--48.53, and iron $-0.30$--$-0.10$ dex.
Their joint covariance is unavailable. We evaluate the Cartesian
grid including the nominal values and supply its pointwise
envelope as a \emph{parameter sensitivity}, not as a joint
confidence interval or posterior. An envelope at all wavelengths
need not correspond to one physical parameter vector; the actual
grid spectra are also distributed.

The published integrated EUV endpoints are $-0.26/+0.37$ dex
in 10--92 nm and $-0.10/+0.20$ dex in 10--50.4 nm.
For display only, these are transported to fixed spectral shapes
in 10--50.4 and 50.4--92 nm. The long-band endpoints are the
differences of the total and short-band endpoints, pairing the
same lower or upper endpoint. This coupling is an explicit
assumption, not a recovered joint probability distribution.
The FITS identifies these bounds separately from conditional
statistical intervals and leaves the unavailable X/EUV standard
deviations undefined instead of inventing a covariance.

\subsection{Ultraviolet data and the IUE/COS discrepancy}
The 2017 COS G130M/1300 LP3 exposures \texttt{ld5b01enq} and
\texttt{ld5b01eqq} span approximately 114.08--143.92 nm. The first
is entirely in orbital night by the solar-altitude criterion,
and most of the second is also at night. Native quality flags and
both supplied error directions are retained; the larger native
error is used in the symmetric conditional fitting/coaddition
approximation. Flux-independent resampling operators propagate
covariance as $A C A^{\mathsf T}$. The 2022 STIS E230H spectrum
\texttt{oee524010}, through the $0.2\times0.2$ arcsec aperture,
provides the 256.9--284.5 nm adopted interval. Order edges are
trimmed and merged using exposure/taper weights. Small unsupported
cells are explicitly flagged when interpolated.

NEWSIPS SWP15716 shows a large false-continuum excess relative
to COS. An independent INES extraction also retains the excess,
so changing extraction alone does not solve it. Scattered light
in late-type SWP spectra, including signal shortward of Ly$\alpha$,
is documented by the IUE archive \citep{Thompson.1998idac.book.....T}. The original
IUEDAC \texttt{SLBOXCAR} software implements a constant detector-FN
pedestal subtraction before flux calibration\footnote{IUEDAC software-change log, entries of 11 and 25 November 1997, including Mark Vincent's \texttt{SLBOXCAR}: \url{https://archive.stsci.edu/iue/rdafnews_10nov98.html}.}.
We use that principle, retaining the NEWSIPS SWET extraction
rather than claiming to reproduce the complete historical
boxcar-extraction algorithm.

Let $R_i=F_i/N_i$ be the native flux-per-FN response, where $N_i$
is the extracted NET array. The corrected spectrum is
\begin{equation}
 F_{i,\rm corr}=F_i-aR_i.
\end{equation}
The response varies with wavelength, so this is not a constant
energy-flux subtraction or a multiplicative IUE/COS scale.
COS is convolved with a 0.6 nm FWHM Gaussian approximation to
the low-dispersion IUE resolution. Generalized least squares
in the common continuum, excluding strong airglow/line intervals,
gives $a=24.324\pm1.878$ FN, conditional on this correction family,
with $\chi^2=125.13$ for 96 degrees of freedom. Covariance includes
the shared COS contribution and the native noise entering $a$.

An independent blind interval at 110--115 nm has a median-filtered
NET mean of 34.87 FN and block-bootstrap dispersion 4.52 FN.
The conservative historical subtraction rule gives 32.39 FN.
These samples have the NEWSIPS flag indicating uncalibrated
\emph{flux}; their detector counts remain a diagnostic, not a
calibrated stellar spectrum. The blind-window correction is
retained as an alternative rather than silently merged with the
COS-constrained value. Leave-band-out residuals in 124--127,
134--136, and 140--143 nm are $+0.28$, $+1.75$, and $-2.77$
times the propagated conditional errors. Thus the dominant
excess is removed, but a perfect transfer-function calibration
is not demonstrated. Possible residual shape and epoch dependence
are not relabelled as an independently measured random error.

The final spectrum uses COS where supported, corrected SWP in
the 128.385--129.780 nm detector gap and from 143.9 to 197.8 nm,
and LWR07349 thereafter outside E230H. E230H-supported 5 nm
bands give a grey LWR scale of $1.11004\pm0.01678$. The band
ratios span 1.021--1.211, with $\chi^2=21.47$ for three degrees
of freedom; this residual variation is retained as sensitivity,
not hidden within the formal mean-scale error. LWR12130 requires
a larger scale of approximately 1.247 and remains an alternative.
The adopted LWR 300--315 nm integral agrees with the
Gaia-anchored NewEra prediction to about 0.6\%, independently
of the E230H normalization fit. No wavelength-dependent tilt
is imposed. Below 250 nm, fixed 5 nm LWR bins preserve the
measured integrated continuum and covariance at its low
signal-to-noise ratio; they are not a synthetic continuum bridge.

\subsection{Ly$\alpha$, airglow, and short-wavelength completion}
The two COS exposures are fitted simultaneously using a Voigt
stellar emission profile with variable Gaussian self-reversal,
single-component H I/D I absorption, the configuration-specific
COS line-spread function, and the empirical airglow template of
\citet{CruzAguirre.et.al.2023ApJ...946...98C}. Each exposure has its own airglow amplitude,
shift, and small baseline. The adopted $D/H=1.5\times10^{-5}$
and $b_D=b_H/\sqrt{2}$ specify the absorption model. The Mg II
LIC velocity from \citet{Nisak.et.al.2025ApJ...983....5N} constrains the relative
stellar/interstellar registration with additional tolerance;
Mg II column density is not converted into an H I measurement.
A contrast regularizer of $1.7\pm0.7$ is an analysis choice,
not a measured target constraint or calibrated population prior.

The nominal intrinsic Ly$\alpha$ flux is
$9.38\times10^{-13}$ erg s$^{-1}$ cm$^{-2}$ and
$\log N_{\rm HI}=17.845$. Six-pixel residual block resampling
produces 127 successful fits from 128 trials; the integrated
16th, 50th, and 84th percentiles are
$(9.28,9.75,10.26)\times10^{-13}$ in the same units.
These are conditional bootstrap diagnostics, not the full
uncertainty in the profile family. The nominal remains the
selected fit, rather than the bootstrap mean or median.
Removing the contrast constraint, removing self-reversal,
masking the core, and using the night exposure alone give
$8.71\times10^{-13}$, $1.36\times10^{-12}$,
$1.20\times10^{-12}$, and $1.14\times10^{-12}$, respectively.
These alternatives are distributed separately.

C II and Mg II diagnostics using \citet{Linsky.et.al.2013ApJ...766...69L} predict
approximately $1.45\times10^{-12}$ and $1.67\times10^{-12}$.
The lower reconstruction differs by about 1.4--1.6 times the
empirical relation scatter, not a formal independent confidence
test. N I and O I controls agree between exposures within their
propagated errors. They remain observed profiles; no unmeasured
metal-line interstellar correction is inferred. The incomplete
130.10--130.65 nm support is not represented as a complete
measured triplet integral. The final Ly$\alpha$ replacement
removes the samples entering its fit from the direct-observation
donor. O V at 1218.34 Angstrom, outside the fitted interval,
is retained as an observed residual over the absorbed and
airglow-contaminated model, with the shared model variation
propagated through the same bootstrap block.

No direct FUSE/EUVE spectrum is adopted for this release. The
92--114.08 nm completion is explicitly a model: the published
coronal/transition-region spectrum supplies non-hydrogen bins,
while higher Lyman lines use the Ly$\alpha$-dependent integrated
relation of \citet{Linsky.et.al.2014ApJ...780...61L}, with a quiet-Sun FISM2 template
\citep{Chamberlin.et.al.2020SpWea..1802588C}
partitioning the line flux. An active-Sun partition is retained
as sensitivity. Shared Ly$\alpha$ and angular-diameter terms
are propagated for hydrogen. No quantified model-error
distribution is assigned to the remaining non-hydrogen bridge.

\subsection{Optical/IR scale and the public product}
Same-star Gaia DR3 XP data, source 777254360337133312, retain the
full coefficient covariance. A PHOENIX NewEra model
\citep{Hauschildt.et.al.2025A&A...698A..47H} is anchored to the Gaia 500--900 nm integral,
sharing its normalization covariance with the observed optical
spectrum. The adopted model parameters are
$T_{\rm eff}=5829\pm95$ K, $\log g=4.328\pm0.032$, and
$[M/H]=0.026\pm0.061$ \citep{Rosenthal.et.al.2021ApJS..255....8R}; local derivatives
propagate these marginal uncertainties under an explicit
independence approximation. They are not a measurement of
atmosphere-model discrepancy. Gaia is used from 336 to 1000 nm,
NewEra from 315 to 336 nm and beyond 1000 nm.

The direct limb-darkened diameter
$\theta=0.774\pm0.073$ mas \citep{vanBelle.and.vonBraun.2009ApJ...694.1085V} sets
$F_{\rm surf}=4F_{\oplus}/\theta_{\rm rad}^{2}$.
Gaia gives $d=13.8875\pm0.0188$ pc. A radius derived from
$d$ and $\theta$ is provided for convenience, but is not an
additional independent surface-error term. The Gaia-anchored
photospheric model implies about 0.833 mas, compatible with
the direct diameter. Surface conversion retains the shared
angular-diameter uncertainty instead of forcing every measured
segment onto a revised photospheric radius.

The available conditional covariance has sparse native,
low-rank normalization/cross-covariance, and joint Ly$\alpha$
sample terms. Native errors, fitted scales, and shared model
anchors are not summed as independent pixel errors. The FITS
columns \texttt{BOUND\_TYPE} and \texttt{COV\_KNOWN} distinguish
conditional errors, parameter envelopes, published-band
transport, and unquantified blocks. A missing error is not zero.
The X/EUV and non-hydrogen FUV blocks prevent the available
covariance from being called a complete total uncertainty.
No arbitrary variability-between-epochs term is introduced.

The signed assembled spectrum is retained. For the nominal
irradiation product, adjacent negative bins are combined with
neighbours, propagating the same linear operator through the
covariance and alternatives. This conserves the total flux
and the adopted broad-band integrals; individual negative
measurements are not clipped to zero. The final product has
27,545 bins and continuous 0.5--5000 nm coverage, including
explicitly labelled model completion. Native uncertainty and
method sensitivity remain distinct. The portable notebook
shows one nominal curve, whole-spectrum and regional plots,
and linear diagnostics. Finite non-positive lower bounds are
filled to the visible floor of a logarithmic plot, with a
continuation marker; this display operation never changes
the scientific bounds. Undefined errors remain unshaded.

\subsection{Final spectrum with available uncertainties}
Figure~\ref{fig:sed:47_UMa} shows the adopted full SED and its available uncertainty bounds.

\begin{center}
\includegraphics[width=\linewidth]{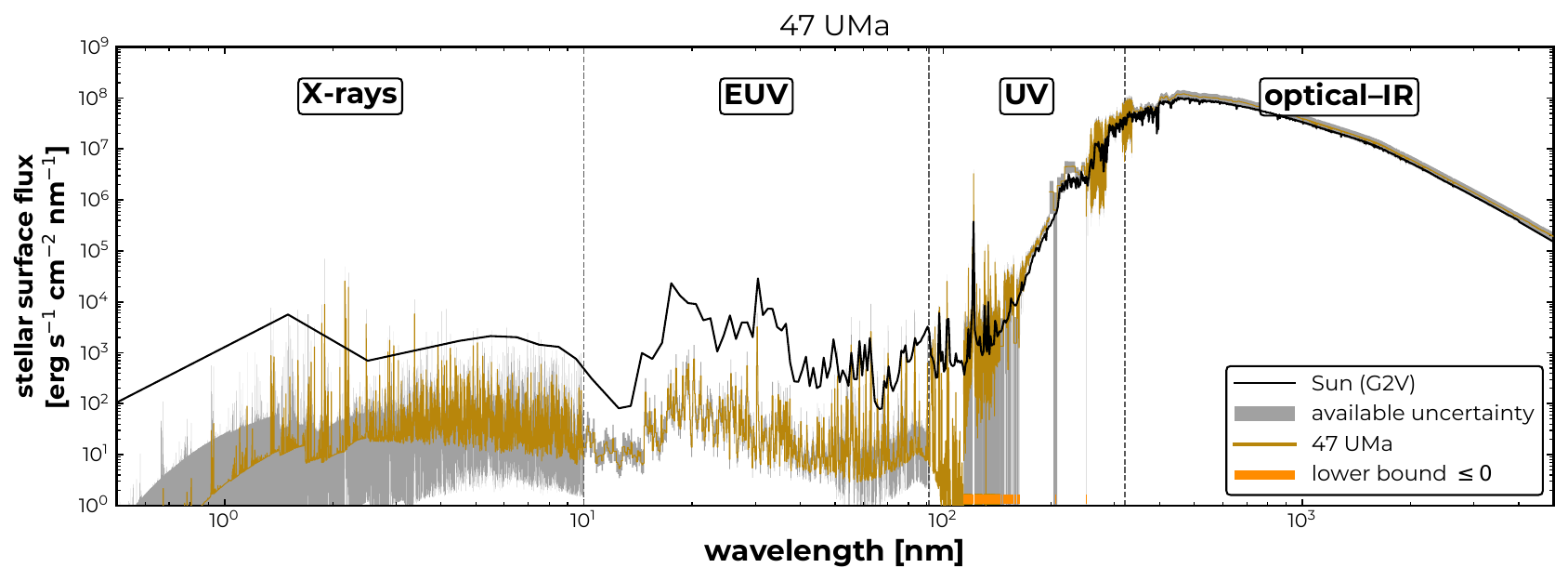}
\end{center}
\captionof{figure}{47 Ursae Majoris: final nominal surface-flux spectrum and available uncertainties. Target colours, shading, solar comparison and nonpositive-bound markers follow Section~\ref{sec:display}. The component origins and uncertainty scope are documented above.}\label{fig:sed:47_UMa}

%% file: sections/iot_Hor.tex
\clearpage
\section{\texorpdfstring{$\iota$ Horologii}{iota Horologii}}\label{star:iot_Hor}
This section follows the assembly of $\iota$ Hor from the adopted
stellar parameters through the X/EUV and ultraviolet reconstructions to
the optical/infrared spectrum. It highlights observing-epoch checks, the
absorbed Ly$\alpha$ core, and the distinction between the nominal composite,
its available covariance, and physical-method alternatives.

\subsection{Stellar parameters and flux convention}
We identified the target with \textit{Gaia} DR3 source
4745373133284418816. Its parallax is $57.61314\pm0.03828$\,mas,
corresponding to $d=17.35715\pm0.01153$\,pc by direct inversion. The small
fractional parallax error makes the inversion adequate here. We adopted
$T_{\rm eff}=6080\pm80$\,K, $\log g=4.399\pm0.022$, and
$R_\star=1.16\pm0.04\,R_\odot$, the Bruntt et al. (2010) values compiled
by \citet{AmazoGomez.et.al.2023MNRAS.524.5725A}. The photospheric abundance is
$[\mathrm{Fe/H}]=+0.19\pm0.06$, from Gonzalez \& Laws (2007) as reported
by \citet{Sanz-Forcada.et.al.2019AA...631A..45S}. These photospheric quantities are distinct from the
element-specific coronal abundances used in the X-ray calculation. The
approximately 625\,Myr asteroseismic age listed by \citet{AmazoGomez.et.al.2023MNRAS.524.5725A}, with
reference to Vauclair et al. (2008), is descriptive metadata; no geological
epoch was assigned to the star.

The primary tabulated quantity is the nominal flux density at Earth per
nanometre. We provide the geometric transformations explicitly:
\begin{equation}
 F_{\lambda,\mathrm{surf}}=F_{\lambda,\oplus}(d/R_\star)^2,
 \qquad
 F_\lambda(a)=F_{\lambda,\oplus}
 \left[\frac{d}{a\,{\rm AU}}\right]^2.
\end{equation}
All integrals use the actual bin widths. Pipeline fluxes per angstrom were
multiplied by ten when converted to per-nanometre densities. Surface-flux
uncertainty includes the common distance and radius factors, rather than
assigning independent geometric errors to individual spectral bins.

\subsection{X-rays and extreme ultraviolet}
The high-energy reconstruction is presented in three steps: verification of
source identity and individual epochs, selection of the nominal coronal
spectrum, and assessment of the EUV normalization and spectral shape.
\subsubsection{Source identity and individual-epoch checks}
We inspected the source, background, redistribution and ancillary-response
files for XMM-Newton observations 0693550701 (MOS1; 2012 December 20) and
0693550501 (pn; 2012 August 9). Propagating the \textit{Gaia} position and
proper motion to each epoch placed the catalogue source centroids within
0.60 and 2.24 arcsec of the star, respectively. The source spectra contained
1127 and 4601 counts before the analysis-channel selection, with effective
exposures of 6549 and 6697\,s.

We refitted these spectra over 0.2--2.4\,keV with the Poisson source/background
statistic in \texttt{XSPEC} 12.15.1. We fixed the column to the approximately
$6.5\times10^{18}\,\mathrm{cm^{-2}}$ inferred from the ultraviolet ISM analysis
and used \texttt{APEC} 3.0.9. The element-specific EPIC abundances of
\citet{Sanz-Forcada.et.al.2019AA...631A..45S} were translated from their Asplund et al. (2009) reporting
convention to the Anders \& Grevesse (1989) standard used by the plasma code.
At the precision relevant here, the respective one- and two-temperature
statistics were $34.27/24$ and $16.49/22$ for MOS1, and $191.81/30$ and
$42.82/28$ for pn (statistic/degrees of freedom). The two-temperature
luminosities over 0.5--10\,nm were approximately $\log L_X=28.76$ and 28.89.
These checks show that the old single-temperature description is inadequate
for at least the pn epoch and is disfavoured with the adopted abundances in
both datasets. They reanalyse observations from the published monitoring;
they are not an independent validation of that observing programme.

\subsubsection{Adopted nominal X-ray spectrum}
For the released mean-state spectrum we recomputed the three-temperature
EPIC model reported by \citet{Sanz-Forcada.et.al.2019AA...631A..45S}:
\begin{align}
 \log T\,[\mathrm K]&=(6.15,\,6.64,\,6.92),\\
 \log\mathrm{EM}\,[\mathrm{cm^{-3}}]&=(50.95,\,51.12,\,50.81).
\end{align}
The corresponding reported marginal uncertainties are
$(0.02,0.01,0.01)$\,dex in temperature and $(0.06,0.02,0.02)$\,dex in
emission measure. The adopted abundance ratios relative to AG89 are
O$=0.562$, Ne$=0.692$, Mg$=1.413$, Si$=0.851$ and Fe$=0.933$; unlisted
elements remain at the code's solar baseline. The recomputed unabsorbed
spectrum gives $\log L_X=28.903$ in the precisely defined 0.5--10\,nm band,
consistent with the 28.91 listed by \citet{Sanz-Forcada.et.al.2025A&A...693A.285S}.

We evaluated 729 combinations of the central, lower and upper marginal
temperature and emission-measure values. Their pointwise envelope records
parameter sensitivity. It is \emph{not} a joint confidence interval, because
the joint parameter covariance is not supplied. Abundances are held fixed
in this calculation. The actual alternative spectra are included alongside
the envelope, so users need not interpret a pointwise envelope as a single
physical plasma. The two observed epochs are retained as activity-state
checks, not converted into an invented Gaussian epoch-error term.

\subsubsection{Target-specific EUV reconstruction and archive consistency}
We retrieved column F68 (HR\,810) from the spectral catalogue associated
with \citet{Sanz-Forcada.et.al.2025A&A...693A.285S}. That work explicitly uses the coronal model of
\citet{Sanz-Forcada.et.al.2019AA...631A..45S} for this target. The model is constrained by RGS and STIS
spectroscopy; it is a reconstruction, not a direct EUV observation.
The archive wavelength unit is 0.1\,nm, and its flux-density unit is
$10\,\mathrm{mW\,m^{-2}\,nm^{-1}}$ at 1\,AU, equal to
$10\,\mathrm{erg\,s^{-1}\,cm^{-2}\,nm^{-1}}$.

An explicit integration exposed an important distinction: F68 gives
$\log L_X=28.615$ over 0.5--10\,nm, whereas the paper's table gives 28.91.
In contrast, its EUV integrals give 29.1356 over 10--92\,nm and 28.6605 over
10--50.4\,nm, reproducing the tabulated 29.14 and 28.66. We therefore used
the stated EPIC model for X-rays and retained F68 for the EUV. We did not
multiply the entire tabulated spectrum by a factor chosen to repair its X-ray
integral; doing so would spoil the correctly normalized EUV bands.

We retained the published EUV band bounds, $29.14^{+0.19}_{-0.18}$ and
$28.66^{+0.23}_{-0.16}$\,dex. To display a spectral interval, we transported
these endpoints to fixed shapes in 10--50.4 and 50.4--92\,nm, requiring the
sum to recover the full-band bounds. Coupling the corresponding endpoints
is an explicit display assumption, not a recovered joint posterior.
No additional X--EUV regression, arbitrary $10^{24}$\,Mx magnetic flux,
or uncertainty taper was applied.

\subsection{Native ultraviolet observations and intrinsic Ly\texorpdfstring{$\alpha$}{alpha}}
The ultraviolet analysis begins with exposure and echelle-order selection,
then addresses the absorption that obscures the Ly$\alpha$ core. The final
step treats other airglow features and the short-FUV intervals that require
model completion.
\subsubsection{Exposure and order selection}
We retrieved the six native STIS \texttt{x1d} products from programmes
15299 and 15512. The adopted FUV spectra are
\texttt{odv501020} and \texttt{odv501030}, obtained in 2019 with E140M and
the $0.2\times0.2$ arcsec aperture, each with 3001\,s exposure. The NUV
segment uses \texttt{odv501010}, E230H with the $0.2\times0.09$ arcsec
aperture and 1899\,s exposure.

The 2018 second FUV exposure showed enhanced transition-region emission,
consistent with the flare-like event discussed by \citet{AmazoGomez.et.al.2023MNRAS.524.5725A}; we retained
it as a comparison. The 2018 NUV file carries acquisition/exposure quality
alerts. Its header comment about absent counts conflicts with a nonempty
extraction, so we did not interpret it as a zero-flux measurement or use it
as the nominal reference. Choosing the 2019 visit also keeps the FUV and NUV
observations contemporaneous with each other.

For each echelle order we retained the native uncertainty array and applied
the pipeline serious-quality mask. We trimmed the outer one percent of each
order and merged overlapping intervals using exposure-time and smooth
geometric edge weights, independent of the noisy measured flux. The linear
overlap operator propagates both variances and the covariance introduced by
redistribution. Geometrically unsupported bins are separately identified;
small repairs carry a DQ-interpolation flag. HSLA coadds were comparison
products, not extra independent observations. The revised E230H selection
extends through 284.4\,nm and therefore restores the observed Mg\,II h and k
region omitted by the old 277\,nm cutoff.

\subsubsection{Two-cloud absorption and the saturated core}
We used the two-cloud solution of \citet{AmazoGomez.et.al.2023MNRAS.524.5725A}, constrained there by
Fe\,II absorption: velocities 7.4 and 15.9\,km\,s$^{-1}$, hydrogen columns
$\log N_{\rm H}=18.09$ and 18.72, and common $b_{\rm H}=11.1$\,km\,s$^{-1}$.
We included deuterium with D/H$=1.5\times10^{-5}$ and thermal/turbulent
width scaling. The intrinsic stellar emission was attenuated, convolved
with the tabulated STIS E140M 1200\,\AA\ line-spread function for the actual
aperture, and integrated over the native pixels. We masked the central
$\pm50$\,km\,s$^{-1}$ in the geocentric frame to exclude the dominant
airglow. The two exposure backgrounds and stellar wavelength offset were
nuisance parameters.

We compared a Voigt family and a two-Gaussian emission family with a
multiplicative central reversal. The two-Gaussian description better
reproduces the observed wings and avoids the negative baseline displacement
favoured by the Voigt tails. Allowing a narrow central reversal to vary
showed that it is effectively unidentifiable behind the saturated
absorption: profiles with very different core depths give nearly the same
fit statistic. We therefore adopted a stated mild-reversal completion,
$\exp[-0.3\exp(-v^2/2\sigma_r^2)]$ with reversal FWHM 60\,km\,s$^{-1}$,
and retained absent and stronger reversals as alternatives. This is a
physical modelling choice, not a measured reversal depth or a flux digitized
from the published figure. Mg\,II is shown as a useful chromospheric
diagnostic, not copied as an intrinsic hydrogen template.

The adopted intrinsic line flux is
\begin{equation}
 F_{\mathrm{Ly}\alpha}=2.122\times10^{-12}\,\fluxunit.
\end{equation}
Three hundred four-pixel residual-block bootstrap fits give conditional
16th--84th percentiles of $(1.995,2.241)\times10^{-12}\,\fluxunit$.
These percentiles condition on the chosen ISM and central-profile recipe;
they are not a total reconstruction error. Repeating the fit after grouping
two, four or six native pixels changes the flux by approximately four to
five percent. Wider/narrower airglow masks have a smaller effect for the
adopted profile.

\begingroup\emergencystretch=1em
The dominant ambiguity is physical rather than numerical. Increasing or
decreasing both cloud columns by 0.10\,dex gives line fluxes of
$(3.821,1.353)\times10^{-12}\,\fluxunit$, respectively. These are chosen
sensitivity tests, since no uncertainty for those literature columns is
specified here. The C\,II and C\,IV relations of \citet{Linsky.et.al.2013ApJ...766...69L} predict
approximately $3.81$ and $4.49\times10^{-12}\,\fluxunit$, respectively,
from the measured diagnostic windows. They exceed the adopted reconstruction
by factors of about 1.8 and 2.1. Their population scatter and the residual
metal-line absorption/DQ limitations do not establish which profile is
correct. We preserve this discrepancy rather than claiming an independent
confirmation or fitting the empirical predictions as exact flux constraints.
The existing published ISM analysis used the same HST observations and is
also not an independent set of photons.\par\endgroup

The released spectrum replaces the stellar \lya\ interval with the intrinsic
model and retains the observed O\,V 1218.34\,\AA\ residual over it. Bootstrap
and model variations propagate through both the intrinsic profile and the
absorbed-profile subtraction at O\,V.

\subsubsection{Other airglow lines and the short-FUV completion}
The 2019 spectra were obtained in \texttt{ACCUM} mode and cannot be divided
into orbital-night event lists. Moreover, the stellar and geocoronal
velocities differ by only about 5\,km\,s$^{-1}$. We therefore examined the
calibrated two-dimensional images: the saturated \lya\ core provides a
slit-filling spatial template, while other bright stellar lines constrain
the point-source profile. No individual N\,I/O\,I exposure gave a
greater-than-$3\sigma$ diffuse-source detection in this approximate
diagnostic. The combined O\,I 1306\,\AA\ result reaches approximately
$2.8\sigma$ and depends on the spatial templates and background treatment.
We retained these weak regions, flagged possible contamination and supplied
one-sided subtraction scenarios instead of claiming a secure sky correction.
The dominant geocoronal \lya\ emission is removed by the reconstruction.

For 92--115\,nm, we retained target-EMD emission outside the hydrogen lines
and replaced the optically thin higher-Lyman predictions with a
\lya-calibrated prescription from \citet{Linsky.et.al.2014ApJ...780...61L}. FISM2, the Flare
Irradiance Spectral Model version 2 \citep{Chamberlin.et.al.2020SpWea..1802588C}, supplies relative
solar line strengths only; an active-Sun alternative is retained. Shared
\lya\ and geometric uncertainty propagates to these lines. The non-hydrogen
model error is not assigned a fictitious distribution. FUV checks also
show that the EMD reproduces some observed transition-region lines much
better than C\,III 1175\,\AA\ or He\,II 1640\,\AA; the full observed STIS
segment takes precedence wherever available. The reconstructed gap is not
advertised as measured or uniformly validated.

\subsection{Optical, infrared and missing ultraviolet intervals}
We recalibrated the target's continuous \textit{Gaia} XP coefficients and
propagated their full covariance through the wavelength response and BP/RP
merge \citep{De.Angeli.et.al.2023A&A...674A...2D}. The resulting fluxes and errors agree with the
archive sampled products to better than 0.05 percent. We used the
PHOENIX NewEra grid \citep{Hauschildt.et.al.2025A&A...698A..47H}, interpolating in temperature,
gravity and metallicity. Its 500--900\,nm normalization is derived from the
same XP data included in the SED. That normalization differs by only about
3.7 percent from the adopted radius/distance prediction.

No same-target LOWLIB entry was identified in the supplied metadata. We did
not substitute an unrelated LOWLIB or NGSL spectrum for an observation of
$\iota$ Hor. The missing 170--257.6\,nm interval uses a coarse NewEra plus
target transition-region template, constrained by mean fluxes over
167--170 and 258--263\,nm. The 284.4--336\,nm interval uses NewEra with
constraints from 281.5--284 and 340--380\,nm. Smooth interpolation of the
logarithmic normalization between these bounding windows constrains the
broad spectral slope; it does not create measured features in the gaps.
Unadjusted-template and altered blue-anchor scenarios quantify sensitivity
to this choice. Native STIS, including Mg\,II, takes precedence inside its
coverage. Gaia supplies 336--1000\,nm and the anchored NewEra spectrum
extends to 5000\,nm.

\begin{table}[ht]
\centering\small
\caption{Adopted contiguous spectral components. Model-filled intervals are
explicitly distinguished from observations.}
\begin{tabular}{@{}ll@{}}
\toprule
Interval (nm)&Adopted source\\\midrule
0.5--10&Published mean-state EPIC 3T, APEC 3.0.9\\
10--92&Sanz-Forcada et al. (2025), target EMD\\
92--115&Target EMD plus higher-Lyman proxy\\
115--121&STIS E140M, 2019\\
121--122&Intrinsic \lya\ plus observed O\,V residual\\
122--170&STIS E140M, 2019\\
170--257.6&Anchored NewEra/transition-region gap estimate\\
257.6--284.4&STIS E230H, 2019\\
284.4--336&Anchored NewEra gap estimate\\
336--1000&Direct Gaia XP\\
1000--5000&Gaia-normalized NewEra\\\bottomrule
\end{tabular}
\end{table}

\subsection{Uncertainty arrays, covariance and released products}
Native HST errors already contain the pipeline's measurement uncertainty;
we did not add an arbitrary instrumental percentage. Linear resampling and
shared normalization produce correlations, which are preserved rather than
treated as independent per-pixel noise. The available covariance is stored
in sparse plus low-rank form,
\begin{equation}
 C=B+UKU^{\mathsf T}+\operatorname{Cov}(\Delta F_{\rm Ly\alpha}),
\end{equation}
where $B$ contains propagated native covariance, the low-rank terms include
the normalization cross-covariance and reported stellar-parameter
derivatives, and the last term embeds the shared reconstruction draws.
The metric $K$ can be indefinite because it represents cross terms; the
combined available covariance passes positive quadratic-form checks.
Unavailable correlations between the reported stellar-parameter marginals
are not fabricated.

The FITS records distinguish conditional statistical components, partial
gap-model components, the X-ray marginal-parameter envelope, and EUV
published-band endpoints. Unknown model uncertainties remain undefined,
not zero. Named physical alternatives are separate arrays and are not
interpreted as random draws. The delivered English note and provenance
record make these distinctions part of the product definition.

We provide both the signed assembled spectrum and a nonnegative irradiation
nominal. The latter groups adjacent bins where noisy measurements are
negative, preserving the integrated energy and propagating the linear
grouping operator. This propagation is conditional on the chosen grouping;
it is not a positivity posterior. Tests verify continuous increasing bins,
flux-unit conversion, FITS checksums, covariance diagonals, and conservation
of the principal band integrals. The integrated surface flux differs from
$\sigma T_{\rm eff}^{4}$ by about three percent, an additional scale check.

The portable public notebook provides spectral and linear uncertainty diagnostics; the final atlas panel follows Section~\ref{sec:display}.

\newpage

\subsection{Final spectrum with available uncertainties}
Figure~\ref{fig:sed:iot_Hor} shows the adopted full SED and its available uncertainty bounds.

\begin{center}
\includegraphics[width=\linewidth]{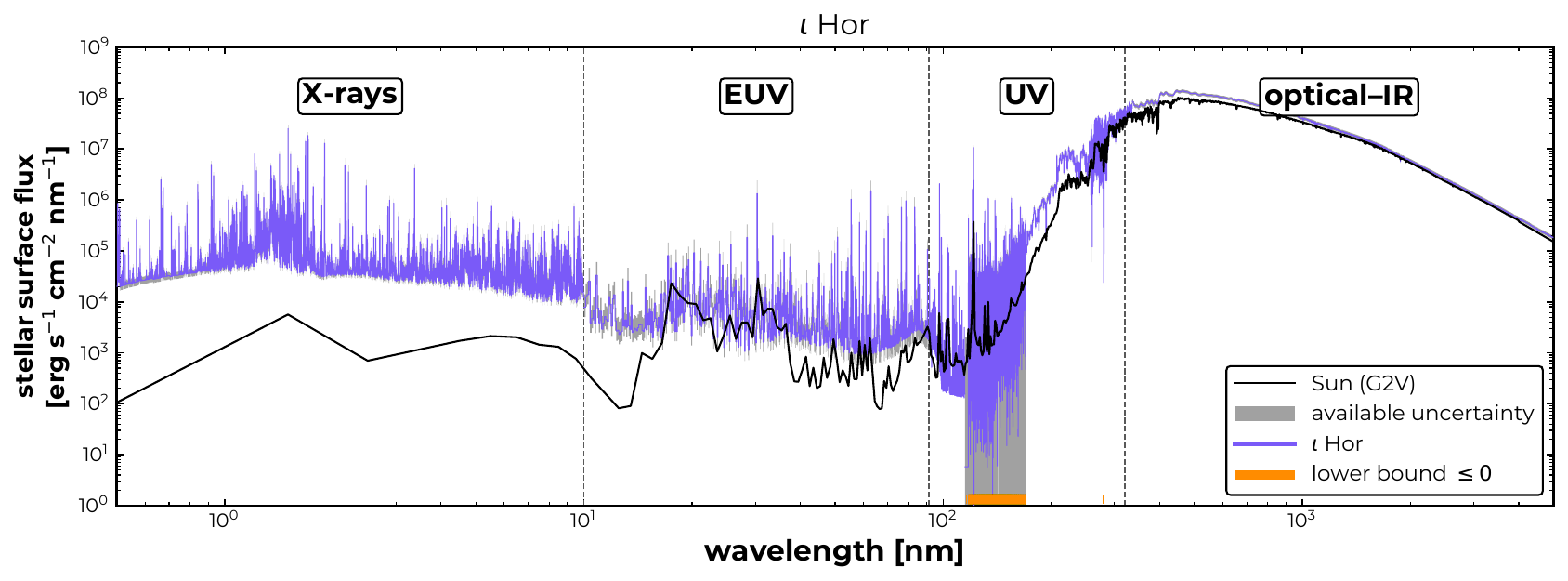}
\end{center}
\captionof{figure}{$\iota$ Horologii: final nominal surface-flux spectrum and available uncertainties. Target colours, shading, solar comparison and nonpositive-bound markers follow Section~\ref{sec:display}. The component origins and uncertainty scope are documented above.}\label{fig:sed:iot_Hor}

%% file: sections/tau_Boo.tex
\clearpage
\section{\texorpdfstring{$\tau$ Bootis A}{tau Bootis A}}\label{star:tau_Boo}
This section documents the component-A spectrum of $\tau$ Boo, starting
with its stellar parameters and spatially resolved Chandra reduction. It then
describes the EUV estimate, STIS visit selection, Ly$\alpha$ and airglow
treatment, and the optical/infrared assembly, together with the uncertainty
representation supplied for the final spectrum.

\subsection{Target parameters, component identity and observations}
The target is the F-type primary of the $\tau$ Boo binary. Gaia DR3 source
1244571953471006720 has parallax $64.04698\pm0.10932$ mas, giving
$d=15.61354\pm0.02665$ pc by direct inversion. We adopt
$T_{\rm eff}=6399\pm45$ K, $\log g=4.27\pm0.06$,
$[\mathrm{M/H}]=+0.26\pm0.03$, and $R=1.42\pm0.08\,R_\odot$
from \citet{Borsa.et.al.2015A&A...578A..64B}. The age attached to this parameter set is
$0.9\pm0.5$ Gyr; an older catalogue age is not silently combined with it.
The radius was originally inferred using a Hipparcos parallax of 64.03 mas.
The change in distance is only about 0.03\%, much smaller than the radius
uncertainty. The adopted Earth-to-surface factor is
\begin{equation}
 G=(d/R)^2=2.37839035\times10^{17}.
\end{equation}
For uncertainty propagation, angular radius and the new Gaia distance are
parameterized separately. The old Hipparcos fractional distance variance
$(0.19/64.03)^2$ is removed from the published fractional radius variance before
propagating the angular-radius marginal. This prevents counting the same distance
term twice. Unpublished covariances among the stellar parameters are not assumed
to have been measured.

The companion contamination test uses $T_{{\rm eff},B}=3580\pm90$ K,
$R_B=0.48\pm0.05\,R_\odot$, and $M_B=0.49\pm0.02\,M_\odot$
from \citet{Justesen.and.Albrecht.2019A&A...625A..59J}. These parameters define a physical comparison spectrum;
they do not establish what fraction of B entered every instrument aperture.

\begin{table}[ht]
\centering\small
\caption{Principal observations and their adopted roles.}
\renewcommand{\arraystretch}{1.15}
\begin{tabular}{>{\raggedright\arraybackslash}p{2.8cm}>{\raggedright\arraybackslash}p{4.3cm}>{\raggedright\arraybackslash}p{7.4cm}}
\toprule
Instrument & Identifiers / epoch & Use \\
\midrule
Chandra LETG/HRC-S & 17715, 20019, 20020; February--March 2017 & Joint A/B extraction, both arms, orders 1--8. \\
STIS G140M & o4ze01010, o4ze01020; 9 June 1999 & Adopted far-UV visit and \lya\ wings. \\
STIS G140M & o4ze02010, o4ze02020; 26 June 1999 & Different flux scale; comparison only. \\
STIS E230H & ob4522010; 14 April 2010 & Near-UV, Mg II and IUE normalization. \\
IUE SWP & SWP06183; 14 August 1979 & Large-aperture far-UV spectrum, native calibration. \\
IUE LWP & LWP14967; 4 February 1989 & Selected near-UV spectrum, constant STIS scale. \\
IUE alternatives & LWP14965 and LWR05348 & Overlap and calibration diagnostics. \\
Gaia DR3 XP & Primary source above & Optical spectrum and photospheric normalization. \\
\bottomrule
\end{tabular}
\end{table}

The relevant LOWLIB metadata contain no match for this target, so no substitute
LOWLIB star is presented as a direct observation of $\tau$ Boo. No direct FUSE or
EUVE spectrum is adopted from the target inventory. Every exported interval
identifies its observation, reconstruction or model origin.

\subsection{Binary-aware Chandra reduction and X-ray inference}
The ordinary extraction aperture contains both stars. We therefore fit two
Moffat components and a sky term to each zeroth-order image, and model the
cross-dispersion profiles of the dispersed events. The measured sky separation
is about 1.46--1.56 arcsec; the fitted companion zeroth-order fraction is
15--19\%. These quantities are not applied as a wavelength-independent
subtraction. The grating-coordinate scale is distinguished from angular sky
separation.

Disjoint primary and companion apertures span $[-1.9,0.75]$ and
$[0.75,3.5]$ arcsec in the adopted grating cross-dispersion coordinate.
Independent background strips span $[-12,-5]$ and $[5,12]$ arcsec.
The spatial fractions are evaluated as a function of wavelength relative to
the standard centered extraction already represented by the archived response.
Native extraction totals are reproduced; two positive-arm cases differ in the
placement of one event at a native channel boundary, without changing the total.
Fits with Moffat indices 1.5 and 2.2 bracket the nominal 1.8 spatial prescription.
These alternatives use the same photons and are method checks, not independent
PSF measurements.

HRC does not sort grating orders by photon energy. We therefore fold orders
1--8 through their archived ARF/RMF pairs, for both arms, as required by the
\href{https://cxc.harvard.edu/ciao/threads/spectra_letghrcs/}{CIAO LETG/HRC-S workflow}.
A first-order-only calculation is retained as a rejected diagnostic, not used
as a valid uncertainty realization. Twelve data sets (three observations,
two arms, two apertures) are fitted jointly with source and background Poisson
likelihoods, using W-statistic in Sherpa/XSPEC. The fitted range is
0.2--2.4 keV. The exported unabsorbed X-ray band is exactly 0.5--10 nm
(approximately 0.124--2.480 keV); model extrapolation outside the fitted range
is thus explicit.

The nominal primary model has two APEC temperatures and a common free metal
abundance. The companion has one temperature with abundance fixed to 0.4 of
the adopted solar mixture. APEC 3.1.3, XSPEC 12.14.0k, the Anders--Grevesse
solar mixture and a fixed absorbing column $\log N_{\rm H}=17.82$ are recorded.
The latter is a Mg-based estimate discussed by \citet{Wood.et.al.2018ApJ...862...66W}, not a precision
direct H I measurement. The best-fit primary temperatures are approximately
0.173 and 0.489 keV, and the global abundance is 0.142 of the adopted solar value.

\begin{table}[ht]
\centering\small
\caption{X-ray model comparisons. Fluxes are primary-only, unabsorbed,
0.5--10 nm, in units of $10^{-12}$ \fluxunit. These alternatives are not
assigned probabilities.}
\begin{tabular}{lrrr}
\toprule
Model & W & Degrees of freedom & Flux \\
\midrule
1T APEC & 6827.797 & 6823 & 3.2552 \\
2T APEC, adopted & 6818.975 & 6821 & 3.2731 \\
3T variable-abundance APEC & 6798.227 & 6813 & 3.2864 \\
2T, abundance fixed to 0.51 & 6855.514 & 6822 & 2.5217 \\
2T, Moffat index 1.5 & 6817.476 & 6821 & 3.2987 \\
2T, Moffat index 2.2 & 6820.320 & 6821 & 3.2494 \\
\bottomrule
\end{tabular}
\end{table}

The extra freedom of the 3T model is not adopted solely because it lowers W;
its integrated flux is nearly unchanged and some abundances are poorly
identified. In contrast, the continuum/abundance assumption changes the inferred
flux appreciably. This dependence remains a named physical alternative.

An attempted local Hessian approximation was rejected because the weak cool
component produced unstable curvature estimates. The adopted conditional
uncertainties instead use the full profiled Poisson likelihood, evaluated
through a temperature/abundance response basis. Against 16 exact XSPEC
calculations, the largest interpolation difference is 0.0052 in W; the
independent likelihood implementation agrees to approximately $10^{-11}$.
Two independently seeded ensembles, each with 32 walkers and 12000 retained
steps after burn-in, give integrated-flux medians agreeing within 0.1\%.
The chain lengths exceed 64 integrated autocorrelation times for every sampled
parameter in the slower ensemble. A linear-parameter-prior comparison gives
nearly the same integrated flux, while the weak cool-component parameters
remain prior-sensitive. The released 4096 correlated spectral realizations
come from the log-parameter-prior ensembles. Their approximate 16th--84th
flux range is $(3.09,3.41)\times10^{-12}$ \fluxunit.
The nominal curve remains the actual best-fit APEC model. These conditional
ensembles weight the profiled W likelihood by the stated parameter priors;
background nuisance intensities are profiled, not Bayesian-marginalized.

An exact-band literature check is essential. Our nominal model gives
$\log L_X=29.002$ in 0.1--2.4 keV, compared with 28.76 in \citet{Wood.et.al.2018ApJ...862...66W}:
the difference is 0.242 dex, not a band-definition artifact. Fixing the abundance
to 0.51 reduces the result to 28.883 but does not eliminate the difference.
\citet{Wood.et.al.2018ApJ...862...66W} reported a weak continuum and adopted an XMM-based abundance for
this target. Our fixed global abundance is a sensitivity benchmark, not a
reproduction of their full abundance pattern or atomic emissivity prescription.
The reductions share the 2017 photons and are not independent
validation. We retain the reproducible fit and its method dependence rather
than forcing its normalization to the literature value or claiming agreement.

\subsection{EUV shape, normalization and correlated uncertainty}
The nominal EUV shape comes from the $\tau$ Boo F93 column accompanying
\citet{Sanz-Forcada.et.al.2025A&A...693A.285S}, based on its target emission measure distribution and
transition-region constraints. The published model uses APEC 3.0.9.
The input flux is at 1 AU; the tabulated unit of 10 mW m$^{-2}$ nm$^{-1}$
equals 10 erg s$^{-1}$ cm$^{-2}$ nm$^{-1}$. Direct integration gives
$\log L_{10-92}=29.2319$ and $\log L_{10-50.4}=28.9641$, reproducing the
published table within 0.01 dex. The X integral of the same column is
0.052 dex below its tabulated X luminosity and is not silently renormalized
across every wavelength.

The XMM coronal constraints on that shape are unresolved A+B. We therefore
use the target-informed \emph{relative shape}, while obtaining the integrated
EUV normalization from the separately extracted primary X-ray flux through
equation (2) of \citet{Sanz-Forcada.et.al.2025A&A...693A.285S} (Eq.~\ref{eq:sf25}).
Here $L_X$ refers to 0.5--10 nm and $L_{\mathrm{EUV,H}}$ to 10--92 nm.
For the adopted primary, $\log L_X=28.9799$ and
$\log L_{\mathrm{EUV,H}}=29.4242$, corresponding to
$F_{\mathrm{EUV,H}}=9.1058\times10^{-12}$ \fluxunit\ at Earth.
This use of a target template is preferable to introducing an unmeasured
total unsigned magnetic flux. No $10^{24}$ Mx fallback is used.

Each X-ray realization is propagated through the relation. The published
0.406 dex residual scatter is drawn once per whole EUV spectrum, rather than
independently at each wavelength. The slope and intercept errors (0.041 and
0.05) are also propagated, assuming independence at the centered pivot because
their covariance is not supplied. The resulting approximate 16th--84th range
is $(3.52,24.43)\times10^{-12}$ \fluxunit. Its asymmetry follows from the
logarithmic predictive distribution; symmetry is not imposed in either linear
or logarithmic plotting space. The large covariance of the normalization is
retained. Quiet- and active-Sun FISM2 shapes \citep{Chamberlin.et.al.2020SpWea..1802588C}, rescaled to the
same integrated EUV total, provide distinct shape alternatives. They are not
additional observations of $\tau$ Boo. The unresolved contribution of B to
the published shape is not claimed to have a measured probability distribution.

Between 92 and 115 nm, the non-hydrogen contribution uses the same target
template and normalization realizations. Higher Lyman lines use the relation
in section 4.1 of \citet{Linsky.et.al.2014ApJ...780...61L} (Eq.~\ref{eq:lyseries}),
with relative line proportions from FISM2. The notation on the right denotes
the solar-radius surface-flux scaling evaluated at 1 AU. The same \lya\ draws
and physical alternatives propagate into these lines. This is a model-filled
interval, with additional shape/proxy discrepancy unquantified, not a recovered
FUSE observation.

\subsection{STIS visit selection and intrinsic Ly\texorpdfstring{$\alpha$}{alpha}}
Native STIS wavelength, flux, error and DQ arrays are used. Flux densities are
converted from per Angstrom to per nm, orders are merged with weights
independent of the noisy measured flux, and the linear rebinning covariance is
retained. DQ gaps are repaired only where necessary for continuity and remain
flagged as interpolations.

The second G140M visit is approximately a factor of two fainter across \lya,
Si III and N V, not only in the geocoronal core. The archived
\href{https://www.stsci.edu/hst/phase2-public/7508.pro}{HST programme 7508 Phase II}
shows acquisition and peak-up for the first visit, but reuse of the saved
offset without a new acquisition/peak-up for the second. This is evidence for
a possible throughput difference, not proof that intrinsic variability was
absent. We adopt 9 June as the nominal epoch and do not silently average the
two scales. The native exposure with a short engineering telemetry gap keeps
its calibrated exposure metadata and quality record.

The stellar \lya\ profile is a Voigt emission function multiplied by a bounded
Gaussian reversal factor. H I and D I absorption are applied before convolution
with the tabulated STIS G140M 1200-Angstrom LSF for the actual $52\times0.1$
arcsec aperture, including its wings and native-pixel integration. The adopted
D/H is $1.5\times10^{-5}$. The geocentric core within 50 km s$^{-1}$ is excluded;
35 and 70 km s$^{-1}$ masks are tested separately. The O V region at
1218.05--1218.85 Angstrom is excluded from fitting \lya\ and its measured
residual is preserved in the assembled intrinsic-line interval.

The interstellar velocity is constrained by the Mg II/Fe II component of
\citet{Malamut.et.al.2014ApJ...787...75M}, $-11.64\pm0.21$ km s$^{-1}$. The H I column and Doppler
width vary. The unequal Mg and Fe widths are not blindly converted into a
unique gas temperature. The nominal fit has $\log N_{\rm H}\simeq17.7975$,
$b_{\rm H}\simeq12.02$ km s$^{-1}$ and integrated intrinsic flux
\begin{equation}
 F_{\lya}=2.7121\times10^{-12}\ \fluxunit.
\end{equation}
Three hundred four-pixel residual-block refits give a conditional 16th--84th
range of $(2.560,3.200)\times10^{-12}$ \fluxunit. These samples condition on
the adopted profile family, bounds, instrumental treatment and first-visit
calibration; they are not a complete posterior for every possible stellar core.

The reversal depth reaches its allowed upper boundary. A non-reversed Voigt
profile gives $3.745\times10^{-12}$ \fluxunit\ with only 3.28 greater in the
penalized squared-residual objective. A depth-one profile gives
$3.017\times10^{-12}$ \fluxunit. Thus the saturated central morphology is
not uniquely identified. These alternatives, different masks, two Gaussian
emission components and different H I columns/widths are retained. Objectives
for different masks have different numbers of data points and must not be
compared as if they were a likelihood-ratio test on the same data. A column
of $\log N_{\rm H}=18.1$ can raise the flux to $5.372\times10^{-12}$
\fluxunit, but worsens the same-mask fit substantially. A proxy prediction
does not justify forcing that column.

An independent-epoch E230H Mg II model gives combined emission flux
$1.180\times10^{-11}$ \fluxunit, with a photospheric-baseline/line-wing
degeneracy. It predicts $F_{\lya}\simeq5.46\times10^{-12}$ \fluxunit\ through
the F5--G9 relation of \citet{Linsky.et.al.2013ApJ...766...69L}; that relation has a 31.2\% RMS scatter.
The IUE C IV feature is only about a $3.2\sigma$ detection and gives a still
higher prediction, while the relevant C II blend is not significantly detected
and does not isolate the calibration component. These proxies therefore do
not establish which hidden core is correct. They are reported as a discrepancy,
not described as independent confirmation. Mg II reversal provides useful
qualitative information but is not equated to a unique \lya\ reversal depth
\citep{Youngblood.et.al.2022ApJ...926..129Y}.

\subsection{Airglow, IUE selection and UV junctions}
The STIS FLT images permit local-sky tests with several cross-dispersion strips.
For the N I triplet the changes are small compared with the native statistical
errors; CalSTIS background subtraction has already removed much of the diffuse
signal. No second unsubstantiated N I correction is applied. \lya\ is treated
through explicit core exclusion and reconstruction. Time-tag diagnostics are
not labelled orbital-night spectra without a corresponding orbital selection.

We also retrieved the IUE SWP06183 SILO image. It shows strong extended
large-aperture \lya\ emission but no comparably convincing extended O I feature.
The continuum-subtracted 1304-Angstrom complex is only a $1.13\sigma$
measurement and the 1356-Angstrom feature is consistent with zero.
Native IUE fluxes and errors are therefore retained and flagged. An extreme
alternative assigns all positive measured line excess to airglow. This is a
decomposition sensitivity, not a fitted correction or confidence bound; a
successful stellar/sky separation at IUE resolution is not claimed.

SWP06183 large aperture is selected for its documented absolute calibration and
longer exposure. The old wavelength-dependent tilt is removed. A narrow STIS
continuum overlap cannot determine a broad SWP correction, and the blind-region
counts do not establish the same additive offset inferred from that small
overlap. In three blue-LWP overlap bands, SWP/LWP differences are below
$1\sigma$ once the very large native LWP endpoint errors are propagated.
This does not prove perfect calibration; it means those noisy bins do not
justify imposing a large wavelength slope on SWP. LWR05348 is retained as
a same-epoch diagnostic rather than forcing all IUE exposures onto one curve.

LWP14967 is normalized to E230H using exact common support in four 5-nm
bands, 258--278 nm, excluding Mg II. The fitted constant is
$1.08556\pm0.00537$ (conditional statistical uncertainty), with
$\chi^2=0.637$ for three degrees of freedom. LWP14965 gives
$\chi^2=12.43$ for the same test and is not selected. The scale is derived
from the same data subsequently used in the SED, so its covariance and
cross-covariance with both IUE and STIS are explicitly propagated. Weak
blue-edge fluctuations and any unresolved junction offset remain visible in
the diagnostics rather than being hidden by a smooth empirical bridge.

\subsection{Optical/IR and final uncertainty representation}
Gaia XP calibration propagates the full BP/RP coefficient covariance.
PHOENIX NewEra models \citep{Hauschildt.et.al.2025A&A...698A..47H} are integrated over the output bins
and interpolated in temperature, gravity and metallicity. The primary grid
brackets 6200--6600 K, $\log g=4.0$--4.5 and $[\mathrm{M/H}]=0$--0.5.
The cool-companion grid brackets 3400--3800 K and $\log g=4.5$--5.0.
All 24 files are verified against the publisher's MD5 list. The old 6250-K,
solar-metallicity fallback is not used.

The primary model is anchored to the measured Gaia integral over 500--700 nm.
The scale is 1.03385 times the prediction from the adopted radius and distance,
corresponding to an effective angular diameter of 0.8601 mas. Model/Gaia shape
comparisons, published parameter marginal errors and the shared anchor
covariance are retained. A full-inclusion companion scenario contributes only
0.261\% in the anchor band, increasing to about 1.68\% near 1 micron. We keep
the primary-source XP nominal and supply the full-B subtraction as a physical
alternative, rather than claiming a measured contamination probability.

The assembled ranges are listed in the machine-readable component table.
Chandra covers 0.5--10 nm; the EUV template 10--92 nm; a labelled completion
covers 92--115 nm. SWP and first-visit G140M provide the far UV, with intrinsic
\lya\ replacing 121--122 nm. LWP and E230H cover the near UV. NewEra bridges
320--336 nm, Gaia covers 336--1000 nm, and the anchored atmosphere supplies
1000--5000 nm. The narrower model bridge is not labelled an observation.

For the available conditional covariance we store
\begin{equation}
 C=B+UKU^{\mathsf T}+\operatorname{Cov}(D_X)
                    +\operatorname{Cov}(D_{\lya}).
\end{equation}
$B$ is the sparse propagated native covariance. $U$ and $K$ encode shared
normalizations, their native-data cross terms and the available stellar
parameter modes. $D_X$ contains correlated X/EUV/short-FUV realizations;
$D_{\lya}$ contains correlated \lya/higher-Lyman realizations. Each block
has explicit row indices. The cross-term metric $K$ need not be positive
semidefinite alone; the full available covariance is. Named method alternatives
are separate arrays and are not assigned a statistical variance without a
specified probability model. Unprovided atmosphere and EUV-shape discrepancies
are not replaced by arbitrary error percentages.

The signed assembled measurements are retained in \texttt{SED\_SIGNED}.
A second table, \texttt{SED}, is the single nonnegative nominal irradiation
spectrum: adjacent bins containing negative noisy estimates are grouped with
exact integrated-flux conservation. This does not clip away negative energy or
alter the original measurement table. The covariance is propagated conditional
on the fixed, data-selected grouping. Users should integrate bin flux densities
times bin widths, not trapezoids across centers of unequal bins.

The release therefore closes the reproducible assembly and available-error
propagation for this target. It does not turn the hidden \lya\ core, unresolved
EUV shape or abundance-dependent X-ray continuum into uniquely measured
quantities. The English document, component table, covariance sidecar,
provenance and executed viewer accompany the public FITS; native reductions,
tests and the executed master notebook remain in the working directory.

\subsection{Diagnostic figure}
\begin{center}
\includegraphics[width=\linewidth]{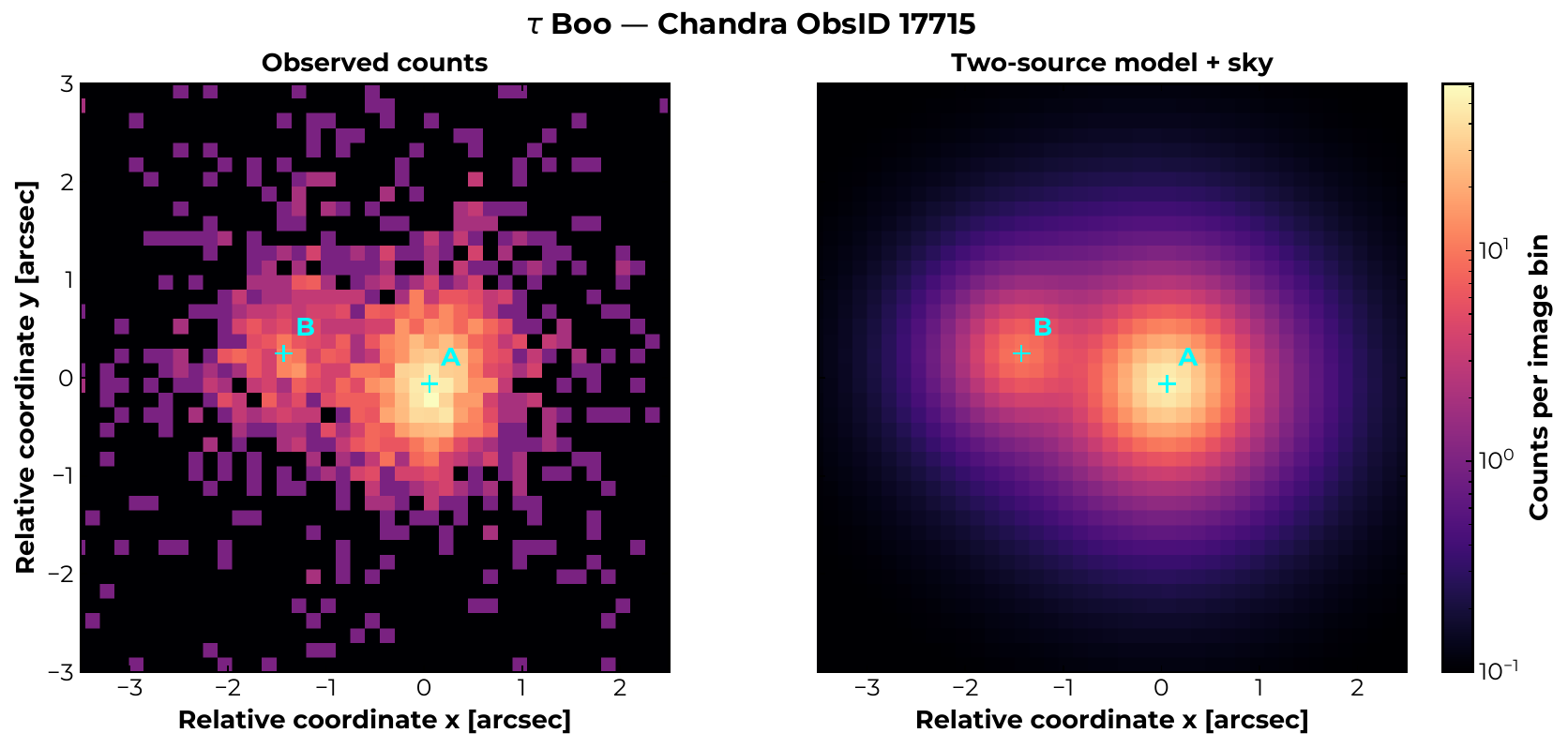}
\end{center}
\captionof{figure}{Chandra observation 17715 of $\tau$ Boo: the recorded zeroth-order counts (left) and fitted two-component Moffat plus sky model (right), on identical coordinate and colour scales. Markers identify the fitted A and B positions; the sky separation is 1.514 arcsec. The roughly 15.3 percent zeroth-order B fraction is not applied as a constant spectral subtraction. The dispersed spectra instead use wavelength-dependent spatial fractions and a joint response-folded model.}\label{fig:diagnostic:tau_Boo}

\newpage
\subsection{Final spectrum with available uncertainties}
Figure~\ref{fig:sed:tau_Boo} shows the adopted full SED and its available uncertainty bounds.

\begin{center}
\includegraphics[width=\linewidth]{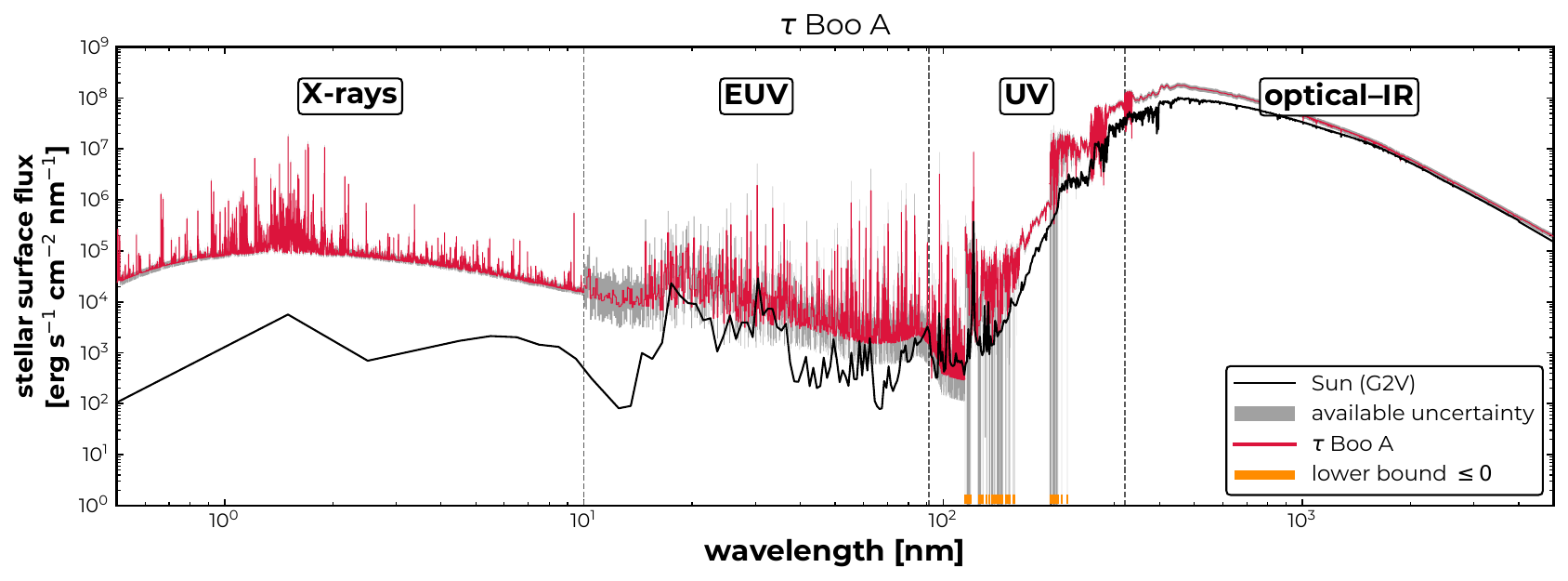}
\end{center}
\captionof{figure}{$\tau$ Bootis A: final nominal surface-flux spectrum and available uncertainties. Target colours, shading, solar comparison and nonpositive-bound markers follow Section~\ref{sec:display}. The component origins and uncertainty scope are documented above.}\label{fig:sed:tau_Boo}

%% file: sections/70_Oph_A.tex
\clearpage
\section{\texorpdfstring{70 Ophiuchi A}{70 Ophiuchi A}}\label{star:70_Oph_A}
This section separates the evidence for 70 Oph A from measurements of
the unresolved binary. It describes the resolved X-ray spectrum, intrinsic
Ly$\alpha$ reconstruction, primary-component EUV estimate, and ultraviolet
and photospheric junctions, and explains the uncertainty information
distributed with the final SED.

\subsection{Target identity, parameters, and flux conventions}
The target is 70 Oph A (HD 165341 A), the brighter component of the visual
binary. Component labels were checked in the native HST headers, programme
information, and the resolved Chandra image. Data assigned to B were not
coadded with A. Historical large-aperture IUE and EUVE observations can include
both components and were evaluated with that distinction preserved.

The adopted spectroscopic parameters are $T_{\rm eff}=5300\pm50$ K and
$[\mathrm{Fe/H}]=+0.04\pm0.05$, from the HARPS analysis of
\citet{Eggenberger.et.al.2008A&A...482..631E}. The limb-darkened angular diameter is
$\theta=1.515\pm0.005$ mas \citep{Boyajian.et.al.2012ApJ...757..112B}. The Gaia DR3 primary-source
parallax is $195.56735\pm0.19639$ mas, giving $d=5.11333$ pc and
$R=0.83289\,R_\odot$. The mass $0.89\pm0.02\,M_\odot$ from
\citet{Eggenberger.et.al.2008A&A...482..631E} implies $\log g=4.5462$ with this radius.
The formal parallax error is retained; Gaia's elevated RUWE is recorded in
the provenance, without inventing an additional astrometric error.

The original GL702A entry in \citet{Yee.et.al.2017ApJ...836...77Y} instead lists
$5407\pm52$ K, inherited from interferometric/bolometric calibration. We do
not attribute a secondary compilation's different temperature to that row.
The 2014 Boyajian bibliography erratum does not alter the measured diameter.
The older bolometric estimate for this binary is not treated as an
independently measured, deblended absolute-flux anchor.

Wavelength bins are in nm and flux densities in
erg\,s$^{-1}$\,cm$^{-2}$\,nm$^{-1}$. Conversion from Earth to surface uses
$G=(2/\theta_{\rm rad})^2=7.41455\times10^{16}$ directly; its relative error is
$2\sigma_\theta/\theta=0.00660$. Consequently distance and radius are not
counted as independent geometric errors. Their known contributions to the
derived gravity are propagated consistently. Irradiation at $a$ AU follows
$F(a)=F_{\oplus}(d\,206264.806/a)^2$.

\subsection{Resolved X-ray spectrum}
Chandra LETG/HRC-S observation 4482, obtained in July 2004, is re-extracted
with disjoint source regions and separate sky strips. A two-source spatial
profile identifies the primary and tests companion wings. The measured
on-sky separation is approximately $4.78$ arcsec; the detector dispersion
coordinates used for extraction are not confused with an on-sky separation.
The A aperture spans $[-2.4,2.4]$ in the recorded cross-dispersion coordinate.
The empirical source-throughput fraction relative to the calibrated standard
extraction is applied after response folding. Reproducing the standard PHA
counts from the events is an independent extraction check.

Both dispersion arms are fitted over 0.2--2.4 keV with source/background
Poisson statistics and response orders 1--8. Background intensities are
profiled using the W statistic. The software is CIAO 4.18, XSPEC 12.14.0k,
and APEC 3.1.3, with the archived matching CALDB 4.9.2 responses. The adopted
column is $N_{\rm H}=10^{18.06}$ cm$^{-2}$; abundances use the
Anders--Grevesse convention. The intrinsic exported X band is exactly
0.5--10 nm. Its low-energy portion below the fitted band is a model
extrapolation, not a separately detected soft-X component.

The adopted three-temperature VAPEC solution has $kT=0.1581$, 0.2725, and
0.5692 keV. C, N, O, Ne, Mg, Si, and Fe vary, with weakly constrained low-FIP
species tied to Fe. Its statistic is $W=2326.30$. A free-S alternative reaches
$W=2320.02$ but places S at zero; it remains a named sensitivity case rather
than being silently discarded. Two-temperature fits and aperture widths
1.6, 2.0, and 3.0 are also retained. The latter change integrated X flux by
approximately $-0.3\%$, $-0.5\%$, and $+3.6\%$. These are method alternatives,
not independent random errors or likelihood-ratio tests across different
extractions.

The nominal flux is $F_X=4.3622\times10^{-12}$ erg\,s$^{-1}$\,cm$^{-2}$.
It is the maximum-likelihood spectrum of the adopted family, not the mean or
median of sampled spectra. An element-resolved response basis reproduces
direct XSPEC calculations to $|\Delta W|<0.004$ in the validation cases.
The 48-walker conditional chain retains 30000 steps, exceeds 56 estimated
autocorrelation times in every parameter, and yields integrated-flux
16th/50th/84th percentiles of $(4.008,4.199,4.385)\times10^{-12}$.
Integrated-flux quantiles agree between the two chain halves to better than
0.1\%. These are conditional posterior summaries with the stated priors;
profiled background nuisances are not claimed to be fully marginalized.

The ROSAT-based primary luminosity tabulated by \citet{Wood.and.Linsky.2006ApJ...643..444W} implies
$6.007\times10^{-12}$ erg\,s$^{-1}$\,cm$^{-2}$ at their distance, whereas our
matched 0.1--2.4 keV model gives $4.578\times10^{-12}$. Their primary value
splits an unresolved earlier system measurement using component information
from Chandra. We therefore retain the approximately 24\% difference as a
band-matched, cross-epoch comparison and do not force it onto the resolved
Chandra calibration.

\subsection{Ly$\alpha$, interstellar absorption, and airglow}
The three native STIS E140M science extensions are fitted with their errors,
serious-DQ masks, pixel integration, and the tabulated 0.2$\times$0.2-arcsec
line-spread function. The geocoronal core is masked in each exposure's native
velocity frame. The intrinsic profile is a Voigt emission component with a
Gaussian-shaped self-reversal of independently adjustable width. Three
H I/D I absorbers use relative velocities constrained by Mg II, with free
hydrogen columns and a common Doppler width. A separate broad H I absorber
represents heliospheric/astrospheric absorption phenomenologically. The Mg II
fits contain structured residuals, so their saturated columns and reversal
depths are not imposed on hydrogen.

Different hidden-core profiles fit the observed wings: the flexible-depth
case gives $1.686\times10^{-11}$, while the unreversed case gives
$2.720\times10^{-11}$ erg\,s$^{-1}$\,cm$^{-2}$. The free reversal reaches its
allowed bound, so the minimum statistic alone does not identify a unique
intrinsic core. We adopt the reconstruction integral
$(2.140\pm0.428)\times10^{-11}$ erg\,s$^{-1}$\,cm$^{-2}$ from
\citet{Wood.et.al.2005ApJS..159..118W}, as tabulated by \citet{Wood.and.Linsky.2006ApJ...643..444W}. These authors analysed the
same STIS observations: this is adopted literature information, not an
independent validation or a second independent likelihood.

The local profile is optimized conditional on that integral. The nominal
fit gives $Q=1284.23$ for 1085 data/prior residuals. Conditional fits along a
positive, truncated-normal representation of the reported 20\% flux marginal
provide correlated spectral realizations. This construction propagates the
adopted integral uncertainty; it is not a new full joint posterior for the
hidden core. Fixed-integral shape variance is not quantified. Reversal,
background, masking, exposure-selection, and absorber alternatives remain
separate method scenarios. The profile normalization and velocity boundaries
are recorded, and no published curve was digitized by eye. The nearby O V
emission is retained as a native STIS residual above the absorbed Ly$\alpha$
prediction rather than overwritten by the reconstructed line.

The selected COS G130M visit of 24 June 2023 is entirely at orbital night.
Independent shallow-night and stricter-night subsets agree within their
native errors in N I 120.0 nm, O I 130.4 nm, and O I 135.6 nm. Spatial STIS
controls do not require a large diffuse-sky correction. We retain the signed
night measurements with their errors, without an unsupported extra airglow
subtraction. Night-only data do not establish zero airglow, and interstellar
attenuation in N I/O I is not represented as a successful intrinsic-line
reconstruction. The 2022 COS visit is excluded because its acquisition header
reports failure and the official programme identifies the 2023 repeat.

\subsection{Primary EUV estimate and the unresolved EUVE observation}
The adopted primary normalization uses equation (2) of
\citet{Sanz-Forcada.et.al.2025A&A...693A.285S}, not the older 2011 relation (Eq.~\ref{eq:sf25}).
The X and EUV bands are 0.5--10 and 10--92 nm. The coefficient marginal
errors, 0.041 and 0.05, and residual scatter 0.406 dex are propagated with the
primary X realizations. The scatter is a common EUV normalization factor,
not independent noise at every wavelength. The adopted integrated EUV flux
is $1.7191\times10^{-11}$ erg\,s$^{-1}$\,cm$^{-2}$, with sampled predictive
16th/50th/84th percentiles $(0.662,1.752,4.476)\times10^{-11}$.

The relative shape is the active-Sun FISM2 template \citep{Chamberlin.et.al.2020SpWea..1802588C},
formed from monthly samples in 2014--2015. The quiet-Sun 2019--2020 template
and the coronal-only APEC extrapolation are supplied at equal integrated EUV
flux as shape alternatives. The solar-like FIP pattern of A reported by
\citet{Wood.and.Linsky.2006ApJ...643..444W} motivates a solar comparison, but does not prove that a
solar template uniquely describes a K dwarf. This is a transferred template,
not a published target DEM or a measured primary EUV spectrum. No arbitrary
$10^{24}$ Mx magnetic-flux fallback is used. The missing short-FUV interval
up to 106.5 nm uses the template for non-H emission and the higher-Lyman
relation of \citet{Linsky.et.al.2014ApJ...780...61L}, sharing the Ly$\alpha$ normalization draws.

The July 1993 EUVE night image is re-extracted with fitted traces, sky strips,
Poisson errors, and a triangular higher-order correction through order 8.
H I/He I attenuation is corrected with the same sight-line column and the
cross sections of \citet{Verner.et.al.1996ApJ...465..487V}; column and aperture alternatives are
stored. The measured 10--40 nm system flux is
$(9.453\pm0.662)\times10^{-12}$ erg\,s$^{-1}$\,cm$^{-2}$ (statistical error).
Short-wavelength higher-order sensitivity changes its integral by less than
0.1\% for the reference binary-amplitude scenario. It remains an A+B product.
The primary proxy gives $1.233\times10^{-11}$ in the same band, approximately
30\% above the earlier system measurement, but with broad predictive scatter.
The two are neither contemporaneous nor measurements of the same component
content. We report the discrepancy; we do not infer a constant A/B EUV ratio
or claim independent confirmation of the primary shape.

\subsection{Ultraviolet joins and optical--infrared completion}
COS G130M and G160M agree in their common supported bands. The older STIS
E140M data fill the G160M detector gap. Native errors are propagated through
the actual order/exposure and wavelength-bin operators. Invalid samples are
flagged; interpolation never acquires the status of a new observation.

For the adjacent E230H settings, the 274--279 nm common interval, excluding
Mg II, gives a scale $1.09684\pm0.00384$ for the 2663 setting relative to the
higher-flux 2862 setting. Its covariance includes shared native-data cross
terms and excess band scatter. The unscaled choice is retained. E230M has no
direct overlap with the 2663 setting, so its native scale is retained.
The 251--253.5 nm gap is explicitly interpolated between measured band means,
with propagated endpoint covariance and a photospheric-shape alternative.
The COS/E230M overlap is noisy and does not justify forcing a slope or scale.

Gaia DR3 XP for the primary provides the absolute optical spectrum with the
full BP/RP coefficient covariance. PHOENIX-NewEra \citep{Hauschildt.et.al.2025A&A...698A..47H} is
interpolated at the adopted temperature, gravity, and metallicity, normalized
through the Gaia 500--700 nm integral, and used for missing near-UV and IR
coverage. The same Gaia data determine the anchor and optical segment, so
their cross-covariance is retained. Parameter derivatives propagate the
reported stellar errors. The unadjusted model overpredicts STIS at 290--298 nm
by approximately 39\%; IUE also favours lower near-UV flux. Therefore the
299.9--336 nm gap uses the model's relative shape with a linear transition
between scale factors measured from STIS at 290--298 nm and Gaia at
340--380 nm. A narrow 336--346 nm anchor is retained as a sensitivity case
to the choice of low-resolution comparison band. Both anchor covariances and parameter derivatives are propagated.
This is explicitly an empirical model bridge; the unadjusted atmosphere is a
separate method alternative. A full-companion-in-XP subtraction scenario uses
cool atmosphere shapes and the resolved Geneva brightness contrast; it is
a sensitivity bound, not a measured contamination fraction. IUE data remain
unresolved diagnostics. No absolute calibration is inferred from the
relative-flux CFLIB spectrum, and no NGSL substitution is made.

\subsection{Product semantics, uncertainty, and reproducibility}
The released FITS contains a nominal irradiation SED, a signed-measurement
SED, a component ledger, and integrated bands. Adjacent signed bins are
grouped only where necessary for non-negative nominal irradiation, conserving
each donor interval's energy. No negative measurements are clipped to zero.
The covariance of the grouped product is conditional on the selected
grouping. Integration uses the supplied bin widths, not spacing between
midpoints.

The covariance sidecar represents
\begin{equation}
C=B+UKU^{\mathsf T}+\mathrm{Cov}(D_{X,\rm EUV})+
\mathrm{Cov}(D_{\mathrm{Ly}\alpha}).
\end{equation}
Here $B$ contains native rebin covariance, $U,K$ encode common scales and
stellar-parameter terms, and the two realization blocks preserve X/EUV and
Ly$\alpha$/higher-Lyman correlations respectively. Shared-data cross terms
mean that $K$ alone need not be positive semidefinite. Named method offsets
are stored separately and are not assigned statistical probabilities.
Coverage flags explicitly distinguish available conditional errors from
unquantified model-shape/decomposition errors. No generic epoch or arbitrary
instrumental percentage is added.

\subsection{Final spectrum with available uncertainties}
Figure~\ref{fig:sed:70_Oph_A} shows the adopted full SED and its available uncertainty bounds.

\begin{center}
\includegraphics[width=\linewidth]{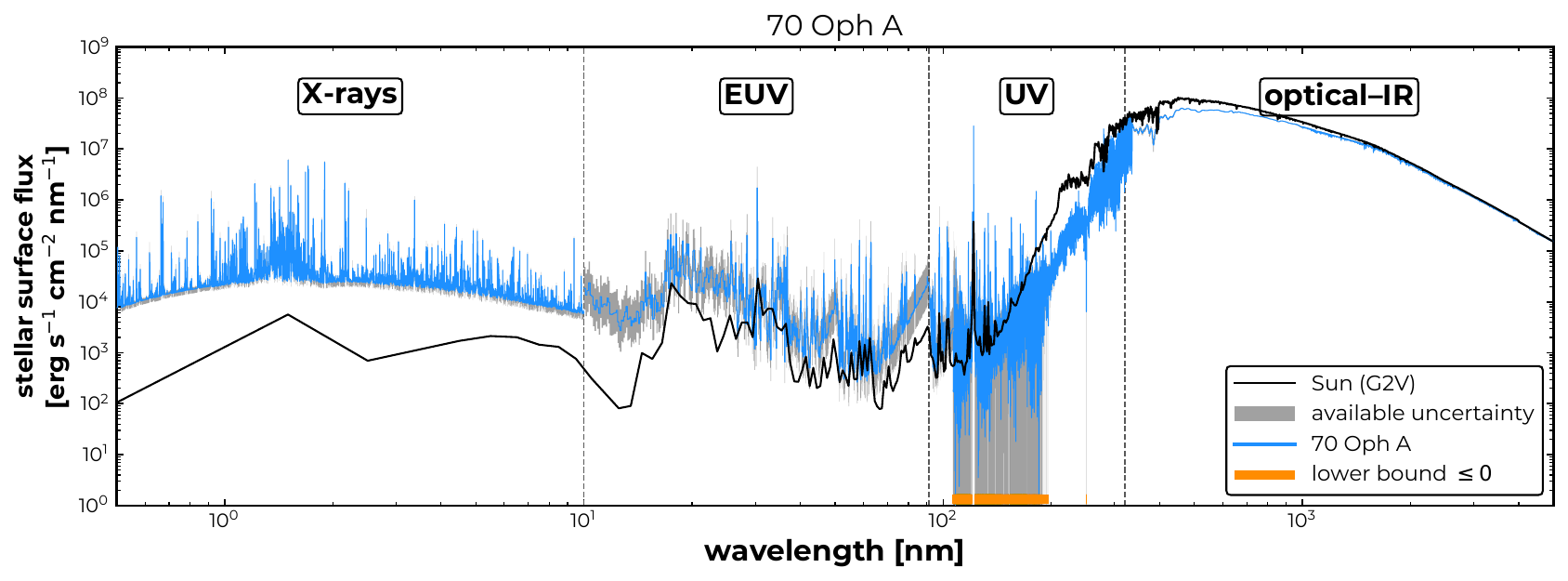}
\end{center}
\captionof{figure}{70 Ophiuchi A: final nominal surface-flux spectrum and available uncertainties. Target colours, shading, solar comparison and nonpositive-bound markers follow Section~\ref{sec:display}. The component origins and uncertainty scope are documented above.}\label{fig:sed:70_Oph_A}

%% file: sections/36_Oph_B.tex
\clearpage
\section{\texorpdfstring{36 Ophiuchi B}{36 Ophiuchi B}}\label{star:36_Oph_B}
This section describes the component-specific reconstruction of
36 Oph B, including the resolved Chandra extraction and the separation of
blended IUE spectra. It then explains the normalization and conditional
profile of Ly$\alpha$, the EUV and short-FUV estimates, and the
optical/infrared completion and delivered uncertainties.

\subsection{Identity, geometry, and adopted parameters}
The target is HD 155885, 36 Oph B, Gaia DR3 source
\texttt{4109030160308317312}, and TIC 1277142190. The nearby primary is
HD 155886. This distinction is essential: every retrieved STIS exposure is
of the primary, although both components occur in the IUE large aperture.
Native target names, acquisition coordinates, binary orientation, observation
notes, and Gaia identifiers were checked rather than inferring component
identity from a directory name. No matching LOWLIB spectrum was found in
the supplied library crossmatch.

We adopt the component-specific parameters compiled or inferred by
\citet{Harada.et.al.2024ApJS..272...30H}: $\varpi=168.1303\pm0.1081$ mas,
$d=5.94777\pm0.00382$ pc, $R=0.74\pm0.02\,R_\odot$,
$T_{\rm eff}=5098^{+48}_{-47}$ K, and an isochrone mass
$M=0.80^{+0.04}_{-0.05}\,M_\odot$. The atmosphere uses the Hypatia mean
$[\mathrm{Fe/H}]=-0.23$ and $\log g\simeq4.60$, derived from $GM/R^2$.
The radius is an SED-derived radius, not an interferometric measurement.
The alternative AMBRE solution identified by the same Gaia source is
$T_{\rm eff}=5092.61$ K, $\log g=4.77$, and
$[\mathrm{Fe/H}]=-0.33$ \citep{Contursi.et.al.2024A&A...690A..97C}.

Earth and stellar-surface flux densities are related by
\begin{equation}
 F_{\lambda,\mathrm{surf}}=G F_{\lambda,\oplus},\qquad
 G=(d/R)^2=1.27087\times10^{17}.
\end{equation}
The radius and distance marginals imply a 5.41\% first-order uncertainty
in this conversion. Temperature, mass, and radius marginals are propagated
conditionally as independent where their joint distribution is unavailable;
the known dependence of atmosphere $\log g$ on mass and radius is retained.
The radius contribution to atmosphere shape and to surface conversion is
therefore a shared mode, not two independent errors.

\subsection{Resolved Chandra extraction and thermal inference}
Chandra observation 4483 (2004 June 1; HRC-S/LETG) provides an exposure of
77.370 ks. The original custom event file included narrow stellar traces
but excluded the off-trace regions needed for background extraction.
The corresponding B background PHA consequently contained zero counts.
Using that PHA as the background caused detector counts to be attributed
to the thermal source. We re-extract source and sky from the full-field
archival event list, using the custom extraction only to verify the B
coordinate transformation.

Exact photon matching checks the transformation from the archived grating
coordinates to B: the wavelength-mapping residual has an RMS of
$1.39\times10^{-4}$ \AA, well below the native 0.0125 \AA\ channel.
The nominal source strip has a cross-dispersion half-width of $2\arcsec$;
two independent sky strips span $[-35,-15]\arcsec$ and
$[15,35]\arcsec$ relative to B. The source-to-background area ratio is 0.1.
Both dispersed arms are retained. Spatial profiles containing two stellar
components and a background estimate the aperture fraction and residual
primary leakage. Alternative half-widths of $1.6\arcsec$ and $2.4\arcsec$
test the extraction dependence. The adopted fit is conditional on the
empirical spatial correction; its alternatives are not folded into a
fictitious Gaussian calibration error.

Responses are generated at the B position using its own aspect information
and include grating orders 1--8 in each arm, as required for the limited
order discrimination of HRC-S. The original, unfiltered pulse-height
selection is retained consistently for source and sky. A later pulse-height
filter is not applied to older gain-calibrated events without the associated
calibration change. The source and background counts enter the Poisson
W statistic; the background is not subtracted before fitting. Four native
channels are grouped for the nominal fit, with an eight-channel test.

The nominal model is three APEC components with a common abundance,
APEC 3.1.3 in XSPEC 12.14.0k, and the Anders--Grevesse solar abundance
reference. Absorption uses \texttt{tbabs} with fixed
$\log N_{\rm H}=17.85$, motivated by the adjacent sightline studied by
\citet{Wood.et.al.2000ApJ...542..411W}. The fit covers 0.15--2.4 keV. The intrinsic spectrum is
integrated over the distinct, exact 0.5--10 nm band required by the adopted
X/EUV relation; wavelengths outside the fitted band are thermal-model
extrapolations.

\begin{table}[ht]
\centering
\caption{Nominal X-ray fit. The emission-measure normalizations use the
standard XSPEC APEC convention. Temperatures are listed in ascending order.}
\begin{tabular}{ll}
\toprule
Quantity & Adopted value\\
\midrule
$kT_1,kT_2,kT_3$ & 0.1764, 0.3409, 0.8129 keV\\
APEC normalizations & $(1.5059,0.9821,0.5197)\times10^{-3}$\\
Common abundance & 0.1425 times the adopted solar reference\\
$W$/degrees of freedom & 3128.21/3095\\
$F_X(0.5$--$10\,\mathrm{nm})$ & $2.1363\times10^{-12}\ \cgs$\\
$F_X(0.1$--$2.4\,\mathrm{keV})$ & $2.2674\times10^{-12}\ \cgs$\\
\bottomrule
\end{tabular}
\end{table}

The two-temperature fits with common and variable abundances, followed by
the three-temperature fit with variable abundances, give, respectively,
$W=3137.50$, 3129.75, and 3125.87 on the same data. The three-temperature
common-abundance model has the lowest AICc among these trials, although
its advantage over the variable-abundance alternatives is small. We retain
those alternatives as distinct spectral scenarios. AICc values from
different extraction widths, energy selections, or groupings are not
compared as if they described the same dataset.

The nominal uses the complete observation. Excluding the modest feature
near 31 ks discussed by \citet{Wood.and.Linsky.2006ApJ...643..444W} lowers the exact-band flux
to $1.963\times10^{-12}\ \cgs$; an initial-8-ks exclusion gives
$2.211\times10^{-12}\ \cgs$. Aperture changes give approximately
$2.095$--$2.165\times10^{-12}\ \cgs$. These are sensitivity tests, not
unmotivated temporal corrections. The earlier ROSAT-based B allocation in
\citet{Wood.and.Linsky.2006ApJ...643..444W}, $\log L_X=27.96$ at 5.99 pc, corresponds to a
0.1--2.4 keV flux only 6.7\% below our result. ROSAT did not independently
resolve the components: the comparison is useful but is not an independent
absolute B calibration, and no forced renormalization is applied.

Conditional parameter draws use the profiled Poisson likelihood and
declared priors: logarithmic temperature and normalization measures,
a linear abundance measure, ordered temperatures, and the fit bounds.
A response-folded APEC interpolation is checked against 13 direct XSPEC
evaluations; the largest statistic discrepancy is $\Delta W=0.011$.
Two independent 30,000-step ensembles agree in their predicted spectral
quantiles: the 95th percentile of their discrepancy is 3.5\% of the
pointwise 16--84\% interval width. Five integrated X-ray bands each span
more than 50 estimated autocorrelation times. Individual thermal labels
mix more slowly because of temperature/emission-measure degeneracies;
the notebook retains those diagnostics rather than claiming convergence
of every parameter. The full-band conditional 16--84\% interval is
approximately $(1.99,2.22)\times10^{-12}\ \cgs$. The released nominal is
the best fit, not the ensemble mean or median.

\subsection{IUE binary separation and flagged ultraviolet regions}
The IUE headers and observer notes show that SWP 39571, 16977, and 48634
contain both components. The observer's identification of the two traces,
the spacecraft roll, and the documented IUE aperture geometry determine
their assignment; the May 1982 orientation reverses the trace order
relative to the September observations. LWR 08882 is likewise a binary
exposure. LWR 13232 deliberately excludes the companion and is not used
as a B spectrum. The IUE Observer Guide and NEWSIPS documentation are
linked in the notebook and provenance.

We fit two pixel-integrated spatial Gaussians and a local background in
the two-dimensional, resampled SILO images. Trace centroid, separation,
and width are constrained across wavelength bands. At each wavelength,
the two amplitudes and the background are solved jointly, retaining
signed values. The native FLUX/NET vector supplies the absolute
flux-number conversion. Pixel-noise normalization is conditional on the
NEWSIPS SIGMA of the combined extraction, and the covariance of the
shared spatial geometry is propagated across wavelengths. This is a
custom decomposition using calibrated native products, not a second
independent observation or a pipeline-provided B error vector.

Gaussian extractions recover approximately 93--96\% of the native
combined continuum; Moffat-profile alternatives recover approximately
96--100\%. The differences, quadratic-background alternatives, and
individual observed epochs are retained as named method scenarios.
They are not silently removed by an absolute rescaling to HST A.
The three SWP exposures are coadded through explicit wavelength-overlap
operators, retaining induced covariance. Native one-dimensional quality
flags and flagged central spatial pixels are rejected.

The LWR blue edge is a non-detection or weak detection after component
separation: mean fluxes in the 198--212 nm sub-bands have signal-to-noise
ratios below unity. Their signed values and errors are preserved. The
nonnegative irradiation product groups adjacent bins with an
energy-conserving operator, including the short SWP/LWR bridge when
necessary. Thus it does not invent fine spectral information from this
low-signal region. The operators identify the constituents of mixed
observation/bridge bins.

Geocoronal intervals around N I, \lya, and O I are excluded from the IUE
coadd. The B Mg II cores in LWR 08882 also contain flagged spatial and
native samples and are not treated as a measured integrated B Mg II flux.
To provide a continuous irradiation SED, these regions use a clearly
identified STIS A template normalized by the measured B C II flux.
The narrow foreground/airglow blends in the A template near
$-28.4\pm12$ km s$^{-1}$ are removed by local interpolation before
transfer. Native template errors and the numerator/denominator covariance
of the C II normalization are propagated. Using the observed C IV
component ratio instead supplies a line-ratio sensitivity test. This
procedure estimates B's missing lines; it neither recovers a direct B
observation nor establishes a full intrinsic metal-line reconstruction.

\subsection{Intrinsic \texorpdfstring{\lya}{Lyalpha}: B normalization and conditional shape}
After local continuum subtraction, the deblended B line fluxes are
\begin{align}
 F_{\mathrm{C\,IV}}&=(1.369\pm0.107)\times10^{-13}\ \cgs,\\
 F_{\mathrm{C\,II}}&=(1.295\pm0.118)\times10^{-13}\ \cgs,
\end{align}
where the quoted errors contain the available conditional extraction
covariance. Both proxy windows have complete valid support in the coadd.
We apply the K0--K5 relations of \citet{Linsky.et.al.2013ApJ...766...69L}, whose fluxes are
defined at 1 AU:
\begin{equation}
 \log_{10}\!\left(\frac{F_{\lya}}{F_{\rm line}}\right)
 =A+B\log_{10}F_{\rm line}.
\end{equation}
For C IV, $(A,B)=(1.508,-0.575)$ and the reported RMS dispersion is 18\%.
The measured proxy flux at 1 AU is 0.2060 in cgs units, inside the
published calibration range. It gives the adopted
$F_{\lya,\oplus}=1.0936\times10^{-11}\ \cgs$.
C II independently as a diagnostic gives
$1.1648\times10^{-11}\ \cgs$, only 6.5\% higher, with a larger published
dispersion. The two empirical relations are not combined as independent
likelihoods. Bad Mg II cores are not used to select a higher normalization.

The predictive C IV relation is represented by a positive common
lognormal residual with $\sigma_{\ln F}=\sqrt{\ln(1+0.18^2)}$ and median
at the regression prediction. This is an explicit distributional
approximation to the published RMS, not a published posterior. Together
with the line measurement, it gives a conditional/predictive 16--84\%
interval of approximately $(0.915,1.314)\times10^{-11}\ \cgs$.
The covariance between the C IV measurement and the derived \lya\ is
preserved; treating them as independent would misstate integrated FUV errors.

There is no directly observed high-resolution B \lya\ profile in the
retrieved data. We therefore fit a conditional shape to A's STIS E140H
exposure, using a Voigt emission profile with allowed self-reversal,
interstellar H I/D I absorption, separate phenomenological heliospheric
and astrospheric absorbers, and the tabulated E140H 0.2-by-0.2 arcsec line
spread function. The A integral $1.42\times10^{-11}\ \cgs$ from
\citet{Wood.et.al.2005ApJS..159..118W} constrains only the A shape fit. The sightline constraints
and stellar velocity follow \citet{Wood.et.al.2000ApJ...542..411W}; the geocoronal core is masked.
The normalized shape is then assigned B's C IV-derived integral.

The best conditional A fit reaches zero self-reversal. Alternative column
densities and a strongly reversed profile remain explicit sensitivity
tests, with their fit statistics visible in the notebook. Their existence
does not imply that a strongly reversed profile fits A equally well.
More fundamentally, the A profile is a template assumption for B, not
an independent B reconstruction or a detection of its hidden core.

\subsection{EUV and short-FUV completion}
We use the current relation of \citet{Sanz-Forcada.et.al.2025A&A...693A.285S}, rather than the older
X/EUV calibration (Eq.~\ref{eq:sf25}),
with $X=0.5$--10 nm and EUV=10--92 nm. Its 0.406 dex residual is a
single predictive normalization uncertainty shared across wavelengths.
The X parameter draws propagate into EUV with the same realization
indices; coefficient marginals are treated independently at the
published pivot because their covariance is unavailable. The nominal
integral is $9.0626\times10^{-12}\ \cgs$, with a broad predictive
16--84\% interval of approximately $(3.50,22.0)\times10^{-12}\ \cgs$.

The nominal shape is an active-Sun FISM2 template \citep{Chamberlin.et.al.2020SpWea..1802588C}.
No suitable resolved B EUV spectrum was established in the archive search;
an EUVE all-sky source identifier is not a resolved stellar spectrum.
Quiet-Sun and coronal-only APEC shapes, normalized to the same EUV
integral, test the shape assumption. The coronal-only extrapolation is
not promoted to a transition-region measurement.

\citet{Ruedi.et.al.1997A&A...318..429R} report a tentative intensity-field signal for B:
an all-lines mean unsigned flux density of $60\pm45$ G, with a
different low-excitation result of $135\pm30$ G. They regard the B
detection as probable, not unambiguous. We therefore evaluate
\citet{Namekata.et.al.2023ApJ...945..147N} shapes at 15, 60, 105, and 135 G using the visible
hemisphere convention $\Phi=2\pi R^2\langle|B|\rangle$.
These measured-field-motivated alternatives replace an undocumented
fixed-$\Phi$ assumption as sensitivity evidence. They are not assigned
probabilities or described as a secure magnetic measurement.
The short-FUV completion below the IUE limit uses the solar template,
with identified higher-Lyman windows scaled from \lya\ according to
\citet{Linsky.et.al.2014ApJ...780...61L}; its shared \lya\ and X/EUV dependencies are retained.

\subsection{Optical/infrared junctions and delivered uncertainty}
The resolved Gaia XP coefficients for B are calibrated with GaiaXPy.
The coefficient covariance is propagated through the calibration
operator, including the BP/RP blending weights. Reproduced fluxes and
errors agree with the archived sampled spectrum to better than 0.05\%.
An interpolated PHOENIX-NewEra spectrum \citep{Hauschildt.et.al.2025A&A...698A..47H} is
anchored to the Gaia 500--700 nm band integral and used beyond 1000 nm.
The shared Gaia normalization introduces optical/IR cross covariance.
The Gaia-to-model ratio over 980--1000 nm is 0.9954. Temperature and
derived-gravity derivatives propagate the available stellar marginals;
the AMBRE atmosphere is retained as a shape scenario. The 330--336 nm
transition uses a NewEra-shaped bridge between measured IUE and Gaia
band anchors, with an unadjusted-atmosphere alternative. The complete
IUE spectrum is not forcibly normalized to the photosphere.

The released covariance is represented as
\begin{equation}
 C=B+UKU^{\mathsf T}+\operatorname{Cov}(D_{X/{\rm EUV}})
                    +\operatorname{Cov}(D_{\lya}).
\end{equation}
Here $B$ contains native statistical covariance and the shared C IV,
C II, and Gaia normalization Jacobians; $U,K$ encode stellar-parameter
and geometry modes. The two realization blocks describe correlated
X/EUV inference and the shared \lya\ predictive residual, respectively.
Named method offsets are provided separately. Their extrema are
sensitivity ranges, not confidence intervals. Unknown template mismatch
or model discrepancy is not assigned an arbitrary Gaussian error.

The public package contains the nominal nonnegative irradiation SED,
the signed pre-grouping estimates, bin edges, component identities,
available uncertainty endpoints, covariance and grouping operators,
band integrals, provenance, this target section, and an executed plotting
notebook. Numerical checks verify finite arrays, ordered endpoints,
covariance diagonals and quadratic forms, continuous wavelength coverage,
and band-integral conservation. The nominal irradiation spectrum is
nonnegative without clipping negative measured bins. On logarithmic
plots, a nonpositive lower endpoint is shaded down to the graph floor
and marked as continuing below it; stored values remain unchanged.
Figures are displayed inline by default, with optional PDF saving.

\subsection{Diagnostic figure}
\begin{center}
\includegraphics[width=\linewidth]{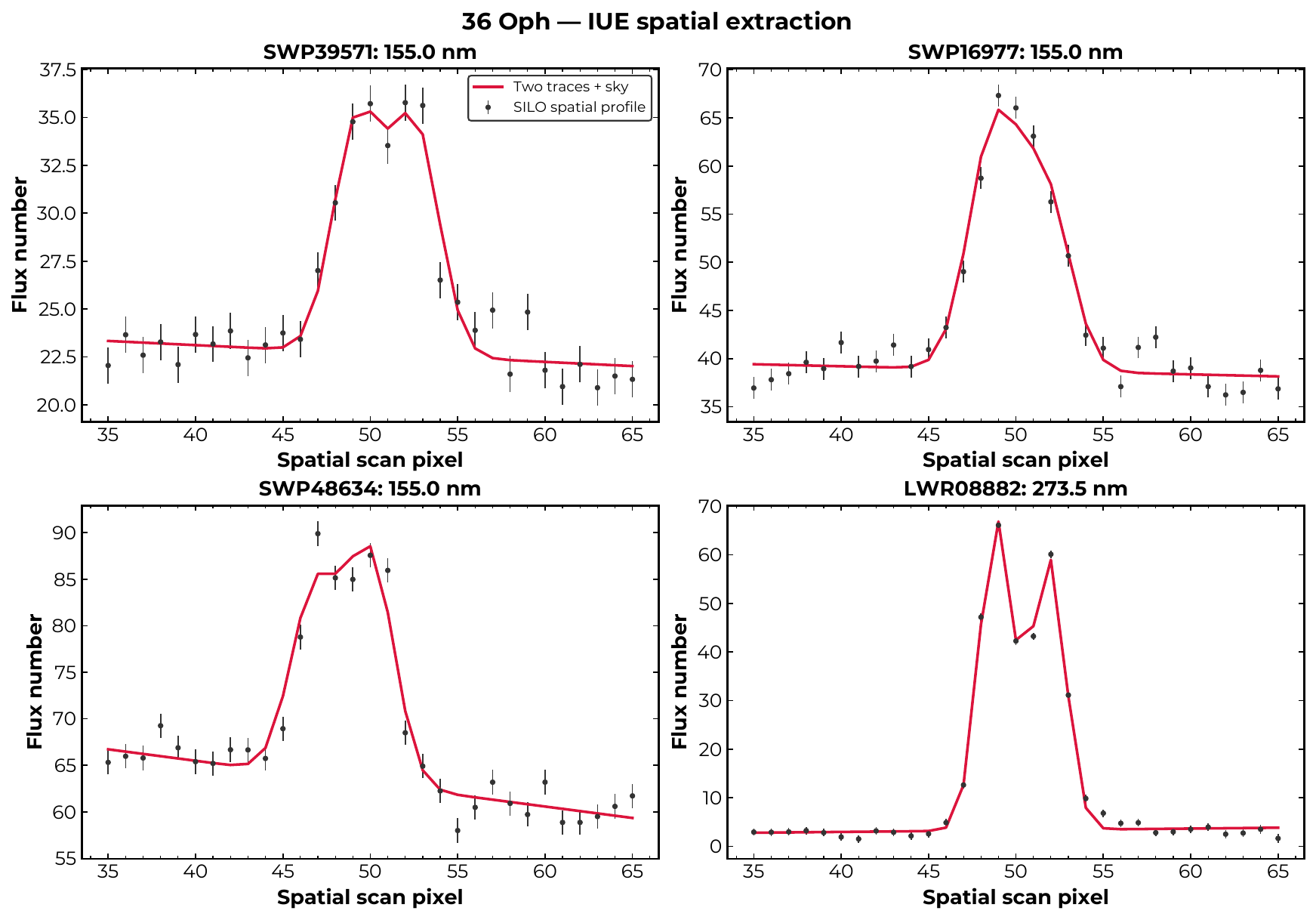}
\end{center}
\captionof{figure}{Examples of the two spatial traces in the IUE SILO images of 36 Oph. Points and errors are the archived conditional spatial-profile inputs; the red curve is the saved two-trace plus background fit. The trace assignment uses observation metadata and orientation, not brightness alone. The spatial order reverses between some visits. The decomposition supplies B estimates with shared spatial-geometry uncertainty; it does not turn HST observations of A into direct B observations.}\label{fig:diagnostic:36_Oph_B}

\subsection{Final spectrum with available uncertainties}
Figure~\ref{fig:sed:36_Oph_B} shows the adopted full SED and its available uncertainty bounds.

\begin{center}
\includegraphics[width=\linewidth]{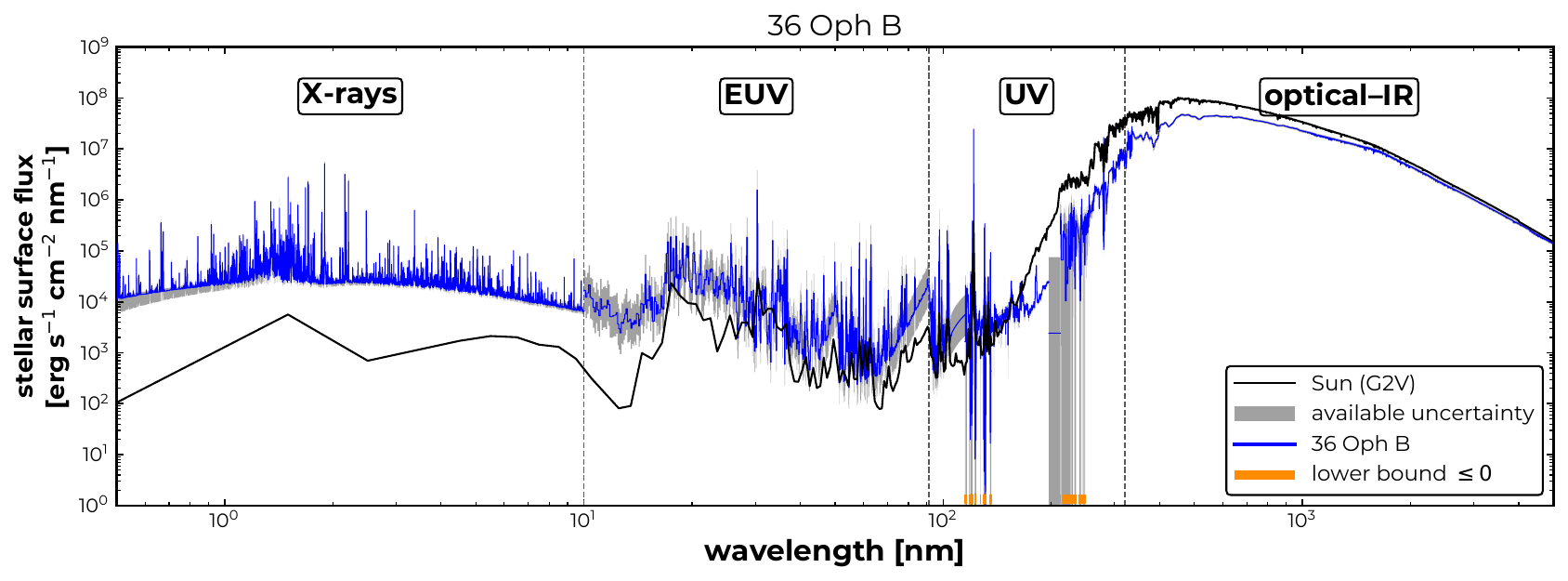}
\end{center}
\captionof{figure}{36 Ophiuchi B: final nominal surface-flux spectrum and available uncertainties. Target colours, shading, solar comparison and nonpositive-bound markers follow Section~\ref{sec:display}. The component origins and uncertainty scope are documented above.}\label{fig:sed:36_Oph_B}

%% file: sections/lambda_Ser.tex
\clearpage
\section{\texorpdfstring{$\lambda$ Serpentis}{lambda Serpentis}}\label{star:lambda_Ser}
This section documents the stellar parameters and available
observations for $\lambda$ Ser, with particular attention to the distinction
between a measured X-ray normalization and an assumed spectral shape.
The subsequent steps cover COS quality and airglow, the Ly$\alpha$ estimate,
the optical/infrared assembly, and the released uncertainty representation.

\subsection{Parameters and observational material}
We adopt $T_{\rm eff}=5901\pm78$\,K, $\log g=4.22\pm0.08$,
$[{\rm M/H}]=+0.04\pm0.07$, and the seismic radius
$R_\star=1.363\pm0.031\,R_\odot$ from \citet{Metcalfe.et.al.2023AJ....166..167M}.
Gaia DR3 source 4430238051199001216 has parallax
$83.9214\pm0.1501$\,mas, giving $d=11.916$\,pc. Its archive flags indicate
that neither sampled nor continuous XP spectra are available. The adopted
radial velocity is $-66.395$\,km\,s$^{-1}$. Ultraviolet wavelengths are
mapped from heliocentric vacuum coordinates to the stellar rest frame.
CFLIB optical air wavelengths, already corrected for radial velocity,
are converted to vacuum using \citet{Ciddor.1996ApOpt..35.1566C}; bin-width Jacobians
preserve energy during coordinate transformations.

The native COS/G140L observation comprises four time-tag exposures in
program 15991, totaling approximately 1976\,s. FITS timing gives
2020 March 2; the September date in the prose of \citet{Metcalfe.et.al.2023AJ....166..167M}
is not propagated into the input provenance. The available IUE files
include three low-dispersion SWP exposures, LWP08049 and two
high-dispersion LWR exposures. Different archive coadds of the same COS
exposures are not counted as independent measurements. No usable direct
EUV spectrum or LOWLIB spectrum of this target was identified in the
inspected inputs and archive inventory.

\subsection{X-rays and EUV: measurement and conditional shape}
\citet{Metcalfe.et.al.2023AJ....166..167M} report
$F_{\rm X}(0.1$--$2.4\,{\rm keV})=(4.0\pm0.4)\times10^{-13}$
erg\,s$^{-1}$\,cm$^{-2}$ from Chandra/HRC-I and ROSAT count-rate
conversions using coronal models. HRC-I imaging does not supply useful
energy resolution for a new thermal spectral fit. We preserve the
published Earth-flux anchor and use a quiet-Sun FISM2 shape
\citep{Chamberlin.et.al.2020SpWea..1802588C}, averaged from 24 monthly samples in 2019--2020,
to distribute that energy with wavelength. The corresponding active-Sun
average from 2014--2015 is a sensitivity case. These solar spectra are
models, not EUV or X-ray observations of $\lambda$ Ser.

The normalization interval is converted exactly to
$0.51660$--$12.39842$\,nm from the energy--wavelength relation.
The nominal flux on the different 0.5--10\,nm interval is
$3.3084\times10^{-13}$\,erg\,s$^{-1}$\,cm$^{-2}$; the active-Sun
conversion gives $3.6020\times10^{-13}$, an 8.9\% difference.
Additional 1.5, 2, 3 and 5\,MK APEC 3.1.3 spectra, using solar abundances,
are normalized on the same published band. These are illustrative
physical-shape alternatives, not fitted temperatures or confidence limits.

For the 10--92\,nm EUV integral we replace the older relation with
\citet{Sanz-Forcada.et.al.2025A&A...693A.285S}, their equation (2) (Eq.~\ref{eq:sf25}).
Here $L_{\rm X}$ explicitly refers to 0.5--10\,nm. The nominal EUV
Earth flux is $1.5281\times10^{-12}$\,erg\,s$^{-1}$\,cm$^{-2}$.
Monte Carlo propagation includes the published X-flux error,
coefficient marginals and a single common 0.406\,dex predictive residual
across the EUV band. Distance is a shared linear mode.
The coefficient covariance is unavailable, so its marginals are treated
as independent at the published pivot. No wavelength-dependent taper is
applied to the uncertainty. A polarimetric or low-order ZDI field is not
substituted for the total unsigned magnetic flux required by the
Namekata prescription; no arbitrary magnetic-flux fallback is inserted.

\subsection{COS quality, airglow and intrinsic Ly$\alpha$}
We reextract the original CORRTAG data with CalCOS 3.6.2 and the
exposure-specific calibration references. The TWOZONE profile reference
contains byte-preserved rows for the relevant configuration. Four
selections are retained: all events, orbital night ($h_\odot<0^\circ$),
strict night ($h_\odot<-10^\circ$), and day. Serious native DQ flags
and invalid error entries are rejected. Flux-conserving overlap operators
propagate the native errors and the resampling covariance. For the linear
covariance, the native upper \texttt{ERROR} approximation is used as a
conservative sigma in the low-count regime.

The O\,I triplet initially appears weaker in the night-only spectrum,
but night data lack valid pixels at 130.56--130.88\,nm. Treating these
missing points as zero would falsely suggest airglow removal.
On identical supported intervals, the day-minus-night difference is
$-0.69\sigma$ for the 130.4\,nm complex and $-0.23\sigma$ for
O\,I 135.6\,nm. We therefore retain the full-exposure spectrum and do
not subtract a statistically unsupported component. The third triplet
core has valid daytime coverage but cannot be checked independently
with these night data. This limitation is retained in the provenance.

Ly$\alpha$ and N\,I lie in the COS detector gap. The long IUE SWP
observations have flagged, airglow-contaminated Ly$\alpha$ cores and
an additive continuum excess relative to COS; they are not rescaled
to manufacture an intrinsic reconstruction. The continuum-subtracted
C\,IV doublet flux is
$(5.938\pm0.235)\times10^{-14}$\,erg\,s$^{-1}$\,cm$^{-2}$,
consistent with the $5.84\times10^{-14}$ value in the published
analysis of the same observation. This checks the reduction, not an
independent epoch.

The F5--G9 relation of \citet{Linsky.et.al.2013ApJ...766...69L},
$\log(F_{\rm Ly\alpha}/F_{\rm CIV})=1.560-0.353\log F_{\rm CIV}$,
with fluxes evaluated at 1\,au, gives
$F_{\rm Ly\alpha}=3.0962\times10^{-12}$\,erg\,s$^{-1}$\,cm$^{-2}$.
We propagate C\,IV errors and the published 22.2\% fractional RMS;
a positive mean-one lognormal is a sampling convention for the latter,
not a published posterior. The independent observable, published
ROSAT-band X flux, gives $3.1331\times10^{-12}$ through the X-ray
relation, with its larger 42.3\% RMS. The agreement supports the
adopted integral without establishing the intrinsic profile.

LWR16313 provides a weak Mg\,II core estimate with native noise.
It has no measured interstellar absorption correction and lies below
the calibrated Mg\,II flux range; it is a diagnostic, not an alternative
precise normalization. The unresolved COS C\,II blend is not substituted
for the isolated 1335.7\,\AA\ calibrator of \citet{Linsky.et.al.2013ApJ...766...69L}.
The adopted Ly$\alpha$ shape remains a conditional FISM2 solar profile
at its native 0.1\,nm sampling, with an active-Sun shape comparison.

For 92--115\,nm we use the \citet{Linsky.et.al.2014ApJ...780...61L} ratio
$F(91.2$--$117\,{\rm nm})/F_{\rm Ly\alpha}=10^{-1.025}$ and its
35\% RMS, with a solar shape normalized on the stated full band.
Only the unobserved subinterval is inserted; COS and EUV data retain
priority elsewhere. In the COS gap, outside the separate Ly$\alpha$
interval, observed continuum sidebands are combined with solar line
excess normalized by C\,IV. These rows, including N\,I, remain
explicitly estimated.

\subsection{Near-UV, optical and infrared}
LWP08049 supplies the 215--330\,nm segment in its large aperture.
The \citet{Bohlin.and.Bianchi.2018AJ....155..162B} correction tables place NEWSIPS on the
CALSPEC scale; the same factors multiply flux and native SIGMA.
Quality-flagged short gaps are interpolated with propagated covariance
and explicit flags. The short blue exposure and noisy SWP continua
are not used to renormalize the well-measured COS spectrum.

The merged CFLIB spectrum alone lacks an uncertainty vector.
Its six individually calibrated exposures retain SIGMA, however
\citep{Valdes.et.al.2004ApJS..152..251V}. We use only those calibrated extensions, avoiding
duplicate count-space versions, and coadd them with flux-conserving
operators. The absolute scale comes from a PHOENIX-NewEra release 3
atmosphere \citep{Hauschildt.et.al.2025A&A...698A..47H}, interpolated in $T_{\rm eff}$,
$\log g$ and metallicity, normalized to the published
$F_{\rm bol}=(4.481\pm0.052)\times10^{-7}$
erg\,s$^{-1}$\,cm$^{-2}$, and matched in the 500--800\,nm integral.
Broad optical continuum windows agree with that model to approximately
2--4\%; this is a calibration consistency check, not an independent
absolute flux measurement. A pure-atmosphere alternative retains the
remaining continuum-shape sensitivity.

NewEra supplies the spectrum beyond 890\,nm. The 330--346\,nm gap
uses its shape constrained by IUE 320--330\,nm and calibrated CFLIB
350--380\,nm anchors. Correlations with both anchors are retained.
The 0.5--5000\,nm integral is
$4.4737\times10^{-7}$\,erg\,s$^{-1}$\,cm$^{-2}$, within 0.2\%
of the bolometric anchor; the wavelength coverage is finite and the
comparison is partly constrained by the normalization itself.

\subsection{Uncertainty representation and reproducibility}
The public SED has one nominal spectrum. A separate signed table
preserves noisy estimates before adjacent, energy-conserving grouping
for nonnegative irradiation use. Group selection is held fixed during
propagation. Surface flux follows $F_{\lambda,\star}
=F_{\lambda,\oplus}(d/R_\star)^2$.

The covariance is represented compactly as
\[
 C=B+UKU^{\mathsf T}
       +\operatorname{Cov}(\Delta_{\rm X/EUV})
       +\operatorname{Cov}(\Delta_{\rm Ly\alpha/FUV}).
\]
$B$ retains native observational and resampling terms, including the
linearized C\,IV-to-proxy dependence. Shared modes include distance,
radius, bolometric normalization and atmospheric-parameter derivatives.
The noisy CFLIB normalization includes negative cross terms in $K$,
as required by the Jacobian of $sf$ with
$s=F_{\rm anchor}/(v^{\mathsf T}f)$; its error is not added as an
independent normalization. Parameter covariances that were not supplied
are not invented. Solar-state and thermal-model changes are named
method sensitivities, not random draws or confidence intervals.

All segments carry the uncertainty components that can be propagated.
Unmeasured stellar-model and absolute-calibration mismatch is distinguished
from those components, rather than described as zero or as a complete
error budget. In nonlinear proxy blocks, the 16th--84th percentile
endpoints need not equal the nominal value plus or minus the covariance
standard deviation. Nonpositive lower endpoints reach the plotting
floor on logarithmic axes; the stored values and linear panels are
unchanged. Numerical checks verify continuous coverage, bin-energy
conservation, covariance symmetry, exported error diagonals and the
surface-flux inverse transformation.

The public package contains the SED, covariance, component and band
tables, provenance, this source and PDF, and an executed portable
plotting notebook. Its default \texttt{SALVAR\_PDF=0} displays figures
without saving them.

\clearpage
{\raggedright

\FloatBarrier
\begin{minipage}{\linewidth}
\subsection{Final spectrum with available uncertainties}
Figure~\ref{fig:sed:lambda_Ser} shows the adopted full SED and its available uncertainty bounds.

\begin{center}
\includegraphics[width=\linewidth]{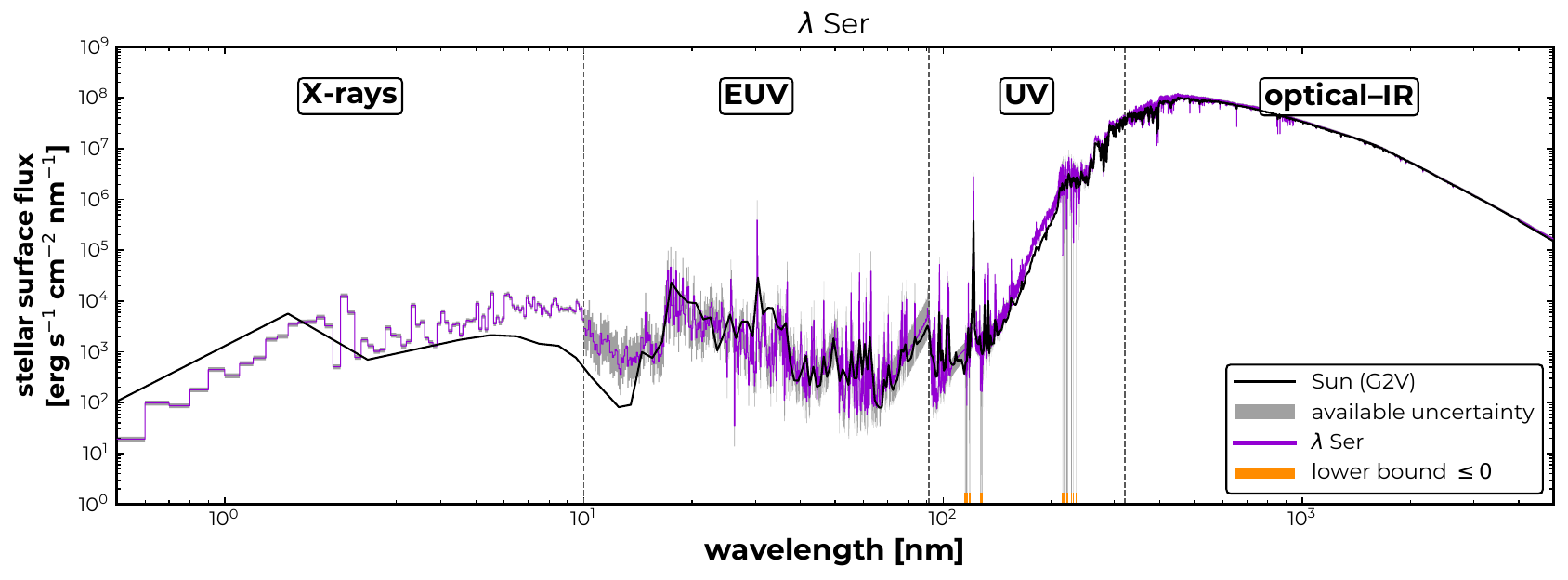}
\end{center}
\captionof{figure}{$\lambda$ Serpentis: final nominal surface-flux spectrum and available uncertainties. Target colours, shading, solar comparison and nonpositive-bound markers follow Section~\ref{sec:display}. The component origins and uncertainty scope are documented above.}\label{fig:sed:lambda_Ser}
\end{minipage}
\FloatBarrier